# Spatial light interference microscopy (SLIM): principle and applications to biomedicine

XI CHEN, MIKHAIL E. KANDEL, AND GABRIEL POPESCU*

*Quantitative Light Imaging Laboratory, Department of Electrical and Computer Engineering, Beckman Institute for Advanced Science and Technology, University of Illinois at Urbana-Champaign, Urbana, Illinois 61801, USA*
*\* gpopescu@illinois.edu*

**Abstract:** In this paper, we review spatial light interference microscopy (SLIM), a common-path, phase-shifting interferometer, built onto a phase-contrast microscope, with white-light illumination. As one of the most sensitive quantitative phase imaging (QPI) methods, SLIM allows for speckle-free phase reconstruction with sub-nanometer path-length stability. We first review image formation in QPI, scattering, holography, and microcopy. Then, we outline SLIM imaging from theory to instrumentation. Zernike's phase-contrast microscopy, phase retrieval in SLIM, and halo removal algorithms are discussed. Next, we discuss the requirements for operation, with a focus on software developed in-house for SLIM that high-throughput acquisition, whole slide scanning, mosaic tile registration, and imaging with a color camera. Lastly, we review the applications of SLIM in basic science and clinical studies. SLIM can study cell dynamics, cell growth and proliferation, cell migration, and mass transport, etc. In clinical settings, SLIM can assist with cancer studies, reproductive technology, and blood testing, etc. Finally, we review an emerging trend, where SLIM imaging in conjunction with artificial intelligence (AI) brings computational specificity and, in turn, offers new solutions to outstanding challenges in cell biology and pathology.© 2020 Optical Society of America under the terms of the OSA Open Access Publishing Agreement

# 1. Introduction

## 1.1. The motivation for label-free imaging

Most biological samples are optically thin and transparent under visible light. They have low contrast under conventional bright-field microscopes. Fluorescent labels are routinely used in combination with optical microscopes to investigate details inside of tissues and cells with high specificity. Although fluorescence microscopy has been used broadly in biomedicine [1], the approach still suffers from important limitations. Phototoxicity and photobleaching modify cellular and fluorophore structures due to the high-intensity illumination, leading to difficulties in live-cell imaging for a long period [2, 3]. Moreover, fluorescent dyes can interfere with cell functions [4]. When multiple fluorescent dyes or proteins are used, the spectra can overlap, making it difficult to distinguish between different structures [5]. Besides, the cost of reagents can add up. Importantly, the fluorescence signal can vary across the specimen when staining is inhomogeneous, resulting in difficulty interpreting the images quantitatively [6]. Finally, whenever using genetic engineering, the transfection process can be complex and time-consuming [7].

Complementary to fluorescence microscopy, label-free imaging is non-destructive and, of course, lacks photobleaching [8]. Imaging unlabeled specimens requires minimal or no sample preparation and provides detailed dynamic and morphological information in live cells [9]. The cells are in their intact, native states, leading to more biologically-relevant studies. Label-free imaging is suitable for long-period live-cell imaging without photobleaching. Multiple cellular characteristics can be measured repeatedly over time in longitudinal studies [10-12]. These capabilities open new opportunities to study long-term cellular events such as proliferation and response to chemical stimulations.

## 1.2. Quantitative phase imaging

Quantitative phase imaging (QPI) is an emerging field built on the foundation of microscopy, holography, and scattering techniques [13-16]. The understanding of the microscopic image as a complicated interferogram established by Abbe has been essential for the development of microscopic techniques, such as phase-contrast microscopy and differential-interference microscopy [17-19]. These techniques greatly enhance the intrinsic contrast without labels and enable insight into the transparent structures. The invention of digital holography opened the door to storing the phase information of optical fields. Unlike the original implementation of holography, which was developed to record the intensity distribution in such a way as to preserve the phase information, QPI is aimed at quantitatively rendering the pure phase distribution, eliminating the intensity dependence [20]. One feature of QPI is the nanoscale sensitivity, enabling important studies of cellular morphology, cell membrane fluctuations, drug response, etc [21-24]. The phase information associated with the sample depends on critical parameters, such as the refractive index (RI) and dry mass of the specimen [25, 26].

The QPI modalities can be divided into interferometric and non-interferometric, depending on whether an interferometer is involved in the phase measurement [27]. The precursors of interferometric phase measurements were performed via single-point, scanning techniques, such as optical coherence tomography (OCT) [28, 29]. Later on, full-field QPI methods were developed based on spatial phase modulation and temporal phase modulation, i.e., off-axis and phase-shifting interferometers respectively (chapter 2) [30-32]. Non-interferometric phase measurements include wavefront sensing, such as the Shack-Hartmann wavefront sensor, which is broadly employed in the adaptive optics field [33, 34]. Other non-interferometric methods include phase retrieval techniques using iterative methods or deterministic methods [35-37]. The well-known iterative methods are Gerchberg-Saxton (GS) algorithm and ptychography [38, 39]. A special case of QPI using deterministic methods of phase retrieval is based on the transport of the intensity equation. In this case, the phase information can be retrieved from the axial gradient of the intensity [40]. Note that, although "non-interferometric" methods lack an interferometer, they still use interference of light as the fundamental process for recording the phase information. For example, the local gradient of the wavefront is captured via the Shack-Hartmann sensor by recording the superposition (interference) of the waves emerging at each aperture. The computational phase retrieval methods exploit the fact that an image is an interferogram. Similarly, the techniques based on the transport of intensity equation, treat the image field as the interference between the incident and scattered field, at several positions around the plane of focus. Interestingly, we can describe the field at each point in the image as the interference between the scattered field and the incident field, which acts as a common reference for a highly parallel interferometry system. This description is fundamental for understanding Zernike's phase-contrast microscopy (section 3.1.1) and spatial light interference microscopy (SLIM) (3.1.2), which is a generalization of this method.

### 1.3. Other label-free methods

Multiphoton microscopy is a nonlinear method widely employed in the biomedicine field, especially for imaging bulk tissues [41]. It includes several label-free methods such as second-harmonic generation microscopy (SHGM) [42], third-harmonic generation microscopy (THGM) [43], and coherent Raman scattering microscopy (CRSM) [44]. The contrast in the SHGM and THGM comes from the variations in a sample's ability to generate harmonics, i.e., $\chi^{(2)}$ and $\chi^{(3)}$ properties. The contrast in CRSM depends on the Raman-active vibrational modes of molecules in the sample. Stimulated Raman scattering (SRS) and coherent anti-Stokes Raman scattering (CARS) are two major techniques in CRSM [45]. Multiphoton microscopy has a large number of applications in cancer studies, cell metabolism, and pharmaceutical research [46-49].

Fluorescence-lifetime imaging microscopy (FLIM) measures the lifetime associated with the fluorophore from a sample, and, in the autofluorescence case, is also a label-free method [50]. The Fluorescence lifetime depends on the micro-environment of the fluorophore, thus, it is very sensitive to pH, chemical species, and viscosity [51-53]. Two-photon microscopy can measure autofluorescence with living tissues up to about 1 mm in thickness [54]. Fourier transform IR (FTIR) spectroscopy is another label-free method that allows for spectroscopic imaging via interferometric imaging [55]. It has found a variety of biological and clinical applications [56, 57].

Imaging techniques such as confocal microscopy and light-sheet microscopy aimed at improving the optical sectioning and larger frequency support can also be used as label-free methods, either using scattered fields or intrinsic fluorophores [58, 59]. Diffuse optical imaging is a label-free method using near-infrared spectroscopy for diffusive samples [60].

Photoacoustic tomography (PAT) combines sound waves and electromagnetic waves to create multiscale, multi-contrast images of biological samples [61]. The penetration depth can go beyond the optical transport mean free path due to the photoacoustic effect, thus enabling imaging from subcellular organelles to organ scales [62].

### 2. Principles of QPI

### 2.1. Scattering

### 2.1.1. First-order Born approximation

The intrinsic contrast generated in QPI is due to light scattering. Scattering is the general term that describes the interaction between a field and the real part of the dielectric permittivity [63]. While the processes involved in QPI are linear, the term scattering includes nonlinear phenomena, such as second-harmonic generation (SHG). In this section, we will restrict ourselves to situations where the response of the object to the incident field is linear and static [64]. In general, solving the wave equations for an arbitrary inhomogeneous object is difficult, with no analytic solutions. However, here we show that, with weak scattering approximation, or the first-order Born approximation, analytic solutions can be obtained. The weak scattering regime occurs wherever the object's refractive index (RI) is very close to the background's RI. In this case, we can derive an expression of the far-zone scattered field using the first-order Born approximation [65].

The light propagation in the medium is governed by the Helmholtz equation

$$\nabla^2 U(\mathbf{r},\omega) + \beta_0^2 U(\mathbf{r},\omega) = -4\pi F(\mathbf{r},\omega) U(\mathbf{r},\omega), \tag{2.1.1}$$

where $\beta_0 = \omega/c$ is the wavenumber in vacuum. $F(\mathbf{r},\omega)$ is the scattering potential defined as

$$F(\mathbf{r},\omega) = \frac{1}{4\pi} \beta_0^2 \left[ n^2(\mathbf{r},\omega) - 1 \right], \tag{2.1.2}$$

where $n$ is the refractive index. Note that, in Eq. (2.1.1), the total field $U$, is present on both of the equation, indicating that any scattered field can be scattered again, generating multiple scattering, and acting as a secondary source. The first-order Born approximation assumes that the field inside of the object is only slightly different from the incident field. This weak scattering approximation dramatically simplifies the work, as it allows us to replace $U$ by $U_i$ on the right-hand side of Eq. (2.1.1). The total field under the first-order Born approximation can be calculated as [65]

$$\begin{aligned} U(\mathbf{r},\omega) &= U_i(\mathbf{r},\omega) + U_s(\mathbf{r},\omega) \\ &\approx U_i(\mathbf{r},\omega) + \int_V F(\mathbf{r}',\omega) U_i(\mathbf{r}',\omega) \frac{e^{i\beta_0 |\mathbf{r}-\mathbf{r}'|}}{|\mathbf{r}-\mathbf{r}'|} d^3 r' \end{aligned} \tag{2.1.3}$$

If we assume the incident field as a plane wave, i.e., $U_i(\mathbf{r},\omega) = e^{i\boldsymbol{\beta}_i \cdot \mathbf{r}}$, and the measurement performs in the far-zone, the scattered field can be further simplified using Fraunhofer approximation, i.e., $|\mathbf{r}-\mathbf{r}'| \simeq r - \mathbf{r} \cdot \mathbf{r}'/r$, as [65],

$$\begin{aligned} U_s(\mathbf{r},\omega) &= \frac{e^{i\beta_0 r}}{r} f(\mathbf{q},\omega) \\ &= \frac{e^{i\beta_0 r}}{r} \int_V F(\mathbf{r}',\omega) e^{-i\mathbf{q}\cdot\mathbf{r}'} d^3 r', \end{aligned} \tag{2.1.4}$$

where $\mathbf{q} = \boldsymbol{\beta}_s - \boldsymbol{\beta}_i$ is the momentum transfer, $\boldsymbol{\beta}_i$ and $\boldsymbol{\beta}_s$ are incident and scattered wave vectors. $f(\mathbf{q},\omega)$ is the scattering amplitude. We can see that the scattering amplitude along a certain scattering direction depends entirely on one and only one Fourier component of the scattering potential, and the scattered field behaves as modulated spherical waves. The scattering potential can be recovered by the inverse Fourier transform of $f(\mathbf{q},\omega)$, i.e.,

$$F(\mathbf{r},\omega) = \int_{V_q} f(\mathbf{q},\omega) e^{i\mathbf{q}\cdot\mathbf{r}} d^3q. \tag{2.1.5}$$

However, the q-domain integration is limited by the Ewald scattering sphere, defined as

$$\Pi[q/(4k_0)] = \begin{cases} 1, \sqrt{q_x^2 + q_y^2 + q_z^2} \leq 2k_0 \\ 0, else \end{cases}$$, whereby the highest possible $q = 2k_0$ is obtained for backscattering. Thus, the reconstructed object from far-zone scattered-field measurement is a low-frequency bandpass version of the true object. Moreover, covering the entire Ewald sphere depends on illuminating the object from all directions and measuring the complex scattered field over the entire solid angle for each illumination direction. This implies that the ideal reconstruction modality requires $4\pi$ illumination and detection [66].

### 2.1.2. Physical significance of phase in transmission & reflection geometries

The phase information obtained by QPI is different in various imaging modalities. In this section, we discuss the interpretation of the phase in transmission and reflection geometries [67]. Let us assume the simplest case where the incident field is a monochromatic plane wave propagating along z, $U_i(\omega) = A(\omega)e^{in_0\beta_0 z}$, where $A(\omega)$ is the spectral amplitude and $n_0$ is the RI of the background (Fig. 1). The wave equation under the first-order Born approximation is

$$\nabla^2 U_s(\mathbf{r},\omega) + n_0^2\beta_0^2 U_s(\mathbf{r},\omega) = -\beta_0^2 \chi(\mathbf{r},\omega) U_i(\mathbf{r},\omega), \tag{2.1.6}$$

where $\chi(\mathbf{r},\omega) = n^2(\mathbf{r},\omega) - n_0^2$ and $n$ is the RI of the object. Taking the 3D Fourier transform of Eq. (2.1.6), we obtain

$$(\beta^2 - k^2)U_s(\mathbf{k},\omega) = -\beta_0^2 A(\omega)\chi(k_\perp, k_z - \beta, \omega), \tag{2.1.7}$$

where $\beta = n_0\beta_0$ and $U_s(\mathbf{k},\omega)$ is the Fourier transform of $U_s(\mathbf{r},\omega)$ with respect to $r$, $k_\perp$ is the transverse spatial frequency. The scattered field in the wavevector space is thus in the form of

$$U_s(\mathbf{k},\omega) = -\beta_0^2 A(\omega)\chi(k_\perp, k_z - \beta, \omega)\frac{1}{2\gamma}\left[\frac{1}{\gamma - k_z} + \frac{1}{\gamma + k_z}\right], \tag{2.1.8}$$

where $\gamma = \sqrt{\beta^2 - k_\perp^2}$. Let us next take the inverse Fourier transform with respect to $k_z$, we have

$$\begin{aligned} U_s(k_\perp, z, \omega) &= -i\beta_0^2 A(\omega)\frac{e^{i\gamma z}}{2\gamma}\chi(k_\perp, \gamma - \beta, \omega)\bigg|_{z\geq 0} \\ &\quad + i\beta_0^2 A(\omega)\frac{e^{i\gamma z}}{2\gamma}\chi(k_\perp, -\gamma - \beta, \omega)\bigg|_{z<0} \\ &= U^+(k_\perp, z, \omega) + U^-(k_\perp, z, \omega). \end{aligned} \tag{2.1.9}$$

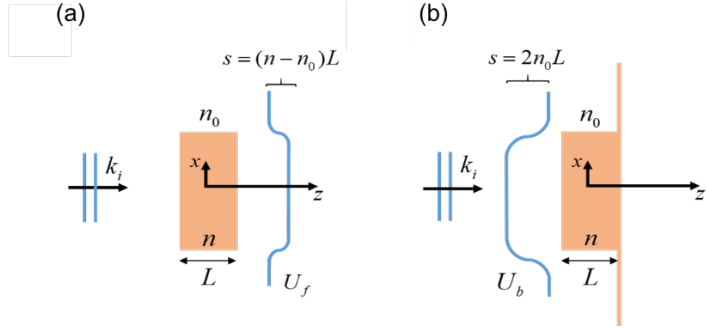

Figure 1 Wavefront changes due to a medium of thickness L and refractive index n for a transmission (a) and reflection (b) measurement.

Here $U^+$ and $U^-$ denotes the forward and backscattering fields. Thus, the forward and back total fields can be calculated under the first-order Born approximation

$$U_f(k_\perp,z,\omega) = A(\omega)\delta(k_\perp)e^{i\beta z} \\ -i\beta_0^2 A(\omega)\frac{e^{i\gamma z}}{2\gamma}\chi(k_\perp,\gamma-\beta,\omega), \quad (2.1.10)$$

$$U_b(k_\perp,z,\omega) = A(\omega)\delta(k_\perp)e^{-i\beta z} \\ +i\beta_0^2 A(\omega)\frac{e^{-i\gamma z}}{2\gamma}\chi(k_\perp,-\gamma-\beta,\omega), \quad (2.1.11) \\ -A(\omega)\delta(k_\perp)e^{-i\beta z}$$

where $\delta(k_\perp)$ is Kronecker delta. To get an intuitive expression for the phase of the total field and connect this quantity to QPI measurements, we limit $k_\perp$ to a small region close to 0, i.e., make a small-angle approximation. Thus, $\gamma = \sqrt{\beta^2 - k_\perp^2} \approx \beta$. As a result, the forward and backscattered fields (Fig. 1) can be approximated as

$$U_f(\mathbf{r}_\perp,z,\omega) = A(\omega)e^{i\beta z}e^{-i\beta_0\left[\bar{n}(\mathbf{r}_\perp,\omega)-n_0\right]L}, \quad (2.1.12)$$

$$U_b(\mathbf{r}_\perp,z,\omega) = A(\omega)e^{-i\beta z}e^{+i\beta_0\int_{-L/2}^{L/2}\left[n(\mathbf{r}_\perp,z,\omega)-n_0\right]e^{i2\beta z}dz} - A(\omega)e^{-i\beta z}. \quad (2.1.13)$$

We can see that in the forward direction, the approximated phase of the total field is the well-known geometrical phase delay $\varphi(x,y) = \beta_0\left[\bar{n}(x,y) - n_0\right]L$. However, the phase of the backscattered field is more complicated, containing two terms. The first term shows the axial projection of the RI contrast weighted by the plane wave $e^{i2\beta z}$. Ignoring transverse features in the object, this expression indicates that the field detected in backscattering consists of a superposition of back-propagating plane waves originating at various depths, z, with respective phases $2\beta z$. The second term is the back-propagating incident field without interacting with the object. This axial integral can be expressed in terms of a z-axis Fourier transform as

$$\phi^-(\mathbf{r}_\perp, z) = \beta_0 \int_{-\infty}^{\infty} \left[ n(\mathbf{r}_\perp, z, \omega) - n_0 \right] \Pi\left(\frac{2z}{L}\right) e^{i2\beta z} dz$$
$$= \beta_0 L \Delta n(\mathbf{r}_\perp, k_z, \omega) \circledv \mathrm{sinc}\left(\frac{L k_z}{2}\right)\bigg|_{k_z = -2\beta}, \quad (2.1.14)$$

where $\Pi\left(\frac{2z}{L}\right)$ is the rectangular function of width L, and $\circledv$ is the convolution operator in the $k_z$ domain. The phase of the scattered field depends on the convolution of RI at the axial frequency $-2\beta$ with a sinc function. The oscillatory behavior leads to speckles in the backscattered quantitative phase images, which relates to the object structure in an intricate manner. In summary, we can approximate the phase in the forward and backward fields as

$$\varphi_f(x, y) = \beta_0 \Delta n L, \quad (2.1.15)$$

$$\varphi_b(x, y) = \arg\left(e^{i\phi^-} - 1\right). \quad (2.1.16)$$

The discussion above is only an approximation of the phase for coherent plane waves but gives us a general interpretation of the phase of the fields in a reflective imaging modality. The contributions to the phase of the total fields include the double transmitted light, backscattered light, and multi-scattered back-propagating light. Using oblique partially coherent illumination or adding a reflective surface on the bottom of the object in epi QPI can minimize the contribution of the back-scattered light and, therefore, enhance the contribution of the double transmitted light [68, 69]. However, in this case, the benefit of capturing high frequencies from the object is lost. In sum, extracting quantitative phase information in a backscattering geometry remains challenging. On the one hand, reflection QPI requires developing techniques for separating the multiple scattering contributions to the phase of the detected field. On the other hand, it needs the theoretical interpretation that includes the coherence properties of the fields [70]. The use of the broadband partially coherent illumination in a reflective imaging modality, such as epi-illumination gradient light interference microscopy (epi-GLIM) [71], can reduce the speckles in phase images, as it provides strong coherence sectioning, of the order of 1 $\mu m$. More generally, white-light interferometry provides optical gating, which minimizes multiple scattering contributions. The scattering of broadband light is discussed next.

### 2.1.3. Scattering of spatiotemporally broadband fields

The coherence properties of light play an important role when working with spatiotemporally broadband source [70, 72-74]. The assumption of the deterministic plane wave is no longer valid. The randomness in the primary sources and propagation media determine the statistical properties of the detected quantities [75-78]. Two important correlation functions to characterize the coherence properties of the fields are the cross-spectral density and mutual coherence function defined as respectively, [70]

$$W(\mathbf{r}_1, \mathbf{r}_2, \omega) = \langle U^*(\mathbf{r}_1, \omega) U(\mathbf{r}_2, \omega) \rangle, \quad (2.1.17)$$

$$\Gamma(\mathbf{r}_1,\mathbf{r}_2,\tau) = \left\langle U^*(\mathbf{r}_1,\tau)U(\mathbf{r}_2,r+\tau)\right\rangle_t, \tag{2.1.18}$$

where the ensemble average is taken over all the different realizations of the fields, the star denotes the conjugate part. According to the generalized Wiener-Khintchine theorem, two functions are Fourier transform pairs:

$$\Gamma(\mathbf{r}_1,\mathbf{r}_2,\tau) = \int_0^\infty W(\mathbf{r}_1,\mathbf{r}_2,\omega)e^{-2\pi i\omega\tau}d\omega. \tag{2.1.19}$$

$$W(\mathbf{r}_1,\mathbf{r}_2,\tau) = \int_{-\infty}^\infty \Gamma(\mathbf{r}_1,\mathbf{r}_2,\omega)e^{2\pi i\omega\tau}d\tau. \tag{2.1.20}$$

For isotropic, statistically homogeneous sources, the correlation function will only depend on the difference of the two vectors $\mathbf{r}_1$ and $\mathbf{r}_2$. The cross-spectral density of the incident field is

$$W_{ii}(\mathbf{r}_1-\mathbf{r}_2,\omega) = \left\langle U_i^*(\mathbf{r}_1,\omega)U_i(\mathbf{r}_2,\omega)\right\rangle, \tag{2.1.21}$$

Under the first-order Born approximation, the correlation of the scattered fields in the far-zone becomes

$$\left\langle U_s^*(r\hat{\mathbf{k}}_s,\omega)U_s(r\hat{\mathbf{k}}_s,\omega)\right\rangle = \frac{k^4}{r^2}V\int_V \left(n^2(r\hat{\mathbf{k}}_s,\omega)-1\right)\left(n^2(r\hat{\mathbf{k}}_s+\mathbf{R},\omega)-1\right)W_{ii}(\mathbf{R},\omega)e^{-i\mathbf{k}_s\cdot\mathbf{R}}d^3R, \tag{2.1.22}$$

where $\mathbf{R}=\mathbf{r}_2-\mathbf{r}_1$, $\hat{\mathbf{k}}_s$ is the unit scattered wave vector, and $V$ is the volume of the scatterer. The correlation between the incident and scattered fields is

$$\left\langle U_i^*(\mathbf{r},\omega)U_s(\mathbf{r},\omega)\right\rangle = \int_V F(\mathbf{r}',\omega)W_{ii}(\mathbf{r}',\omega)\frac{e^{ik_0|\mathbf{r}-\mathbf{r}'|}}{|\mathbf{r}-\mathbf{r}'|}d^3r' \tag{2.1.23}$$

The propagation of the random fields is governed by the correlation propagation equations, known as the Wolf equations [65],

$$\nabla_1^2 \Gamma(\mathbf{r}_1,\mathbf{r}_2,\tau) = \frac{1}{c^2}\frac{\partial}{\partial\tau^2}\Gamma(\mathbf{r}_1,\mathbf{r}_2,\tau), \tag{2.1.24}$$

$$\nabla_1^2 W(\mathbf{r}_1,\mathbf{r}_2,\omega) + k^2 W(\mathbf{r}_1,\mathbf{r}_2,\omega) = 0. \tag{2.1.25}$$

$\nabla_1^2$ is the Laplacian operator with respect to the position $\mathbf{r}_1$. We can see that the propagation of the correlation functions is similar to the deterministic case when measuring at two independent points. However, if the two points of interest are not independent, the propagation of the correlation functions is more complicated, due to the two extra terms from the Laplacian operator [79]. To study broadband light propagation into biological samples or dynamic live-cell scattering, the statistical coherence theory is required to retrieve accurate results [80, 81].

## 2.2. Holography

### 2.2.1. Gabor (in-line) Holography

Compared to conventional imaging, holographic imaging captures both the intensity and the phase information of light, using interference. Dennis Gabor first introduced the principle of holography in 1948 [82], involving two steps, writing and reading the holography as illustrated in Fig. 2.

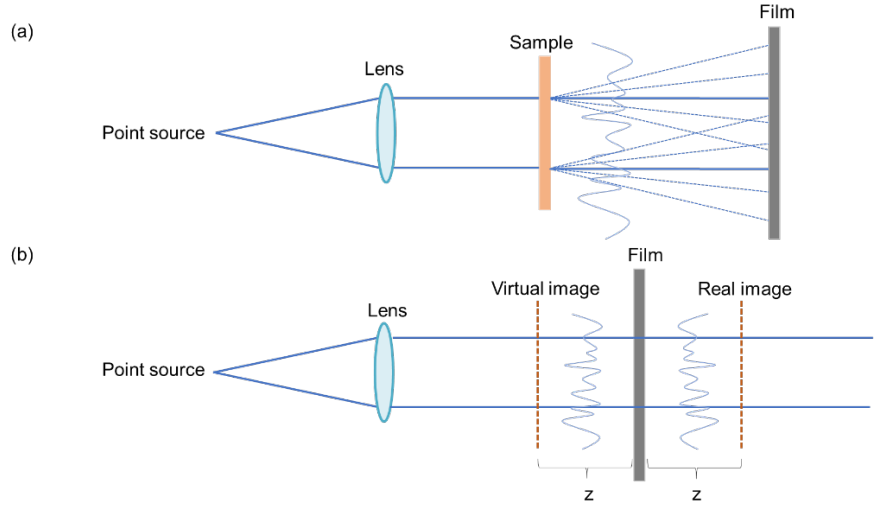

Figure 2 (a) Diagram for in-line writing a Fresnel hologram. (b) Reading an in-line hologram.

A point source is placed at the focal point of the lens illuminating the object. The film records the Fresnel diffraction of the total field after the object. The intensity of this interference pattern can be expressed as

$$I(x,y) = |U_0 + U_1(x,y)|^2$$
$$= |U_0|^2 + |U_1(x,y)|^2 + U_0 U_1(x,y) + U_0 U_1^*(x,y) \quad (2.2.1)$$

If the response of the detecting film is linear, we can express its transmission function as

$$t(x,y) = a + bI(x,y), \quad (2.2.2)$$

where $a$ and $b$ are constants. The film records the transmission function, which contains complete information about the field.

The next step is reading the hologram. Illuminating the film with the same field $U_0$, the scattered field is the product between the incident field and the transmission function,

$$U(x,y) = U_0 t(x,y)$$
$$= U_0 \left(a + b|U_0|^2\right) + b U_0 |U_1(x,y)|^2 + b|U_0|^2 U_1(x,y) + b|U_0|^2 U_1^*(x,y) \quad (2.2.3)$$

The first term is a constant. The second term, for weak scattering, is negligible compared to the other terms. The last two terms refer to the real and virtual images generated by the hologram, resembling the original object field $U_1(x,y)$ or $U_1^*(x,y)$. However, the two images are both along the optical axis, forming "twin" images, degrading the signal to noise of the reconstruction. To get rid of the DC term and the twin term, several techniques were developed such as high pass filters, combinations of two or more holograms with stochastic changes in the object speckles, and phase-shifting holography [83]. Importantly, off-axis holography can overcome this problem, as discussed in the next section [84, 85].

### 2.2.2. Leith and Upatnieks' (off-axis) holography

The principle of writing and reading an off-axis hologram is illustrated in Fig. 3. The object is illuminated by a plane wave $U_0$, the transmitted fields $U$ are Fresnel propagating to the film at distance z. The Fresnel diffraction pattern, $U_F$, is a convolution between the transmitted fields and the Fresnel diffraction kernel (prefactors are omitted.)

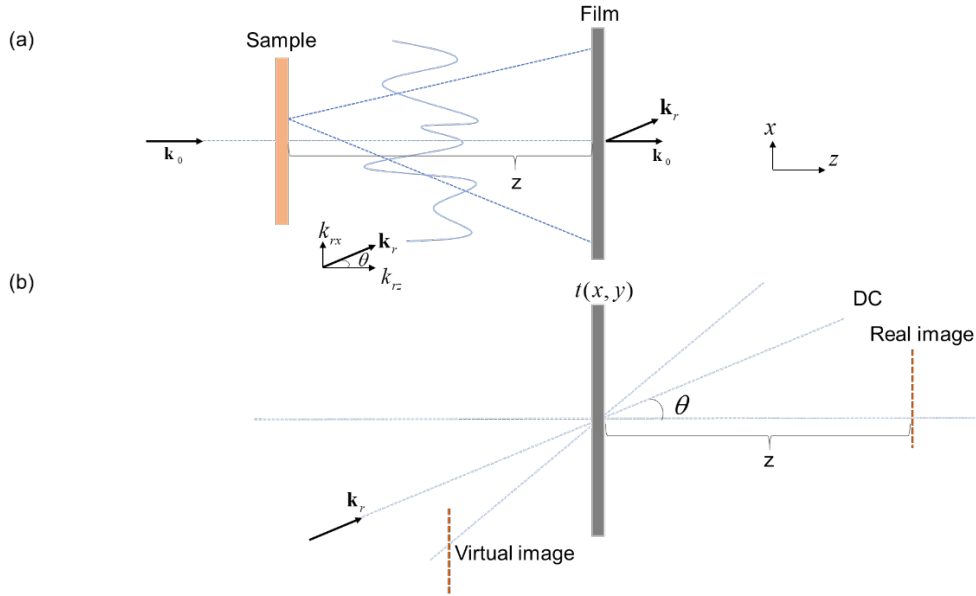

Figure 3 (a) Writing an off-axis hologram. (b) Reading the off-axis hologram.

$$U_F(x,y) = U(x,y) \otimes e^{\frac{ik_0(x^2+y^2)}{2z}} \qquad (2.2.4)$$

The reference field $U_r$ is an off-axis plane wave with the angle $\theta$ with respect to the z-axis. The total filed arriving at the film can be expressed as

$$U_t(x,y) = U_F + |U_r| e^{i(k_{rx}x + k_{rz}z)}, \qquad (2.2.5)$$

where $k_{rx} = k_0 \sin\theta$ and $k_{rz} = k_0 \cos\theta$. Since the z-component of the reference field provides a constant shift $k_{rz}z$ at the film plane, we will omit this constant. The total transmission function associated with the hologram is proportional to the intensity and can be written as

$$t(x,y) = |U_F(x,y)|^2 + |U_r|^2 + U_F(x,y)|U_r|e^{-ik_{rx}x} + U_F^*(x,y)|U_r|e^{ik_{rx}x}. \tag{2.2.6}$$

For the reading process, a reference plane wave $U_r$ illuminates the hologram at the same angle $\theta$. The total field at the film plane becomes

$$\begin{aligned} U_h(x,y) &= |U_r|e^{i\mathbf{k}_r\cdot\mathbf{r}}t(x,y) \\ &= |U_F(x,y)|^2 |U_r|e^{i\mathbf{k}_r\cdot\mathbf{r}} + |U_r|^3 e^{i\mathbf{k}_r\cdot\mathbf{r}} + U_F(x,y)|U_r|^2 + U_F^*(x,y)|U_r|^2 e^{i2k_{rx}x}. \end{aligned} \tag{2.2.7}$$

The first two terms are the fields propagating along the direction $\mathbf{k}_r$, i.e., the DC signals. The third term contains the information of the complex field $U_F$, meaning the observer can see the object along the axis z. The last term, involving the conjugate object field is propagating with twice the frequency $k_{rx}$ along x, indicating the observer can see a virtual image of the object along the direction with a larger angle than $\theta$. In this way, the two images of the object are conveniently separated (Fig. 3(b)).

### 2.2.3. Digital holography

Conventional, analog holography cannot be performed in real-time due to the cumbersome physical and chemical process, largely limiting the applications. Digital holography replaces the photosensitive film with digital writing using a camera, and the optical reading process with numerical reconstruction based on diffraction theory [20].

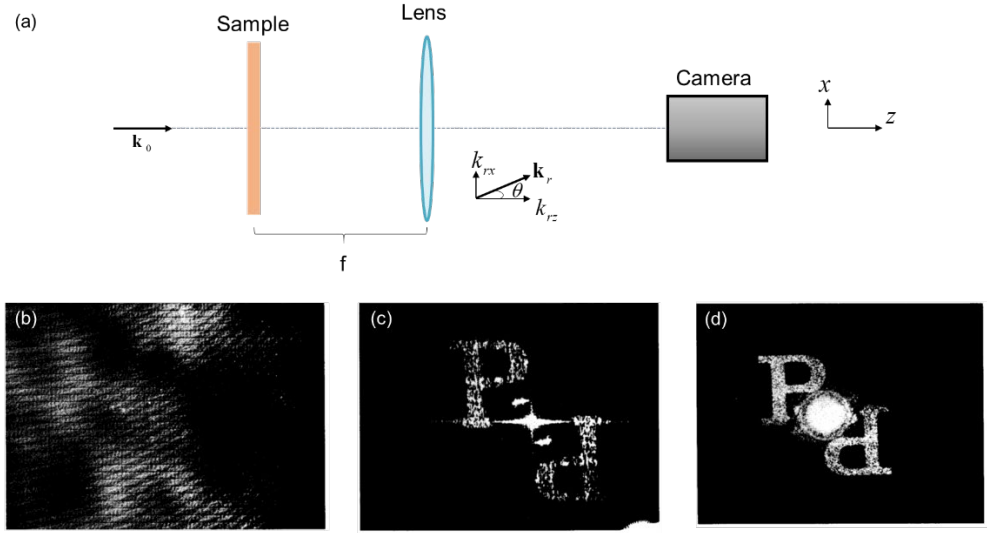

Figure 4 (a)Digital recording of a Fourier hologram. (b) Hologram stored in computer memory. (c) Image formed by digital computation. (d) Image obtained optically from a photographic hologram. From "Goodman, Joseph W., and R. W. Lawrence. "Digital image formation from electronically detected holograms." Applied physics letters 11.3 (1967)"

The digital recording process is illustrated in Fig. 4. The sample is illuminated by a plane wave. The field after the sample $U$ is Fourier transformed by the lens at its back focal plane, where the camera is located. The off-axis reference field, $U_r$, is incident on the camera at an angle $\theta$. The total field on the camera, $U_h$, assumed to be at the Fourier plane of the lens, is

$$U_h(x',y') = \tilde{U}(k_x,k_y) + |U_r|e^{i\mathbf{k}_r \cdot \mathbf{r}'},$$
$$k_x = 2\pi x'/\lambda f; \quad k_y = 2\pi y'/\lambda f;$$
(2.2.8)

where $\tilde{U}$ is the Fourier transform of $U$. The intensity on the camera is in the form

$$I_H(x',y') = |\tilde{U}(k_x,k_y)|^2 + |U_r|^2 + \tilde{U}(k_x,k_y)|U_r|e^{-ik_{rx}x'} + \tilde{U}^*(k_x,k_y)|U_r|e^{ik_{rx}x'}.$$
(2.2.9)

After removing the DC signal (first two terms), the field after the sample $U$ can be obtained by the numerical inverse Fourier transform as illustrated in Fig. (c). Notice that the last two terms are shifted symmetrically with respect to the origin in the camera plane. If the camera is placed at an arbitrary distance from the lens, the Fresnel propagation transformation is used instead of the Fourier transform [86].

## 2.3. Full-field QPI methods

### 2.3.1. Spatial phase modulation: off-axis interferometry

Quantitative phase information can be retrieved via spatial phase modulation or temporal phase modulation. In this section, we discuss the first case, where phase modulation is performed by an off-axis reference wave. The experimental setup is shown in Fig. 5 in the section of off-axis holography. The intensity at the detector is

$$I(x,y) = |U_r|^2 + |U_i(x,y)|^2 + 2|U_r||U_i(x,y)|\cos[k_{rx}x + \phi(x,y)].$$
(2.3.1)

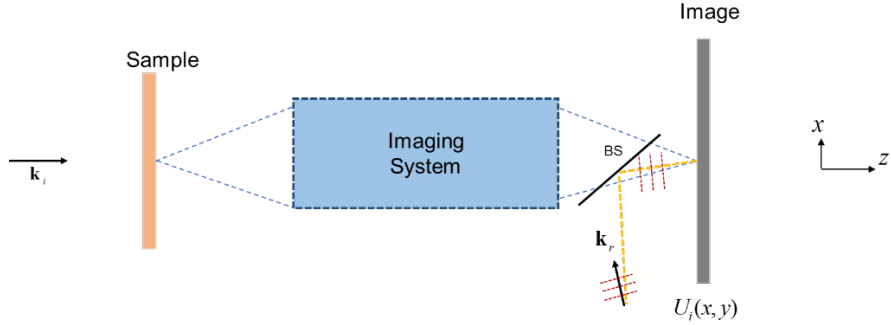

Figure 5 Diagram for off-axis interferometry. BS beam splitter.

Applying a high-pass filter in the frequency domain can remove the DC term, thus the *cosine* term can be obtained, which is the real part of the complex correlation function. Hilbert transform of the real part yields the imaginary part of the correlation function

$$\sin[k_{rx}x + \phi(x,y)] = P\int \frac{\cos[k_{rx}x' + \phi(x',y)]}{x-x'}dx', \quad (2.3.2)$$

where $P$ indicates the principal value integral. The highly wrapped phase can be retrieved as the argument of the complex correlation function

$$\phi(x,y) + k_{rx}x = \arg[\cos(k_{rx}x+\phi), \sin(k_{rx}x+\phi)]. \quad (2.3.3)$$

The $k_{rx}$ term can be calculated with the known reference tilt angle. Thus, the final phase map $\phi$ is obtained by subtracting the modulation frequency term. Conventional off-axis interferometry has lower space-bandwidth coverage, meaning that either the resolution or the field of view must be compromised. However, slight off-axis interferometry has the problem of overlapping of the DC and AC signals. Techniques such as introducing a second color in the interferometry can resolve this problem [87].

### 2.3.2. Temporal phase modulation: phase-shifting interferometry

On-axis interferometry offers temporal phase shifts between the object field and the reference field, which preserves the space-bandwidth product, at the expense of the time-bandwidth product. [88]. In Fig. 6, the intensity on the detector can be expressed as

$$I(\bar{\omega}\tau) = |U_r|^2 + |U_i(x,y)|^2 + 2|U_r||U_i(x,y)|\cos[\bar{\omega}\tau + \phi(x,y)], \quad (2.3.4)$$

where $\bar{\omega}$ is the central frequency of the reference field. To retrieve the information of $\phi$, one can control $\bar{\omega}\tau$ to be different values and solve the equations for $\phi$. Most commonly phase-shifting methods use four phase shifts with the increment of $\pi/2$. Therefore, the phase becomes

$$\phi = \arg[I(0) - I(\pi), I(3\pi/2) - I(\pi/2)]. \quad (2.3.5)$$

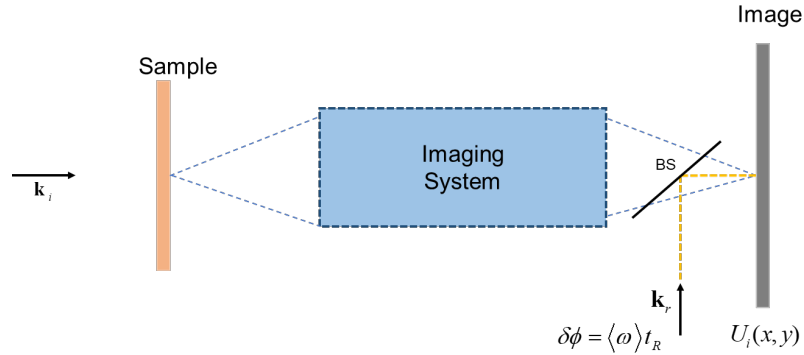

Figure 6 Diagram for phase-shifting interferometry. BS beam splitter.

Phase-shifting QPI methods have demonstrated their capability for biological studies [89, 90]. The imaging modalities using phase-shifting interferometry include Fourier phase microscopy (FPM) [91], spatial light interference microscopy (SLIM) [92], and optical quadrature microscopy [93], etc.

### 2.3.3. QPI figures of merit

The main figures of merit of QPI are acquisition rate, transverse resolution, temporal phase sensitivity, and spatial phase sensitivity [15]. The acquisition rate of QPI depends on the modality used for phase retrieval. The single-shot measurement is only limited by the camera, which can exceed 1,000 frames per second at megapixel resolution. In this context, phase-shifting methods are slower since more frames are required, although possible approaches to achieve framerate parity include piezoelectric or electro-optical modulators.

Defining a proper measurement of transverse resolution in QPI is not trivial. It needs to consider the coherence properties of the system and also relies on different QPI methods. The common-path modality and the phase-shifting interferometry is more likely to preserve the diffraction-limited resolution. Off-axis methods reduce the information content of the hologram to about one-quarter of the pixel count, resulting in the lower transverse resolution in the phase.

To assess the temporal stability experimentally, one can perform successive measurements of no-sample images. The histogram of the optical path length (OPD) can be obtained for the entire stacks of data, which yields the standard deviation of the data, defined as.

$$\sigma_t = \sqrt{\left\langle \left[ \delta\phi(t) - \left\langle \delta\phi(t) \right\rangle_t \right]^2 \right\rangle_t}, \qquad (2.3.6)$$

where $\delta\phi(t)$ is the temporal phase fluctuation. Another way to describe the temporal phase noise is the temporal power spectrum by computing the Fourier transform of the no-sample stacks along time t

$$\left| \delta\phi(\omega) \right|^2 = \left| \int \delta\phi(t) e^{i\omega t} dt \right|^2. \qquad (2.3.7)$$

Similarly, the spatial phase sensitivity can also be calculated by taking no-sample images. The standard deviation for the entire field of view is defined as

$$\sigma_r = \sqrt{\left\langle \left[ \delta\phi(x,y) - \left\langle \delta\phi(x,y) \right\rangle_{x,y} \right]^2 \right\rangle_{x,y}}. \qquad (2.3.8)$$

Analog to the temporal power spectrum, the spatial power spectrum has the expression

$$\left| \delta\phi(x,y) \right|^2 = \left| \iint_A \delta\phi(x,y) e^{i(k_x x + k_y y)} dxdy \right|^2. \qquad (2.3.9)$$

Fig. 7 shows the histogram of the OPD in the spatial and temporal domain. One can get the temporal and spatial standard deviations by fitting the Gaussian curves. The spatiotemporal power spectrum is illustrated in Fig. 7 [94]. If the signal of interest lies in a certain frequency band, filtering can be used to significantly improve the signal to noise ratio (SNR) of the measurement.

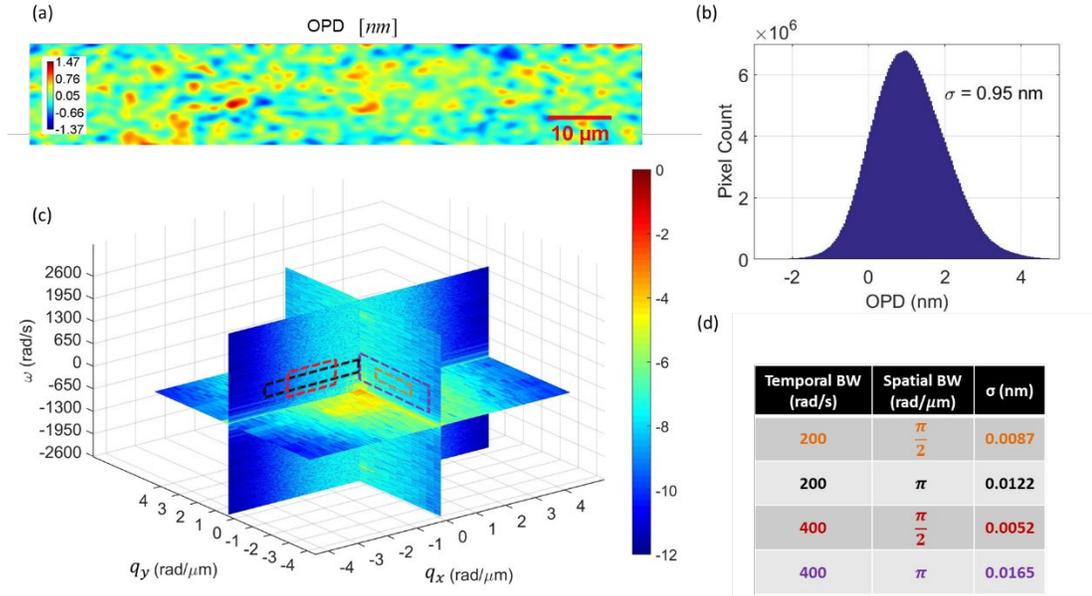

Figure 7 Analysis of spatiotemporal stability of the MISS microscopy system. (a) A 256 x 1500 pixels no-sample OPD image with color bar in nm. (b) The histogram of the OPD stack acquired at 833 fps. (c) Plot showing the noise content of each spatial and temporal frequency component along three different planes in 3d frequency space. Color bar is in log scale with units of nm2/(rad2/um2).(rad/s). (d) Band-pass filtering over the spatio-temporal bands shown in (c) results in noise values orders of magnitude less than the total noise of 0.95 nm. From Majeed, Hassaan, et al. "Magnified Image Spatial Spectrum (MISS) microscopy for nanometer and millisecond scale label-free imaging." Optics express 26.5 (2018): 5423-5440.

The approaches to enhance the temporal phase sensitivity include passive stabilization, active stabilization, differential measurements, and common-path interferometry [95-97]. The spatial phase sensitivity can be improved by using the white light source and keeping the optics pristine [98].

### 2.3.4. Tomographic methods based on QPI

Tomography is one of the most important studies in QPI, aiming at solving the inverse scattering problem [99-101]. A large number of tomographic methods suitable for different scales and distinct types of biological samples are developed utilizing different parts of the spectrum of the electromagnetic fields [102-104]. X-ray computed tomography is a tomographic method based on Radon transform with a large number of applications in medicine and industry [105].

In 1969, Emil Wolf developed the theory, known as optical diffraction tomography (ODT), to reconstruct the object using the scattered fields [106]. ODT depends on either scanning the illumination angles or sample rotations to reconstruct the 3D object information [107-109]. The procedures of obtaining reconstructed tomograms with illumination rotation and cell rotation are illustrated in Fig. 8 and 9. However, to have the isotropic resolution in the reconstructed object, one needs to cover all the frequencies in the Ewald sphere by scanning the illumination angle through the entire solid angle $4\pi$ and taking the measurements over the entire solid angle (see details in Fig. 10) [110]. Most imaging modalities cannot cover all the frequencies in the Ewald sphere, a problem known as the *missing cone problem* [111]. Two categories of solutions were proposed to solve it. One is through the various iterative algorithms after the acquisition process such as edge-preserving, total variation regularization, and the

Gerchberg-Papoulis algorithm [112]. Another solution is through the hardware improvement before the acquisition process such as 4pi microscope, label-free light sheet microscope, confocal microscope, and cell rotations by dielectrophoretic forces [113-116]. Machine learning can potentially mitigate the *missing cone problem*, the ground truth to train the network, however, is key to solving this problem [117, 118].

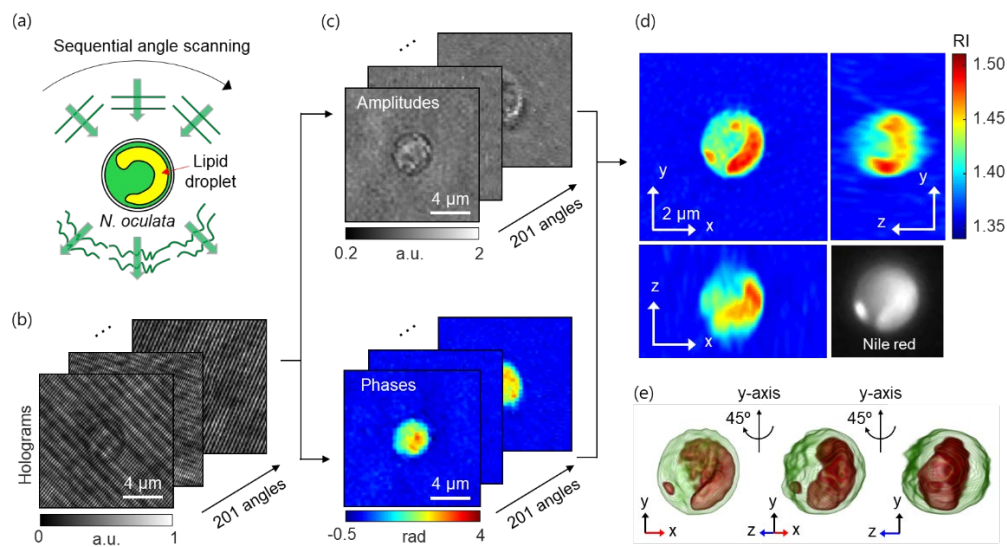

Figure 8 Schematic diagrams of the label-free identification of lipid droplets in individual N. oculata cells using ODT (a) The sample is consecutively illuminated by a plane wave at various incident angles. (b) The holograms are recorded at 201 incident angles. (c) Retrieved amplitudes and phases of the optical fields diffracted by the sample. (d) Tomograms of the reconstructed 3D RI distribution of N. oculata in the x-y, y-z, and x-z planes. The Nile red fluorescence image of the same cell is shown in the lower right corner for comparison. (e) The 3D rendered iso-surface image of the reconstructed RI distribution at various viewing angles. From Jung, JaeHwang, et al. "Label-free non-invasive quantitative measurement of lipid contents in individual microalgal cells using refractive index tomography." Scientific reports 8.1 (2018): 1-10.

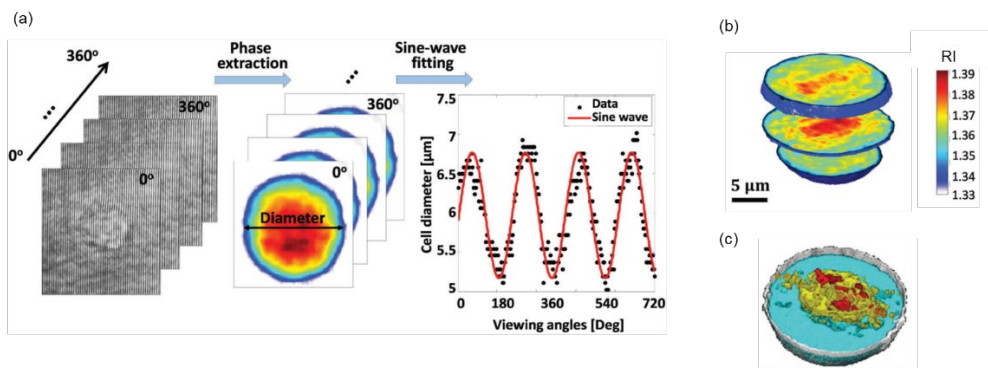

Figure 9 (a) Detection of the rotation cycle time and evaluation of the angle of the present point of view are done by fitting the cell diameter in the quantitative phase map during cell rotation to a sine wave. (b,c) 3D rendering (b) and rendered iso-surface plot (c) of the refractive-index map of an MCF-7 cancer cell. From Habaza, Mor, et al. "Rapid 3D refractive-index imaging of live cells in suspension without labeling using dielectrophoretic cell rotation." Advanced Science 4.2 (2017). Reprinted with permission from AS.

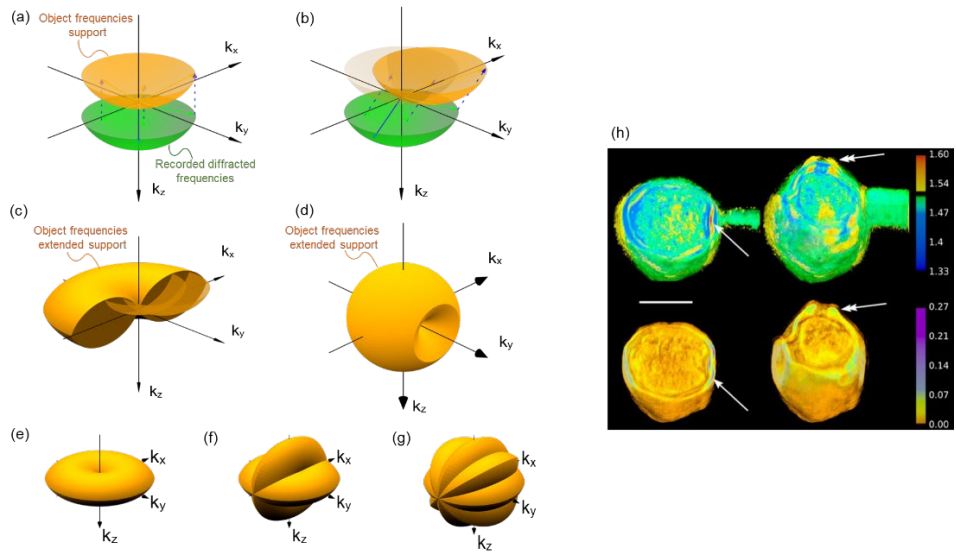

Figure 10 Construction of the OTF for various configurations of TDM. (a) Digital holographic microscopy. The recorded k-vectors are shifted back according to momentum conservation to provide object vectors: the OTF depicts a cap of sphere of a large lateral, but limited, longitudinal extension. (b) When using inclined illumination in TDM-IR, the same positions of the recorded vectors provide new object vectors. (c) A large set of illuminations results in a filled and extended OTF: TDM-IR provides improved-resolution, higher quality 3D images. Note, however, the presence of a so-called "missing cone" along the optical axis z, limiting longitudinal resolution and sectional capabilities. (d) OTF for TDM-SR. An almost completely filled sphere is obtained, but of lesser extension than in the previous case. (e) OTFs obtained for TDM-IRSR when combining TDM-IR with 0°, (0°; 90°) and (0°; 45°; 90°; 135°) specimen rotations [displayed at half-scale compared with (a)–(d)]. A missing-cone-free and extended support is obtained, showing that TDM-IRSR can deliver 3D, isotropic- and improved-resolution images. (h) 3D complex RI of betula pollen grain observed with TDM. First row: real part of RI, second row: imaginary part of RI (absorption). Note the higher index of refraction of the pollen walls, especially near the pores (double-headed arrow), and the double-layer outer wall (arrow). Scale bar: 10 μm. From Simon, Bertrand, et al. "Tomographic diffractive microscopy with isotropic resolution." Optica 4.4 (2017): 460-463.

The original ODT was formulated with coherent illumination. However, wide-field microscopes with coherent illumination often suffer from speckles and other coherent artifacts, resulting in degraded contrast and resolution. Later on, several ODT techniques with partially coherent illumination were proposed [119-121]. We discuss two ODT techniques with partially coherent light, termed as white-light diffraction tomography (WDT) [122] and Wolf phase tomography (WPT) [79] in section 4.3.

As we mentioned in section 1.2, tomographic methods such as OCT [123] and optical Doppler tomography [124] are interferometric techniques. Several non-interferometric tomographic methods based on the transport of intensity were also developed [125-127]. Fourier ptychographic tomography (FPM) and other synthetic aperture methods aim at improving the resolution in the 3D reconstruction [128-130]. Ghost tomography is based on the correlation between the known structured illuminating patterns and the total integrated intensity, the 3D object information can be reconstructed with Ghost imaging algorithms such as iterative cross-correlation via the Landweber algorithm [131, 132].

To improve the penetration depth into the biological samples, photoacoustic tomographic methods combine the utilization of the electromagnetic fields with sound waves [133]. Diffuse optical tomography (DOT) uses diffusive photons to reconstruct the samples based on the transport equation of near-infrared light [134]. The forward model is based on the radiative transfer equation, however, the diffuse equation is often used because of the high computational load. The reconstruction algorithms are categorized in linearization approaches based on Born or Rytov approximations and nonlinear iterative approaches [135].

## 3. Principles of SLIM

### 3.1. Theory

#### 3.1.1. Zernike phase-contrast microscope

The phase information of the object is hidden in bright-field microscopic images. In the 1930s, Zernike solved this problem by inserting a $\pi/2$ phase retarder in the objective pupil plane, introducing extra $\pi/2$ phase delay between the incident and scattered fields [18]. As a result, the information of the phase object can be retrieved quantitatively from four intensity images. We can illustrate this based on the scattering theory discussed in section 2.1.2. Recall that, for phase objects, the total field of the forward scattering under the first-order Born approximation and paraxial approximation has the form [67]

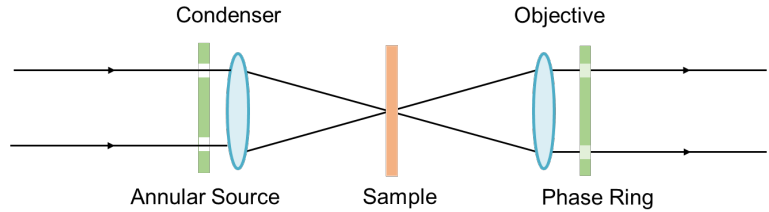

Figure 11 Diagram of Zernike phase-contrast microscopy.

$$U_f(r_\perp, z, \omega) = A(\omega)e^{i\beta z}\{1 - i\beta[n(r_\perp, \omega) - n_0]L\}. \quad (3.1.1)$$

Thus, the intensity can be calculated as

$$I = U_f U_f^* \\ \simeq A^2(\omega). \quad (3.1.2)$$

We can see that the information of the weak phase object is lost in the conventional bright-field microscopy. After inserting a phase π/2 retarder to the incident field, the forward scattering field becomes

$$U_{PC}(r_\perp, z, \omega) = A(\omega)e^{i\beta z}\{i - i\beta[n(r_\perp, \omega) - n_0]L\}. \\ = A(\omega)e^{i\beta z}i\{1 - \beta[n(r_\perp, \omega) - n_0]L\}. \quad (3.1.3)$$

Now the intensity has the expression
$$I_{PC}(x, y) = U_{PC}U_{PC}^* \\ \simeq A^2(\omega)\{1 - 2\beta[n(r_\perp, \omega) - n_0]L\}^2. \quad (3.1.4)$$

It can be seen that the object information appears as a linear term in the intensity of the phase-contrast microscope, resulting in a much higher contrast for phase objects. The diagram of Zernike's phase-contrast microscope is presented in Figure 11. In the objective pupil plane, Zernike introduced a phase retarder to give a $\pi/2$ shift to the unscattered field. This filter also attenuates the unscattered field to further decrease the background light. In commercial microscopes, the pupil function is designed to match the annular illumination given by

$$P(r) = \begin{cases} 1 & r < R_i \\ \pm ai & R_i \leq r \leq R_o \\ 1 & R_o \leq r \leq R \end{cases}, \quad (3.1.5)$$

where $R_i$ and $R_o$ are the inner and outer radius of the ring retarder, $R$ is the radius of the aperture, the $\pm$ sign corresponds to positive and negative phase contrast.

### 3.1.2 Phase retrieval in SLIM

SLIM is implemented as an add-on module to the commercial phase contrast microscope, it combines the spatial uniformity associated with white light illumination and the stability of common-path interferometry [92, 136]. The schematic set up is illustrated in Fig. 12(a). The spatial light modulator (SLM) in the add-on module provides further phase shifts in the pupil plane with increments of $\pi/2$, and the active pattern on the SLM is calculated to precisely match the size and the position of the objective phase ring image. As a result, the phase delay between the scattered and unscattered fields is controlled and the four images corresponding to each phase shift are recorded on the camera.

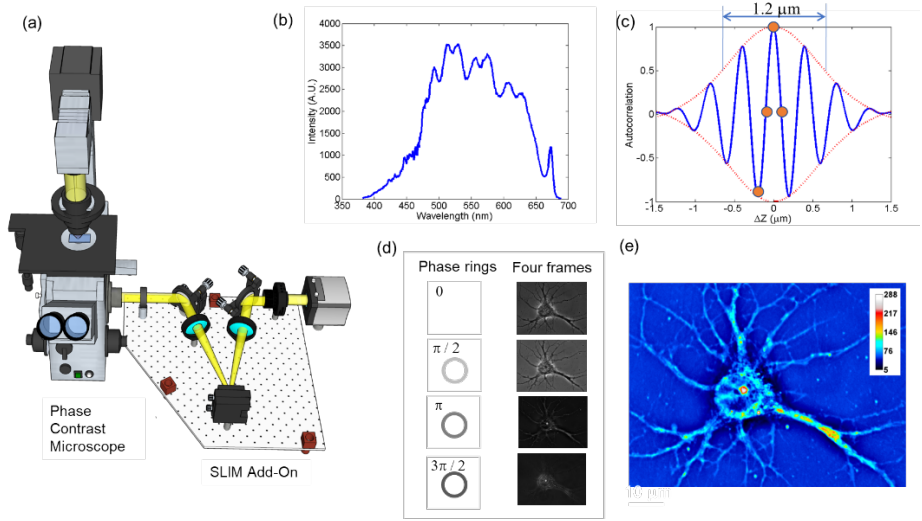

Figure 12 SLIM principle. (a) Schematic setup for SLIM. (b) Spectrum of the white light emitted by the halogen lamp. The center wavelength is 552.3 nm. (c) The autocorrelation function (blue solid line) and its envelope (red dotted line). The 4 circles indicate the phase shifts produced by SLM. (d) The phase rings and their corresponding frames recorded by the camera. (e) SLIM quantitative phase image of a hippocampal neuron. The color bar indicates optical path-length in nanometers. From Wang, Zhuo, et al. "Spatial light interference microscopy (SLIM)." Optics express 19.2 (2011). Reprinted with permission from OSA.

The white light illumination in SLIM can be considered as spatially coherent, but temporally partially coherent fields. The phase information in SLIM can be understood as that of an effective monochromatic field oscillating at the average frequency of the broadband fields. For broadband fields, the cross-spectral density of the incident and scattered fields are defined as

$$W_{is}(\mathbf{r},\omega) = \langle U_i^*(\omega) U_s(\mathbf{r},\omega) \rangle, \tag{3.1.6}$$

where $U_i$ and $U_s$ are the incident and scattered fields, $\mathbf{r}$ is the spatial coordinate, $\omega$ is the frequency of the light. The total field on the image plane is thus

$$\begin{aligned} U(\mathbf{r},\omega) &= U_i(\omega) + U_s(\mathbf{r},\omega) \\ &= |U_i(\omega)| e^{i\phi_0(\omega)} + |U_s(\mathbf{r},\omega)| e^{i\phi_1(\mathbf{r},\omega)}. \end{aligned} \tag{3.1.7}$$

$\Delta\phi(\mathbf{r},\omega) = \phi_1(\mathbf{r},\omega) - \phi_0(\omega)$ is the phase delay between the incident and scattered fields. For most transparent specimens of interest here, we consider the dispersionless case, i.e., $\phi$ independent of $\omega$. With the mean frequency $\omega_0$ of the broadband fields, the cross-spectral density can be expressed as

$$W_{is}(\mathbf{r},\omega - \omega_0) = |W_{is}(\mathbf{r},\omega - \omega_0)| e^{i\Delta\phi(\mathbf{r})}. \tag{3.1.8}$$

The temporal autocorrelation function is obtained by taking the Fourier transform of Eq. (3.1.8) with respect to $\omega$,

$$\Gamma_{is}(\mathbf{r},\tau) = |\Gamma_{is}(\mathbf{r},\tau)| e^{i[\omega_0 \tau + \Delta\phi(\mathbf{r})]}. \tag{3.1.9}$$

The phase map retrieved from phase-shifting measurements is equivalent to that of coherent monochromatic light at frequency $\omega_0$. The intensity in the plane of interest is thus a function of the time delay as

$$I(\mathbf{r},\tau) = I_i^2 + I_s^2 + 2|\Gamma_{is}(\mathbf{r},\tau)| \cos[\omega_0 \tau + \Delta\phi(\mathbf{r})]. \tag{3.1.10}$$

The magnitude of the correlation function $|\Gamma_{is}(\mathbf{r},\tau)|$ around $\tau = 0$ can be assumed to vary slowly at each phase shift. Therefore, the phase delay between the incident and scattered fields can be calculated as

$$\Delta\phi(\mathbf{r}) = \tan^{-1}\left[\frac{I(\mathbf{r},\tau_3) - I(\mathbf{r},\tau_1)}{I(\mathbf{r},\tau_0) - I(\mathbf{r},\tau_2)}\right], \tag{3.1.11}$$

where $\tau_j = j\pi/2$, $j = 0,1,2,3$. If we define $a(\mathbf{r}) = |U_s(\mathbf{r})|/|U_i|$, then the phase delay between the incident and the total fields can be reconstructed as

$$\phi(\mathbf{r}) = \tan^{-1}\left[\frac{a(\mathbf{r})\sin[\Delta\phi(\mathbf{r})]}{1 + a(\mathbf{r})\cos[\Delta\phi(\mathbf{r})]}\right]. \tag{3.1.12}$$

From four successive intensity measurements for each phase shift (Fig. 12(d)), the phase information of the object is retrieved (Fig. 12(e)). The spectrum of the halogen lamp is presented in Fig. 12(b). The real part of the autocorrelation function $\Gamma_{is}$ (blue solid line) and its magnitude (red dotted line) are depicted in Fig. 12(c). The 4 circles show the phase shifts produced by the liquid crystal phase modulator (LCPM).

Fig. 13 compares the spatial accuracy of SLIM and atomic force microscopy (AFM) by imaging an amorphous carbon film deposited on glass. The topography measurements by SLIM and AFM, respectively, are presented in Fig. 13 (a) and (b). The two types of measurement agree within a fraction of a nanometer. Unlike AFM, SLIM is non-contact, parallel, and faster by more than 3 orders of magnitude. Thus, SLIM can optically measure an area of 75 × 100 μm$^2$ in 0.5 s compared to a 10 × 10 μm$^2$ field of view measured by AFM in 21 minutes. To further compare to the diffraction phase microscopy (DPM), an off-axis laser-based technique that was interfaced with the same microscope, the background images (i.e., no sample) from SLIM and DPM are shown in Fig. 13 (d) and (e). SLIM's spatial uniformity and accuracy for structural measurements are substantially better than DPM's due to the lack of speckle effects granted by its white light illumination. To quantify the spatio-temporal phase sensitivity, a 256-frame stack of background images was obtained using SLIM. Figure 13 (f) illustrates the spatial and temporal histograms associated with the optical path-length shifts across a 10 × 10 μm$^2$ field of view and over the entire stack. These noise levels, 0.3 nm, and 0.03 nm represent the limit in optical path-length sensitivity across the image and between frames, respectively.

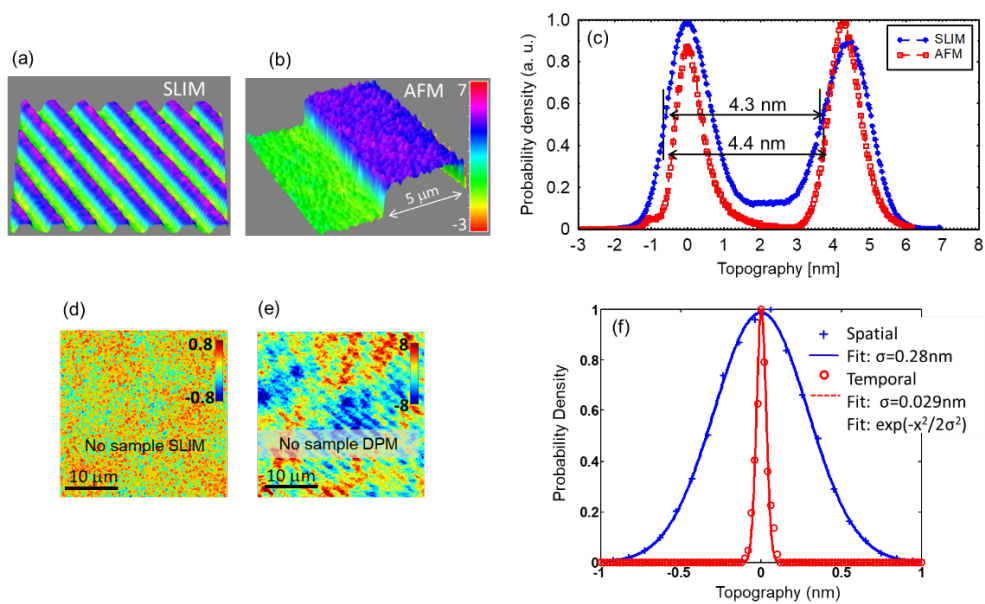

Figure 13 SLIM Figures of merit. (a) SLIM image of an amorphous carbon film (40X/0.75NA objective). (b) AFM image of the same sample. The color bar indicates thickness in nm. (c) Topographical histogram for AFM and SLIM, as indicated. (d) No-sample background of SLIM. (e) No-sample background of DPM. (f) Optical path-length noise level measured spatially and temporally. The solid lines indicate Gaussian fits, with the standard deviations as indicated. From Wang, Zhuo, et al. "Spatial light interference microscopy (SLIM)." Optics express 19.2 (2011). Reprinted with permission from OSA.

### 3.1.3. Halo removal

Because the illumination in phase-contrast microscopy is not perfectly spatially coherent, the images in SLIM are affected by a coherent artifact, known as the phase-contrast "halo" which resembles a glow around the edges of the cell. In SLIM, as in phase-contrast microscopy, a phase ring at the pupil plane is used to delay controllably the transmitted light relative to the scattered light. The final result is a greatly improved sensitivity to optical path length shifts [18].

(a) Traditional Interferometry

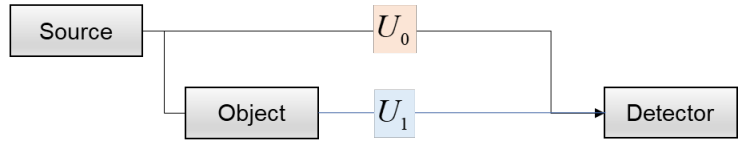

(b) Common-Path Interferometry

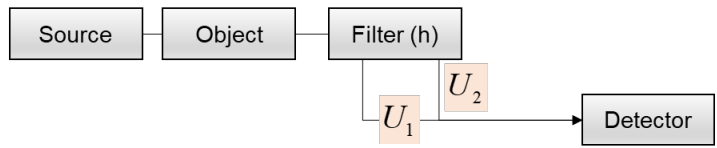

Figure 14 In traditional interferometry, interference occurs between the scattered field (U1) and reference field (U0) while in common-path configurations the reference field U2 is generated from the sample.

This ring illumination creates a spatial coherence area that is generally smaller than the field of view. As such, in reality, the description of image formation that assumes the imaging instrument can unambiguously separate "scattered" and "transmitted" components is an idealization (Fig. 14) [120]. In phase-contrast microscopy and SLIM, actual components of the modulated field are determined by the shape of the phase contrast pupil as well as the coherence properties of the illumination. In practice, the illuminating pupil cannot be made too small [137], thus practical designs lead to the introduction of cross-talk between the scattered and transmitted fields [120]. In effect, a low-resolution version of the object is imparted into the reference field which leads to an unwanted halo-like glow around the sample (Fig. 15(a)). This effect is particularly acute for low-frequency structures such as flat semiconductors while being almost absent in intracellular details like nucleoli or mitochondria. As the SLIM image is a faithful measurement of the field associated with the phase-contrast microscope it, too, suffers from halos.

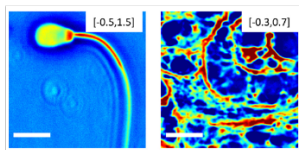
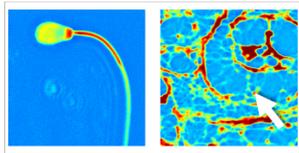
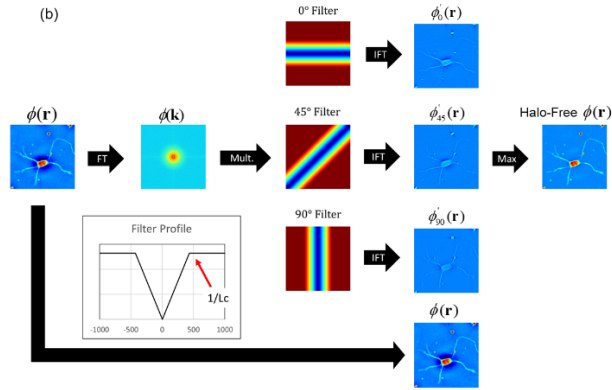

Figure 15 (a) The halo artifact appears as unwanted glow around the specimen (sperm and tissue biopsy, 40x/0.75 SLIM). (b) The halo artifact can be partially-corrected by using a nonlinear computational algorithm. In the direct halo removal algorithm, a series of directional derivative images are combined with the original image using a pixel wise maximum. (c) The resulting SLIM images highlight details that were previously submerged by the halo (white arrow).

Unlike phase contrast, where the amplitude is coupled to phase, performing phase-interferometry, SLIM recovers the deterministic signal associated with the optical field which, in turn, provides a computational strategy to remove the halo artifacts [138]. The most complete model for halo formation is presented in [139], where the authors use a variation of the transmission cross-coefficients (TCC) to model image formation (Hopkin's TCC) [140, 141]. In general, a TCC-based approach is difficult to invert, motivating the authors to simplify image formation to [142],

$$\phi_m(\mathbf{r}) = \phi(\mathbf{r}) - \arg\left[e^{i\phi(\mathbf{r})} \circledv_\mathbf{r} h(\mathbf{r})\right]. \qquad (3.1.13)$$

Here, $\phi_m$ is the measured SLIM image, $\phi$ is the desired phase associated with the object, and $h(\mathbf{r})$ is an impulse response like function related to the condenser and illumination shape that captures the spatial incoherence of the system. Noting that the Fourier transform of $h(\mathbf{r})$, $h(\mathbf{k})$ resembles the physical aperture, we see that as $h(\mathbf{k})$ approaches a pinhole, $h(\mathbf{k}) \approx \delta(\mathbf{k})$ in Eq. (3.1.13), we recover the phase without error: $\phi_m(\mathbf{r}) \approx \phi(\mathbf{r})$. In practice, $h(\mathbf{k})$ is built into the microscope. While $h(\mathbf{k})$ can be adjusted via the condenser aperture in brightfield instruments, in phase-contrast microscopes, the apertures do not permit easy manipulation as they are matched to the phase-rings inside the objectives. This approach was extended to 3D imaging, by approximating the halo as a linear high-pass filter [143].

Using the observation that the halo artifact mostly corrupted low-spatial frequencies, a non-linear filtering technique was proposed, using directional derivatives [142] (Fig. 15). In this method, a series of images are collated by taking the pixel-wise maximum of the derivative images and the original halo-corrupted image. Importantly, this approach is non-iterative and can be applied to real-time operations without the need to measure complicated impulse responses.

### 3.2. Instrumentation

### 3.2.1. Alignment and calibration

In SLIM, an active modulating element introduces controlled phase shifts at the pupil plane, modulating the delay between the scattered and transmitted light. The resulting implementation resembles an external form of phase-contrast with a tunable phase-ring. When compared to off-axis methods [144], by using a series of temporal modulations to acquire the complex field, SLIM trades temporal bandwidth (more images) for spatial bandwidth (better use of the camera sensor) in a way that improves image quality.

The light throughput advantage comes from the use of phase rather than amplitude modulating elements, which makes better use of the photon budget. This is especially important in applications where high illumination intensities can compromise sample viability [2] or when sharing a camera for phase and fluorescence imaging (low light) [145]. The improvement in image quality is the result of a reduction of "coherent noise" related artifacts [146, 147].

SLIM addresses these concerns by performing modulation on a pupil plane conjugate to the back focal plane of the objective (phase ring of the objective). This reduces fixed pattern noise as the fringes are generated at the pixel level, in time, by modulating a retarder rather than spatially, with an off-axis reference field. At this plane, misalignment introduces directional shading rather than a grid-like pattern. In practice, misalignment rarely occurs as the ring-like illumination and attenuation from the phase-contrast objective provide a convenient fiducial marker for alignment (Fig. 16).

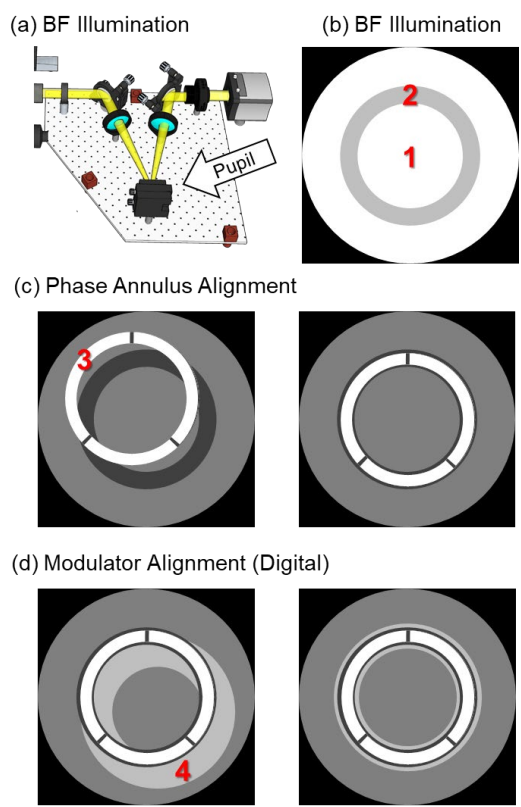

Figure 16 Alignment of a commercial SLIM system. (a) The SLIM add-on interferometer is implemented as a 4-f system with a reflective spatial light modulator manipulating the pupil plane of a commercial phase contrast microscope. For alignment, an additional lens with integrated analyzer is positioned after the pupil plane (b) SLIM alignment begins by configuring the microscope into brightfield illumination. Here #1 is the bright background due to a fully open condenser and #2 represents the attenuation due the phase ring typical of phase contrast objectives. The square-root of the average intensity values between #1 and #2 is a per-objective attenuation constant used during the image reconstruction process. (c) Next, the microscope's condenser is configured for phase contrast illumination. #3 shows the illumination ring which is then aligned to the match the objective's phase ring. (d) Lastly the modulation (#4) is aligned to the phase ring by digitally adjusting the pattern on the spatial light modulator. In general, this procedure must be performed for each objective, and in some cases a zoom lenses immediately before the add-on module is used to adjust the location of the pupil plane on a per objective basis.

By far the most popular method to modulate a ring shape involve the use of a spatial light modulator (SLM) [148]. This device contains a digitally addressable grid of pixel-like variable retarders, each of which uniquely modulates the polarization state of one polarization relative to another. For SLIM imaging, these are mostly reflective devices, although at least one attempt was made to use a lower cost twisted nematic transmission SLM [149]. Other authors have noted that the same SLM can be multiplex for optical trapping [150].

The relationship between the instrument's defined voltage units or "gray-levels" and the imparted phase shift depends on the illumination spectrum and detector response. In practice, this relationship is determined on a per-instrument basis by recording a sequence of amplitude images with increasing phase modulation. In the case of SLMs, this is accomplished by placing the modulator between crossed polarizers.

The resulting amplitude curves can be analyzed using a Hilbert transform technique where a complex analytic signal is constructed from the measured data by performing a Hilbert transform [92] (Fig. 17(b-c)). The argument of this complex analytic signal yields the instantaneous phase response at the configured gray-level. In practice, the Hilbert transform is calculated by way of the fast Fourier transform, which requires continuity and periodicity assumptions to be met [151]. Further, SLMs have fixed modulation levels leading to discretization errors. This problem is particularly acute as the number of actual gray-levels is only a fraction of the total addressable range. For example, a common SLM may provide 50 gray-levels over an addressable range of 256 values (8-bit). One proposed solution to increase the number of gray-levels is to write a checkboard style dithering pattern where neighboring pixels are averaged to achieve an intermediate discrete phase-shift value [139] although a more robust method involves refining the reconstruction formulas to match the discrete phase-shifts [71].

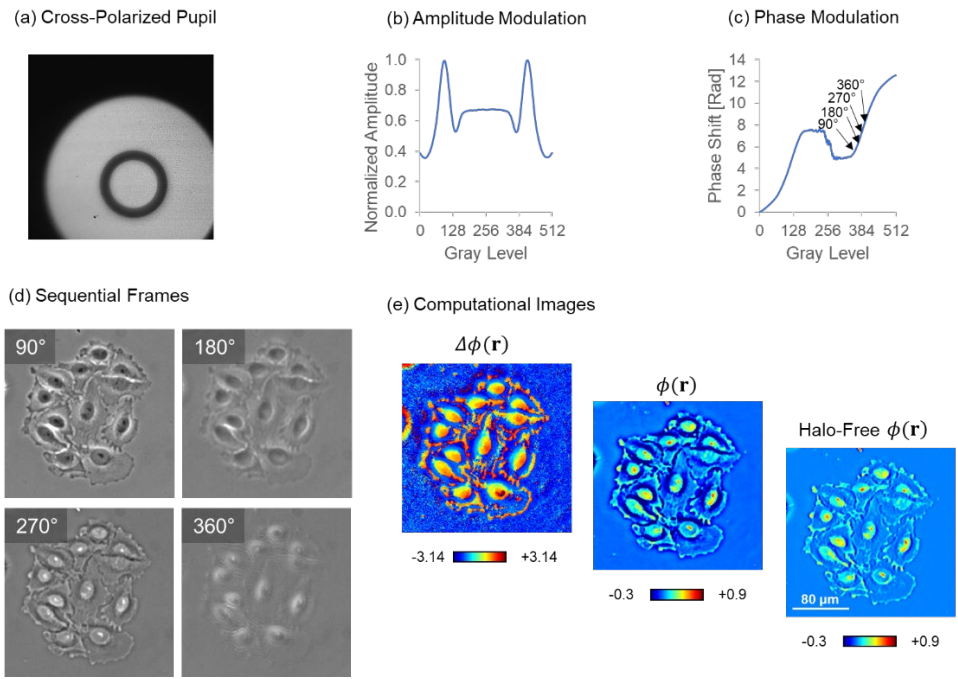

Figure 17 Spatial light modulators consist of a gird of addressable variable retarders capable of imparting a phase shift at each pixel. The relationship between the digitally controlled phase value and the actual achieved phase delay depends on the spectral properties of illumination and is usually calibrated for each light source. (a) One calibration strategy involved inserting a lens containing a polarizer to visualize the pupil plane. (b) A series of images are acquired with increasing phase modulation producing an amplitude curve (Meadowlark XY series, shown). (c) a Hilbert transform is used to find the instantaneous phase shift associated with each gray level. (d) The captured four frames after calibration. (e) The reconstructed phase maps.

As SLIM is rate-limited by the speed of the SLM, some authors have explored using mirrors attached to piezo-electronics to improve frame rates [152, 153]. For example, in [152] the authors used a three-step phase-shifting algorithm to achieve 50 Hz imaging (reported as 150 Hz with "interlacing"). So far, these efforts have fallen short of the potential imaging rates due to trouble synchronizing acquisition with camera exposure. For example, these attempts used software triggering which uses an extra readout step and effectively halves frame rates when compared to continuous acquisition. Further, most implementations wait for the SLM to stabilize (stop-and-go) instead of performing modulation simultaneous with camera acquisition (bucket integration) [100]. An unsolved challenge with mirror-based approaches is the need to adjust the modulating element on a per-objective basis. In a parallel development, overdrive techniques have pushed SLM switching times to the kilohertz regime where light budget concerns begin to dominate [154]. These advances make SLM based approaches more competitive.

### 3.2.2. High-throughput acquisition

Besides the SLM, the throughput of a SLIM system depends on the extent to which the acquisition process is parallelized [155]. In general, the SLIM acquisition process involves translating the stage or focus, introducing a phase shift, exposing the camera, reading out the image, phase retrieval, rendering the result, and saving the data. Fig. 18 presents three possible acquisition schemes; a serial version, the implementation in [145], and a theoretically optimal but, yet, unrealized variant. Unsurprisingly, the optimal version performs 10x faster than the serial version.

(a) Serial Scheme

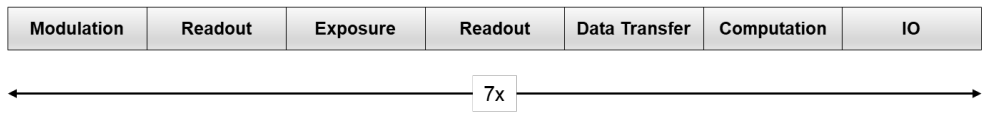

(b) Phase-Stepping with Software Triggering

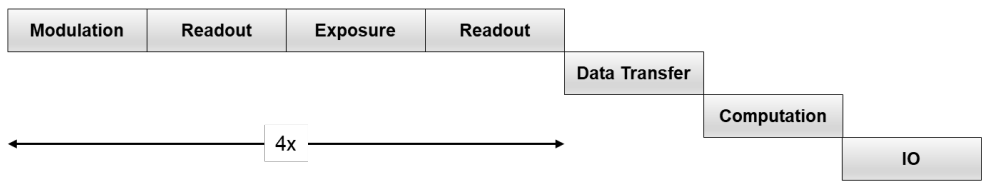

(c) Integrated-Bucket with Hardware Synchronization

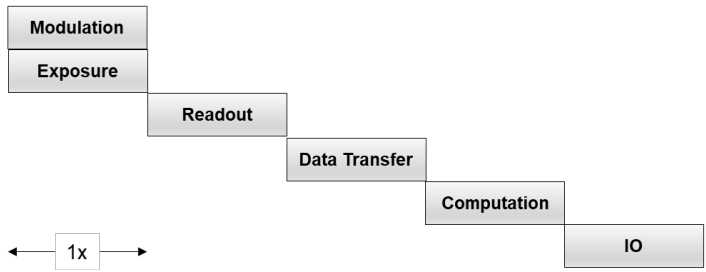

Figure 18 (a) A serial SLIM acquisition is the most popular computational imaging scheme for home-built instruments. In that scheme, all steps of the acquisition process are performed in series. (b) The commercial instrument implement a more advanced scheme where hardware events are overlapped with computation. The principal limitation of this scheme is due to the need to discard extra charge during software triggering. In software triggering modes the camera must discard the charge on the detector to ensure correct exposure time by performing a charge readout, which effectively halves frame rates. The advantage of software triggering and hardware analogs is that a variable amount of time can elapse before the image is recorded. Thus, this mode is preferred when the microscope is expected to move before each acquisition. (c) A potentially faster but unrealized approach exists where camera exposure is overlapped with modulation.

In early implementations of the SLIM design, the rate-limiting factors were found to be computational [156], yet in more modern revisions, the rate-limiting steps are reported to be exposure time or SLM stability [145]. In a broader context, the shift from computational back to optical limitations highlights a trend in imaging where computational techniques have advanced faster than the optical elements. This is likely to remain the case as computing bottlenecks (GPUs or hard drives) can be scaled by adding more computing hardware while similar approaches do not apply to the construction of an optical crystal. Notably, the theoretically optimal performance remains unrealized. While this would appear to be strictly due to deficient software implementation, in reality, it is difficult to implement bucket integration with a digitally controlled device. This is because a sinusoid rather than a fixed level signal must be supplied to the SLM elements. Achieving a smooth sinusoidal modulation, rather than a unique-wave form for each modulation [154] requires a yet to be realized calibration procedure or some other fundamental change in the existing SLM hardware.

In general image acquisition and computation can occur in parallel, meaning that the current image can be processed while the next one is acquired. An important but often understated limitation is that not all programming languages are well suited for task-parallel operations. For example, C++, C#, or LabView have individually addressable threads but those constructs are more challenging to use in Python or MATLAB. This design limitation is evident by the choice of language used in some SLIM implementations. In [92], LabView was used to control the modulator while stage and camera control was performed by the microscope control software (Zeiss, AxioVision). In [156] the authors used a combination of C++ (backend) and C# (frontend), while the Cell Vista Pro (PhiOptics Inc) is written in C++ and uses Qt as the widget kit to facilitate parallelized rendering [145].

### 3.2.3. Whole slide imaging

Multiscale experiments such as high-content phenotypic screening [145], or 4D imaging of mesoscopic structures [157] present challenges to image acquisition and data storage [158]. As a point of reference, a free-running 5 MP camera is capable of producing approximately 4 TB of data in an hour [159]. To obtain ample storage most authors have preferred to use a combination of high-speed networking and large hard drive arrays [160]. While such strategies are often able to meet total data storage requirements, throughput is often difficult to achieve, especially when data redundancy is required [161]. For example, RAID 6 parity reduces throughput by six times. Further, achieving optimal performance requires using multiple threads to saturate the write cache, and ensure that the hard drives are constantly writing/reading. An alternative strategy for burst imaging is to use solid-state based storage [162], which affords more throughput at a lower cost but comes at a fraction of the storage capacity [163]. Thus, to achieve both high-throughput and total capacity some authors have preferred a combination of SSDs and external storage. In this case, acquired data are written onto an SSD disk and a multithreaded copy (robocopy.exe) is used to perform data transfer. This approach is well suited to storage computers running Windows, which, unlike Linux (Samba), supports multithreaded data transfer [164]. Further, using an intermediate local drive introduces a measure of tolerance for network disruptions.

In addition to challenges due to computer storage constraints, it was found that a purpose-built graphical user interface and digitization strategies are crucial for SLIM's operation. One difficulty was in maintaining focus when imaging large surfaces such as microscope slides or multi-well plates (128 x 85 mm). In [155] the authors developed an axial scanning technique where the plane of best focus is determined by finding the point where the variance of the phase within mid-range frequency bands was maximized (Fig. 19). This strategy is typically unworkable as a large number of axially scanned samples need to be acquired, yet it was found that in most cases the plane of best focus resembled a tilt (due to the sample being slightly tilted) which in turn improved the speed of the autofocusing algorithm. Further, with phase imaging and interpolation, it was possible to reduce the number of axial samples to five. To scan samples with discrete regions (such as high well multiwells) a graphical interface was developed to represent the plane of best focus as a collection of "focus points". These points were then used to construct a Delaunay triangulation for interpolating the focus at each mosaic tile [145].

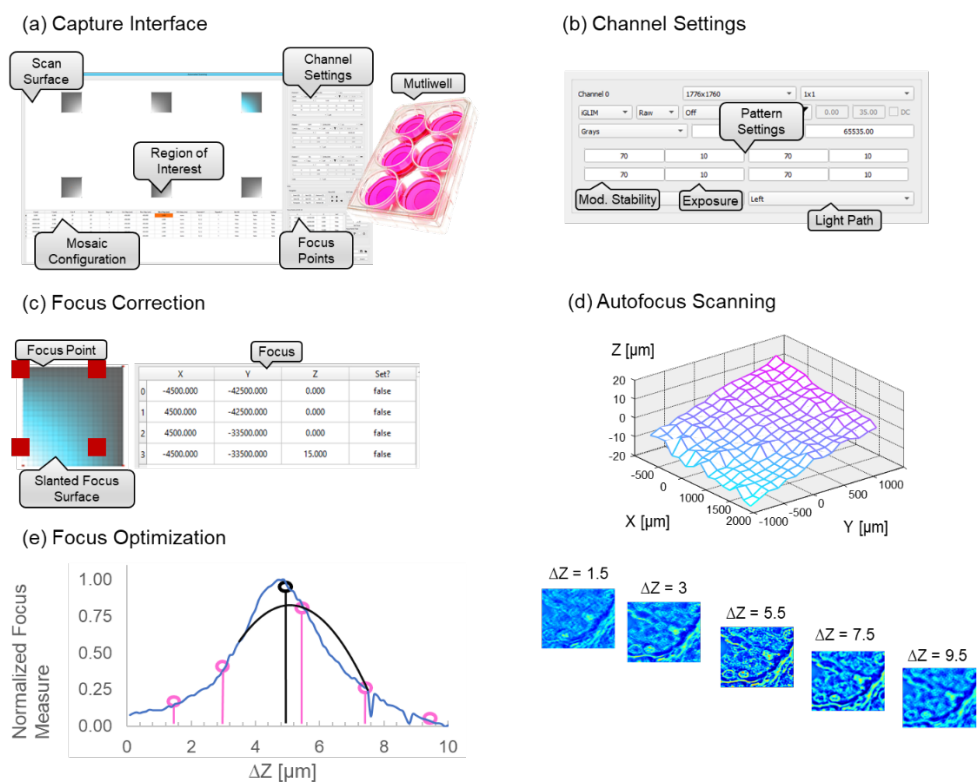

Figure 19 Graphical user interface to configure complicated imaging experiments (such as those involving multiwells) include SLIM imaging specific features. (a) The capture interface enables scanning multiple regions of interests with separate focus focusing points. (b) Channels such as fluorescence microscopy are presented side-by-side with phase imaging specific features such as control with exposure and modulator stability on a per-pattern basis. (c) To account for variations in the plane of best focus, a Delaunay triangulation constructed from "focus points" is interpolated to determine the ultimate coordinates of each mosaic tile. (d) Some slide scanning instruments include an autofocusing features where through focus stack is small through focus stack is acquired offset from the manually configured plane-of-best focus. (e) Following a focus optimization scheme the most in focus position (black) is selected from a series of sub-optimally focused images (pink, shown). In practice this procedure is relatively quick with less than a second used to optimize each mosaic tile.

### 3.2.4. Mosaic tile registration

To acquire samples larger than the field of view, SILM relies on a stage scanning strategy, where a series of high-resolution mosaic tiles are composited to form a larger image.

While this strategy has the advantage that individual frames do not suffer from motion blur and that samples much larger than the objective can be acquired [165], the motion of the microscope stage introduces rigid misalignment that must be compensated through digital methods. The most popular method to perform rigid image registration relies on identifying the peak values in the cross-correlation [166] and merging disagreements between neighboring tiles using a least-squares approach [167]. It was found that this algorithm was well suited to GPU computation and that the rate-limiting factor was disk access, motivating a caching strategy to avoid redundant reads [155]. A challenge with phase correlation is that dense regions of the sample produce spurious or unwanted peaks in the cross-correlation between neighboring regions. This problem is addressed by searching for peaks within a limited window which is adjusted iteratively [139]. Further error suppression occurs by performing phase correlation on background-corrected images, with the background generated by averaging a large number of images acquired during the experiment.

### 3.2.5. SLIM with a color camera

Most pathology applications rely on colored stains such as H&E to introduce specificity for cellular structures such as nuclei and cytoplasm. To facilitate co-localized QPI and histopathological imaging a variant of SLIM was developed using a brightfield objective and color camera [168]. This setup enabled the authors to acquire co-localized grayscale SLIM and stain images such as H&E (Fig. 20) [169]. While H&E images are rendered in full color, the SLIM image contains a single channel and represents the composite of the three colors on the detector. Importantly, it was found that the microscopes illumination spectrum was, on average, green which lies between the red and blue of the H&E stain, reducing dispersion related errors. An alternative strategy is to treat each color channel independently producing a three-color phase map from a single phase-shifting sequence [168].

In most color cameras, the field is sampled by a specialized bayer-mask consisting of a chromatic filter (RGB) at each pixel [170]. As only one color is detected at each position, a demosaicing interpolation procedure is performed to estimate the missing color values from neighboring elements. This procedure introduces additional computational considerations as industrial camera vendors often do not supply adequate demosaicing algorithms [171]. For example, the authors in [169] implemented a variation of the "high-quality linear" algorithm to preserve the resolution of the analog to digital converter [172]. Further, as interpolation is used during demosaicing, color SLIM instruments require a factor of $\sqrt{2}$ denser sampling at the image plane compared to their gray-scale counterparts.

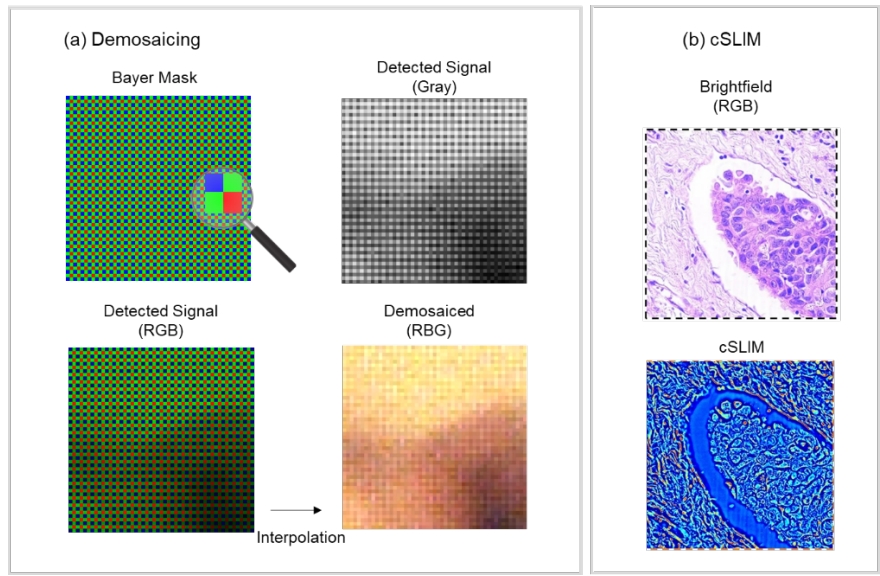

Figure 20 (a) Most color imaging sensors consist of a Bayer mask where every pixel has a preferential spectral sensitivity. The acquired data contains a single gray level at each pixel which can be interpreted as a color value (Detected Signal). As part of routine processing the missing color information is interpolated so that each pixel contains three values corresponding to red, green, and blue (Demosaiced). (b) The color imaging instrumentation was used for cSLIM where ring illumination was used in conjunction with a brightfield objective to form what resembled a conventional brightfield image when the SLM acted as a mirror. As outlined in that work the three-color channels were reweighted to produce a gray-scale image which was subsequently used for phase-reconstruction.

## 4. SLIM applications

### 4.1. Basic science applications

#### 4.1.1. Cell dynamics

SLIM is an ideal candidate to study cellular dynamics for a long period ranging from seconds to days because of its extremely low spatial noise (0.3 nm) and temporal stability (0.03 nm). As an early example, the dynamics of mixed glial-microglial cell culture are presented in Fig. 21 based on 397 SLIM images over 13 minutes. The comparison of phase-contrast and SLIM images are presented in Fig. 21 (b). We can see that the cell is bigger in the phase-contrast image due to the halo around the edge of the cell. Fig. 21 (c) shows the path-length changes due to both membrane displacements and local refractive index changes caused by cytoskeleton dynamics and particle transport at two arbitrary points on the cell. It reveals an interesting, periodic behavior. Moreover, the rhythmic motions have different periods at two locations inside of the cell, which may indicate different rates of metabolic or phagocytic activity. The probability distribution of pathlength displacements between two successive frames was retrieved with a dynamic range of over 5 orders of magnitude (Fig. 21 (d)). This distribution can be fitted very well with a Gaussian function up to path-length displacements $\Delta s = 10$ nm, at which point the curve crosses over to exponential decay. The normal distribution suggests that these fluctuations are the result of numerous uncorrelated processes governed by equilibrium. On the other hand, exponential distributions indicate the deterministic motions, mediated by metabolic activity.

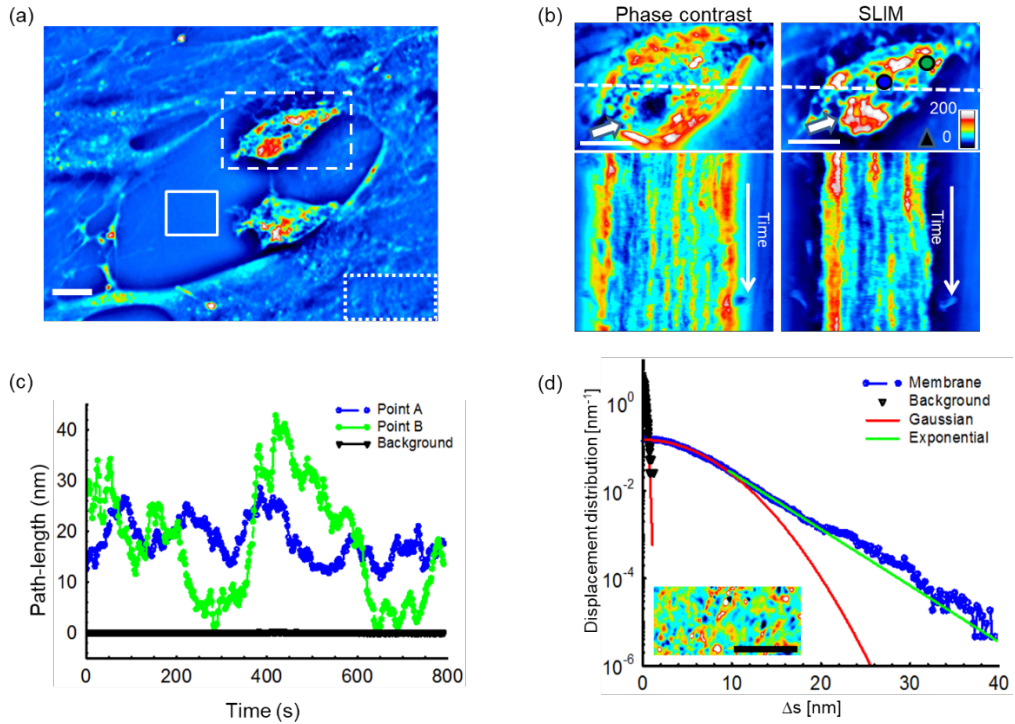

Figure 21 SLIM dynamic imaging of mixed glial-microglial cell culture. (a) Phase map of two microglia cells active in a primary glial cell culture. Solid line box indicates the background used in (d), dashed line box delineates a reactive microglial cell used in (b) and dotted line box indicates the glial cell membrane used in (d). (b) Phase contrast image and SLIM image of the cell shown in (a). psuedocoloration is for light intensity signal and has no quantitative meaning for phase contrast. Registered time-lapse projection of the corresponding cross-section through the cell as indicated by the dash line in (b). The arrows in (b) point to the nucleus which is incorrectly displayed by PC as a region of low signal. (c) Path-length fluctuations of the points on the cell (indicated in b) showing intracellular motions (blue- and green-filled circles). Background fluctuations (black) are negligible compared to the active signals of the microglia. (d) Semi-logarithmic plot of the optical path-length displacement distribution associated with the glial cell membrane indicated by the dotted box in (a). The solid lines show fits with a Gaussian and exponential decay, as indicated in the legend. The distribution crosses over from a Gaussian to an exponential behavior at approximately 10 nm. The background path-length distribution, measured from the solid line box, has a negligible effect on the signals from cells and is fitted very well by a Gaussian function. The inset shows an instantaneous path-length displacement map associated with the membrane. Scale bars, 10 μm. From Wang, Zhuo, et al. "Spatial light interference microscopy (SLIM)." Optics express 19.2 (2011). Reprinted with permission from OSA.

As another example of cellular dynamics, SLIM can be used to examine the diameter and axonal mass transport of the neurons [173] as the phase is correlated with the dry mass as following

$$M(x,y) = \frac{\lambda}{2\pi\gamma}\phi(x,y), \tag{4.1.1}$$

where λ is the center wavelength; γ=0.2 ml/g is the refractive increment, and $\phi(x, y)$ is the measured phase. The reconstructed SLIM image of an axon is shown in Fig. 22 (a). The average diameter and average phase of axons treated with different drugs are monitored in Fig. 22 (b) and (c) over time. Disrupting actin filaments resulted in an increase (Fig. 22 (b) red) in average diameter while disrupting microtubules (Fig. 22 (b), blue) led to a decrease in average diameter after 60 minutes of drug treatment. SLIM images revealed that the average phase increased when actin was disrupted (Fig. 22 (c), red). The average phase remained unchanged upon microtubules disruption (Fig. 22 (c), blue) or Y-27632 treatment (Fig. 22 (c), cyan).

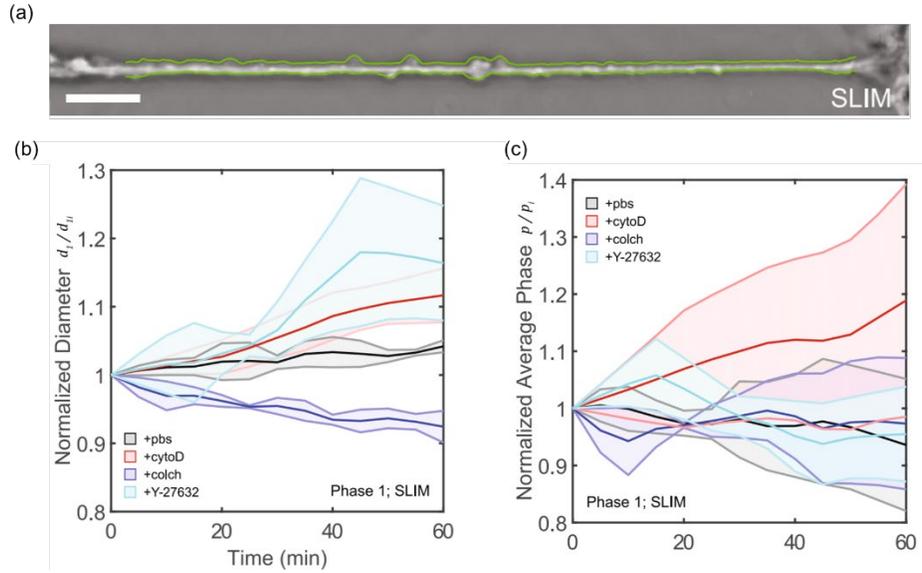

Figure 22 (a) Reconstructed SLIM image of a cleaned axon. Green lines labeled the boundaries determined by the analysis algorithm. Scale bar at 10 microns. (b) SLIM measurements of average diameter over time of axons treated with PBS (grey), cytoD (red), noco/colch (blue), and Y-27632 (cyan). (c) Average phase measured by SLIM of axons treated with pbs (grey), cytoD (red), colch (blue), and Y-27632 (cyan). The average density of the cytoplasm and the cytoskeletal components increase with time with disruption of actin, but not with microtubules. All shaded regions indicate error bar in standard deviation. Unpaired two-sample t-test used to obtain p-values. From Fan, Anthony, et al. "Coupled circumferential and axial tension driven by actin and myosin influences in vivo axon diameter." Scientific reports 7.1 (2017).

### 4.1.2. Cell growth

SLIM can monitor cell growth and proliferation, thus, can serve as an imaging tool to study cell-growth related problems [174]. As an example, Fig. 23 (a) presents the dry mass growth curves for a family of E. coli cells using SLIM. As a control, fixed cells were measured under the same conditions, from which we retrieved SD of 19.6 fg. Fig. 23 (b) shows the growth rate of 22 single cells as a function of mass, $dM(t)/dt$. The average of the data (black squares) shows that the growth rate is proportional to the mass, $dM(t)/dt = \alpha M(t)$. This indicates the growth is exponential [174].

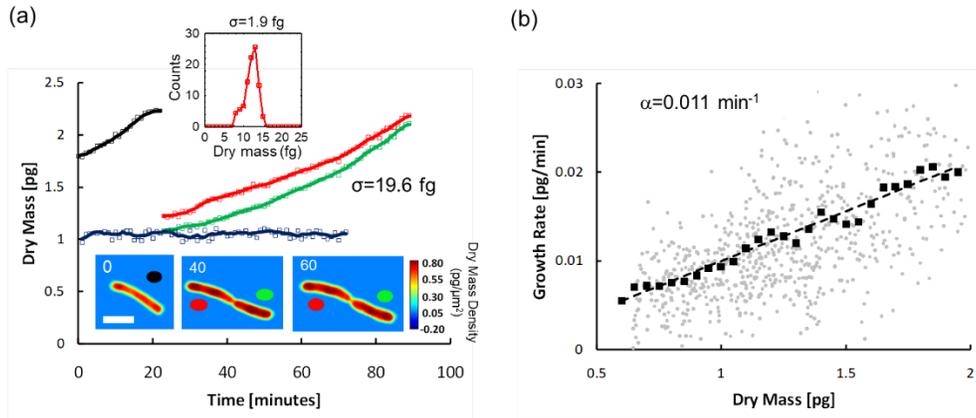

Figure 23 SLIM measurements of E. coli growth. (a) Dry mass vs. time for a cell family. Growth curves for each cell are indicated by the colored circles on the images. Images show single cell dry mass density maps at the indicated time points (in minutes). (Scale bar: 2 μm.) (Inset) Histogram of the dry mass noise associated with the background of the same projected area as the average cell (SD σ = 1.9 fg is shown). The blue line is a fixed cell measurement, with SD of 19.6 fg. Markers indicate raw data, and solid lines indicate averaged data. (b) Growth rate vs. mass of 20 cells measured in the same manner. Faint circles indicate single data points from individual cell growth curves, dark squares show the average, and the dashed line is a linear fit through the averaged data; the slope of this line, 0.011 min−1, is a measure of the average growth constant for this population. The linear relationship between the growth rate and mass indicates that, on average, E. coli cells exhibit exponential growth behavior. From Mir, Mustafa, et al. "Optical measurement of cycle-dependent cell growth." Proceedings of the National Academy of Sciences 108.32 (2011): 13124-13129.

As another example of cell proliferation, we performed long imaging of cell growth to estimate the "influence" of cellular clusters on their neighbors [175]. We analyzed epithelial and fibroblast cultures imaged over several days. Fig. 24 shows cell growth and division based on SLIM images. We can see that the time-lapse images of cells resemble "genealogical" trees, with cells splitting into branches after each division.

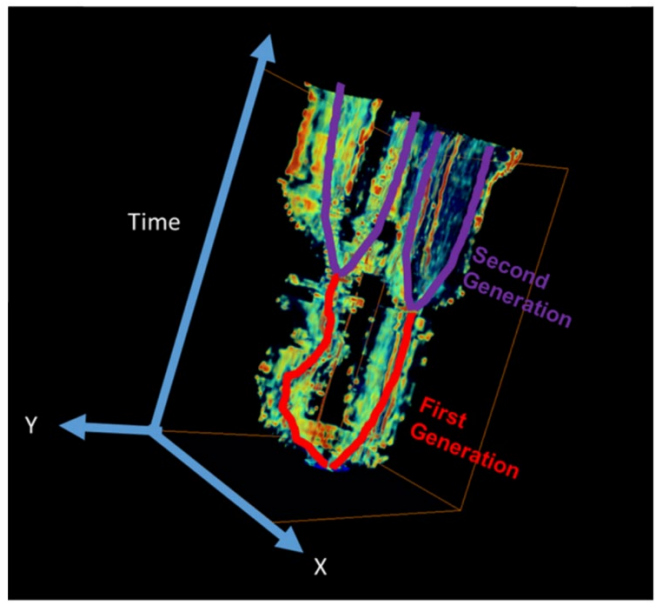

Figure 24 Cell growth resembles a genealogical tree when time is taken as the 3rd dimension, with two daughter cells after the first division (red), and four daughter cells (purple) after the second division. From Kandel, Mikhail E., et al. "Cell-to-cell influence on growth in large populations." Biomedical optics express 10.9 (2019): 4664-4675.

We can define a growth rate as the dry mass doubling time, $b$, from the fit of the exponential equation $M(t)/M(0) = 2^{bt}$. The covariance between distance and growth rate is $\text{cov}(r, b^{-1})$, such that the Pearson correlation coefficient between growth and distance is,

$$\rho = \frac{\text{cov}(r, b^{-1})}{\sqrt{\sigma_r \sigma_{b^{-1}}}}, \qquad (4.1.2)$$

where $\sigma_r$ and $\sigma_{b^{-1}}$ are the variances of the variable $r$ and $b^{-1}$. The distributions of the correlation coefficients for clusters of fibroblasts (NIH/3T3, ATCC CRL-1658) and epithelial (HeLa, ATCC CCL-2) cells are presented in Fig. 25 (b, d). We can see that there is a significant number of cells characterized by a medium correlation between their growth rate and distance (modulus of the Pearson coefficient between 0.25-.5). While a small percentage of cells exhibit strong correlations, which we label as "influencer" cellular clusters.

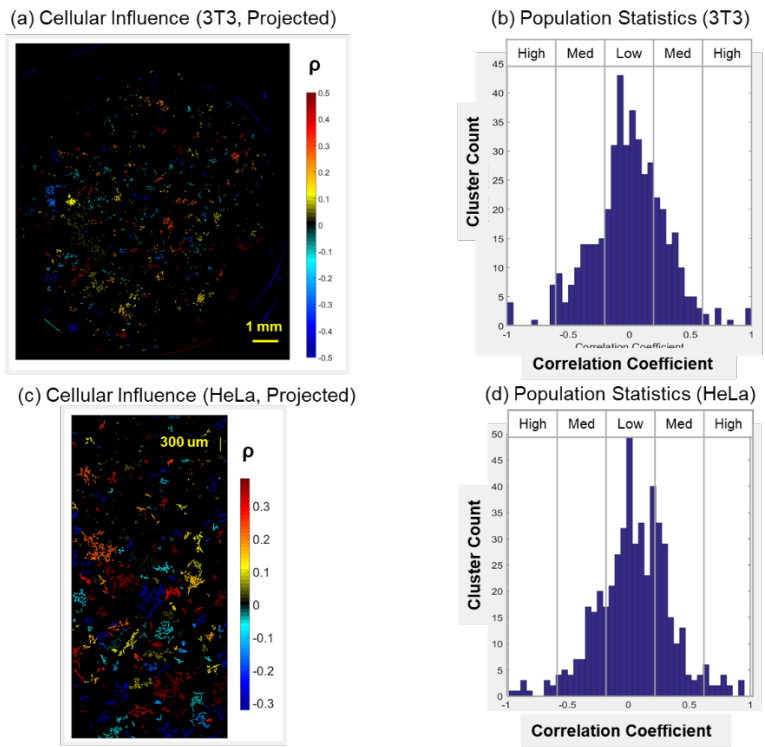

Figure 25 Projection of cellular influence. (a) Correlation coefficients for the 3T3 culture are projected onto the segmentation map at the end of the experiment. This gives a spatial distribution of correlation coefficients. (b) Histogram of the correlation coefficient for all 3T3 clusters. (c) Correlation coefficients for the HeLa culture are projected onto the segmentation map at the end of the experiment. (d) Histogram of the correlation coefficient for all HeLa clusters. From Kandel, Mikhail E., et al. "Cell-to-cell influence on growth in large populations." Biomedical optics express 10.9 (2019): 4664-4675.

### 4.1.3. Cell migration

Cell growth and motility are both crucial parts for understanding a proliferating cellular system. To study cell motility, we used SLIM to track single cells [176]. The trajectories of the attached (red) and motile cells (black) are shown in Fig. 26 (a). Clearly, the cell associated with the red trace is attached to the glass surface, as the motion exhibits very limited motion. The dry mass analysis for the attached and motile cells is illustrated in Fig. 26 (b). We can see that cells with lower mean square displacements (MSDs) (< 10 µm$^2$) present low or negative dry mass growth. The MSDs for these two cells are depicted in Fig. 26 (c). Fig. 27 (a) shows that the MSD increases by four orders of magnitude between the 1$^{st}$ and 4$^{th}$ generations. Fig. 27 (b) compares the maximum MSD for each cell with its fitted linear growth rate, indicating that the MSD and growth rate are related exponentially.

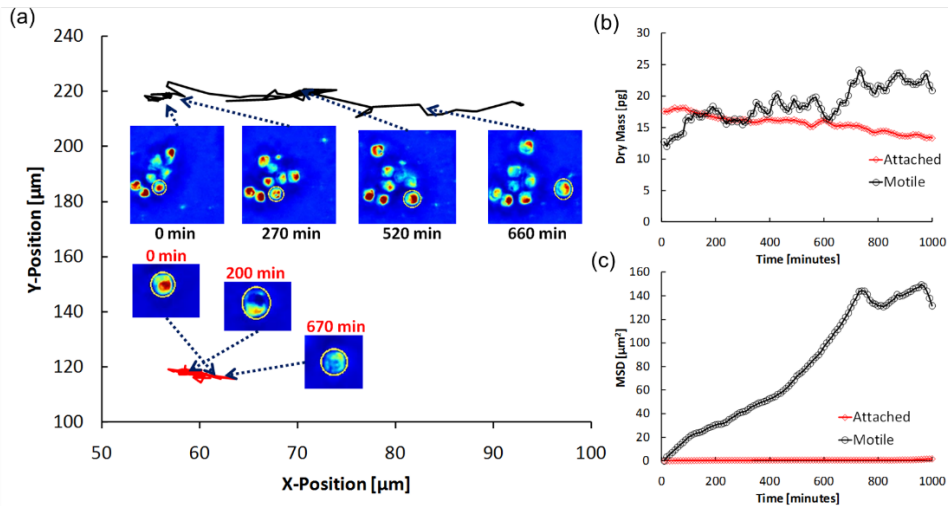

Figure 26 (a) Trajectories of attached (red line) and motile (black line) cells. Time-stamped insets show the tracked cell at various time points. The motile cell exhibits a clear directional motion over time whereas the adherent cell is jostling in place. (b) The dry mass growth of the two cells shown in a, the attached cell exhibits no growth whereas the motile cell approximately doubles its mass. (c) MSD for the two cells shown in (b). From Sridharan, Shamira, Mustafa Mir, and Gabriel Popescu. "Simultaneous optical measurements of cell motility and growth." Biomedical optics express 2.10 (2011): 2815-2820.

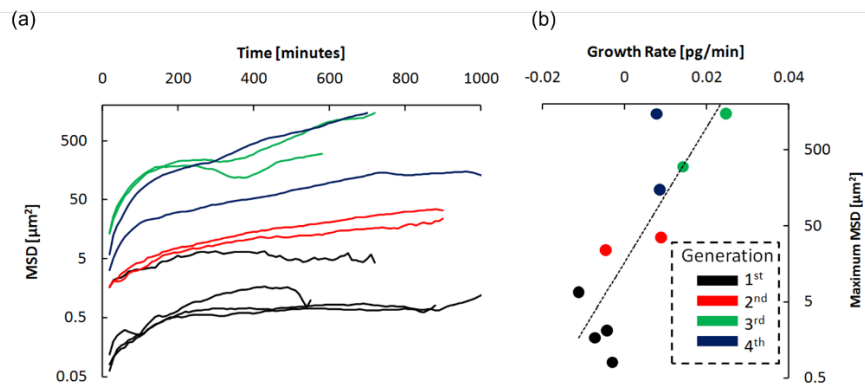

Figure 27 (a) Semilogarithmic plot MSD vs. time for all the individual cells tracked. It can be seen that the MSD increases by 3-4 orders of magnitude between the 1st and 4th generations (b) Semilogarithmic plot of the maximum MSD vs. the approximated linear growth rate for each cell. From Sridharan, Shamira, Mustafa Mir, and Gabriel Popescu. "Simultaneous optical measurements of cell motility and growth." Biomedical optics express 2.10 (2011): 2815-2820.

### 4.1.4. Intracellular transport

Dispersion relation phase spectroscopy (DPS) is a technique to characterize mass transport using SLIM [162, 177]. It starts with the assumption that the dry mass density satisfies the diffusion-advection equation, namely,

$$\left（-Dq^2 + i\mathbf{q}\cdot\mathbf{v} - \frac{\partial}{\partial t}\right）\eta(\mathbf{q},t) = 0, \qquad (4.1.3)$$

where $D$ is the diffusion coefficient and $\mathbf{v}$ is the velocity of the advection motion, and $\mathbf{q}$ is the spatial frequency, the Fourier conjugate to the 3D spatial coordinate. Under the assumption that the advection velocity distribution is Lorentzian with width $\Delta v$ and mean advection velocity $v_0$, the spatial Fourier transform of the temporal autocorrelation of dry mass density is in the form of

$$\langle g(\mathbf{q},\tau)\rangle_v = \exp(iqv_0\tau)\exp\left[-\left(Dq^2 + \Delta vq\right)\tau\right], \qquad (4.1.4)$$

where $\langle\ \rangle_v$ denotes the ensemble average over the velocity distribution. Therefore, the mean advection velocity produces a modulation frequency $qv_0$ to the temporal autocorrelation, whose envelope decays exponentially at a rate

$$\Delta\omega(q) = Dq^2 + \Delta vq. \qquad (4.1.5)$$

This relationship describes the dispersion relation associated with intracellular mass transport. The detailed steps are illustrated in Fig. 28. Through the time-lapse phase measurement (Fig. 28 Step 1) using SLIM, the Fourier transform of the temporal autocorrelation function (Fig. 28 Step 2) can render the decay rate $\Delta\omega(q)$ with the bandwidth estimation (Fig. 28 Step 3). The diffusion coefficient $D$ and the velocity $\Delta v$ are obtained by fitting the $\Delta\omega(q)$ function (Fig. 28 Step 4). Note that the diffusion coefficients and applicable spatial scales ($\lambda_{min}, \lambda_{max}$) are expected to vary between individual specimens, due to the diversity of transport phenomena in living cells.

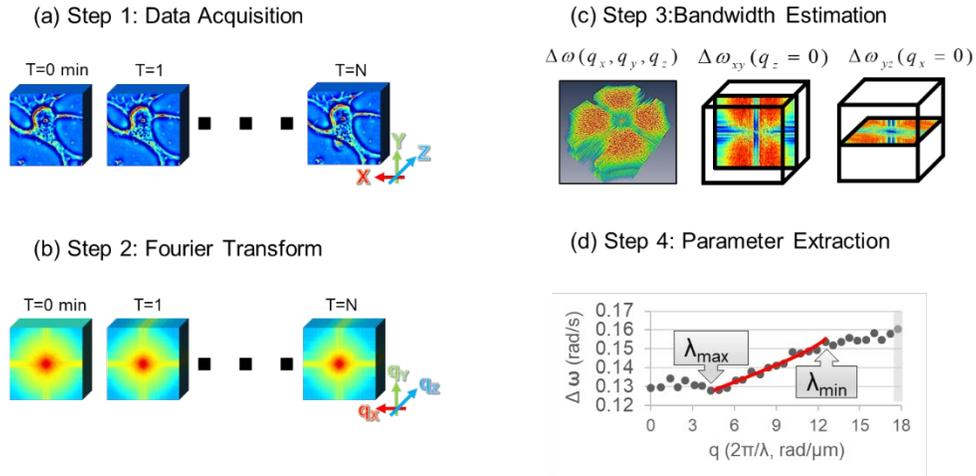

Figure 28 Dispersion relation calculated from the bandwidth of the Fourier transformed volumes. (a) The DPS transport assay consists of four steps. First, time-lapse tomograms are acquired, the Fourier transform is taken of each volume. (b) To estimate the dispersion relation at each spatial mode, we perform a forward difference which reduces the 3D data to a single cube. (c) This cube is then reduced to two lines by taking the radial average along the XY and YZ dimensions. (d) Yielding relations for horizontal and vertical motion. From Kandel, Mikhail E., et al. "Three-dimensional intracellular transport in neuron bodies and neurites investigated by label-free dispersion-relation phase spectroscopy." Cytometry Part A 91.5 (2017): 519-526.

As an example, we used the dispersion relation to study microtubule motility [178]. Figure 29 shows the DPS procedure for analyzing full-field, time-lapse SLIM images of microtubules. The value of $\Delta v$ considerably decreases linearly in time after approximately 20 min, most likely due to ATP depletion. This result implied that the velocity distribution narrows down, i.e., the probability of having high speeds decreases over time. This linear decrease with time reveals a deceleration in microtubule activity commensurate with a reduced availability of ATP.

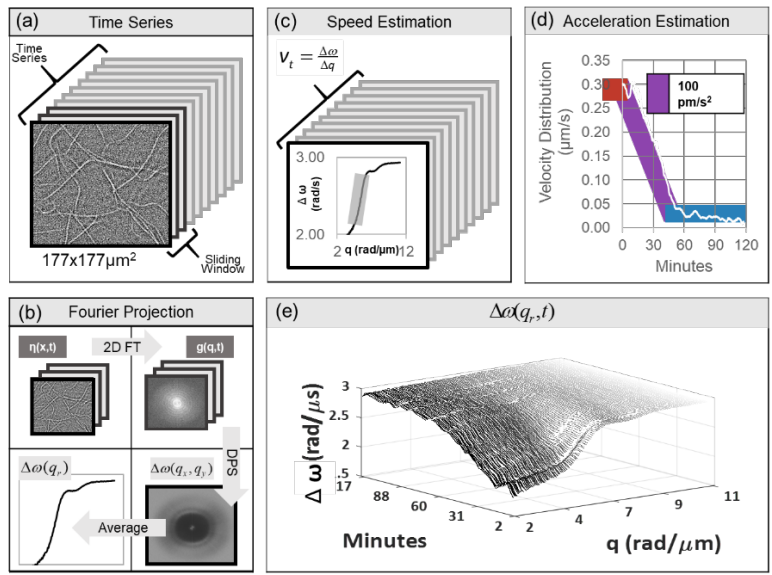

Figure 29 Long-term imaging reveals deceleration of microtubules. (a) Illustration of the time-resolved SLIM image stacks. b) The dispersion relation is computed over a temporal window of 128 frames for each time point in the series. In short, the method involves taking the 2D Fourier Transform of each image and computing the decay (or temporal bandwidth) at each spatial frequency. After the isotropic assumption, the volumetric data is reduced to a single dimension. When the window is advanced one of the old Fourier transforms is discarded and the bandwidth is recomputed. (c) Microtubule gliding velocity standard deviation was calculated using DPS, on a rolling basis over approximately 15,000 frames, taken 0.475 s apart. (d) In this run, after 20 min, the spread of the velocity distribution begins to decrease, with virtually no significant motion after the 60 min mark. (e) DPS signals vs time shows continuous change in slope. From Kandel, Mikhail E., et al. "Three-dimensional intracellular transport in neuron bodies and neurites investigated by label-free dispersion-relation phase spectroscopy." Cytometry Part A 91.5 (2017): 519-526.

### 4.1.5. Applications in neuroscience

The emergent of a neuronal network in a developing nervous system is a complicated process involving a multitude of chemical, mechanical and electrical signals. Studying neuronal networks is essential to understand brain connectivity and the mechanisms involved in central nervous system disease.

SLIM can measure several fundamental properties of neural networks from the sub-cellular to the cell population level [179]. For instance, SLIM can calculate the correlations between trends in the growth, transport, and spatial organization of neural networks [180]. It was found that cell density (confluence) affects significantly both the growth rate and mass transport [181].

The topological properties of the neuronal culture networks were analyzed to elucidate how neurons connect through time-lapse imaging with SLIM [182]. Figure 30 (a)-(c) show the SLIM images for three different timeframes. The zoomed portions of the middle region of the neurons are presented in Fig. 30 (d)-(f). The reconstructed neuronal culture networks are depicted in Fig. 30 (g)-(i) with the tracing algorithm. Different colors represent the different identifications for each neuron and neurite. After constructing the adjacency matrices from the tracing and segmentation algorithm, the visualization of the network layouts are illustrated for different timeframes in Fig. 30 (j)-(l).

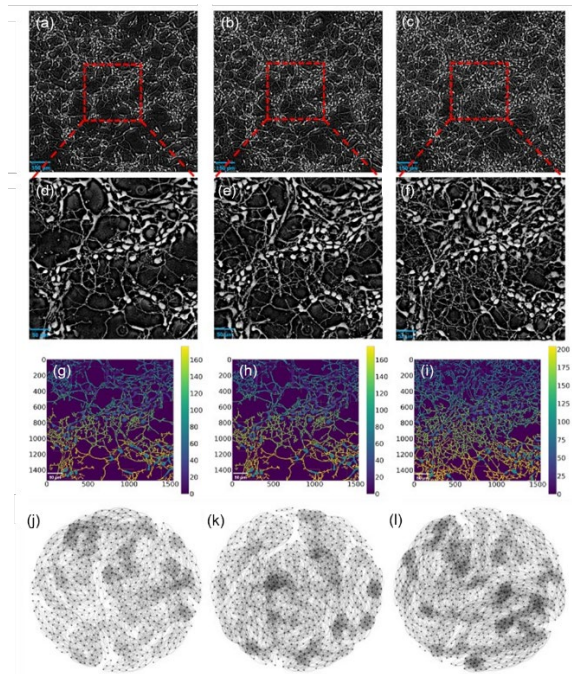

Figure 30 Layouts for neuronal culture networks at three representative time points. Neurons at the start of the experiment at time t = 0 h (a), t = 7 h (b) and the end of the experiment t = 14 h (c). The magnification zoom of the neurons at t = 0 h (d), t = 7 h (e) and t = 14 h (f). (g–i) show the identified neurons and their connections obtained with our algorithm (see "Methods" section on "Cell segmentation and neural tracing") for the three corresponding time points (each neuron and neurite is identified by a unique color). After constructing the adjacency matrices from the tracing and segmentation algorithm, the visualization of the network layouts at t = 0 h (j), 7 h (k) and 14 h (l) are presented. From Yin, Chenzhong, et al. "Network science characteristics of brain-derived neuronal cultures deciphered from quantitative phase imaging data." Scientific reports 10.1 (2020): 1-13.

## 4.2. Clinical applications

### 4.2.1. Cancer screening

The current practice of surgical pathology relies on external contrast agents to reveal tissue architecture, which is then qualitatively examined by a trained pathologist [183]. The diagnosis is based on the comparison with standardized empirical, qualitative assessments. Moreover, the analysis of the stained tissue is affected by staining strength, color balance, and imaging conditions.

SLIM is a label-free approach to pathology with unstained biopsies, which provides quantitative optical path length measurements that are sensitive to the nanoscale tissue architecture [184-186]. With our highly parallelized, dedicated software algorithms for data acquisition, SLIM allows us to image at a throughput comparable to that of commercial tissue scanners [155]. Based on the measured phase information, we implemented software tools for autofocusing during imaging, as well as image archiving and data access. To illustrate the potential of our technology for large volume pathology screening, we established an "intrinsic marker" for the colorectal disease that detects tissue with dysplasia or colorectal cancer and flags specific areas for further examination, potentially improving the efficiency of existing pathology workflows [155]. Figure 31 presents the automated processing workflow for SLIM images, and the procedure for tissue classification. In short, cancer is detected through these two parameters solidity ("glandular solidity") and median phase value. We use a support vector machine (SVM) to determine their optimal fusion (Fig. 32). We can see that the flagged regions based on the SVM model agree well with the classifications by the pathologist in Fig. 33.

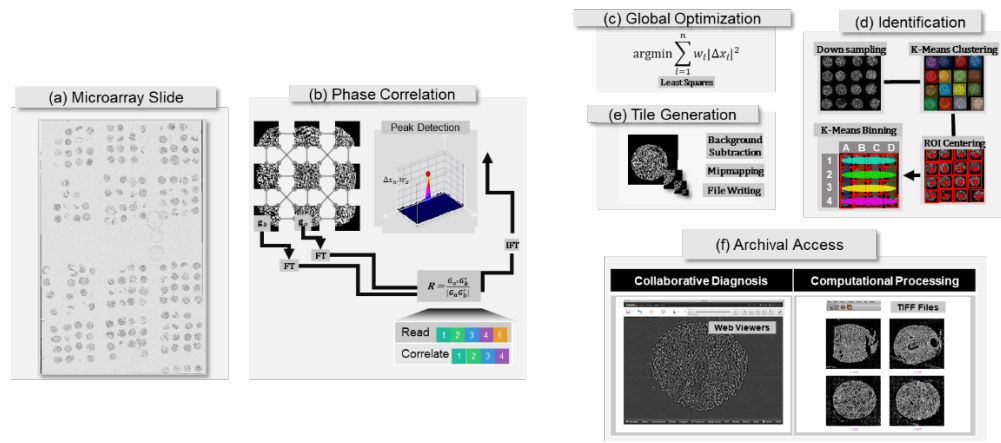

Figure 31 Automated processing workflow for gigapixel SLIM images. (a)Typical samples used in this paper consist of tens of thousands of tiles. (b) The tiles are assembled using a "phase correlation" scheme where the optimal displacement between neighboring tiles is determined by the location of the peak in the correlation image. (c) In our implementation disk access is overlapped with the correlation procedure, with performance dependent on the longer disk read and write operations. Disagreements between estimated tile positions are resolved with a least-squares fit (d) and the result positions are used to generate image pyramids typically used for archival access. (e) Optimally, regularly

spaced tissue microarray cores are cropped and labeled by way of a thresholding technique. (f) Resulting images are ready for computational processing and collaborative diagnosis. From Kandel, Mikhail E., et al. "Label-free tissue scanner for colorectal cancer screening." Journal of biomedical optics 22.6 (2017): 066016.

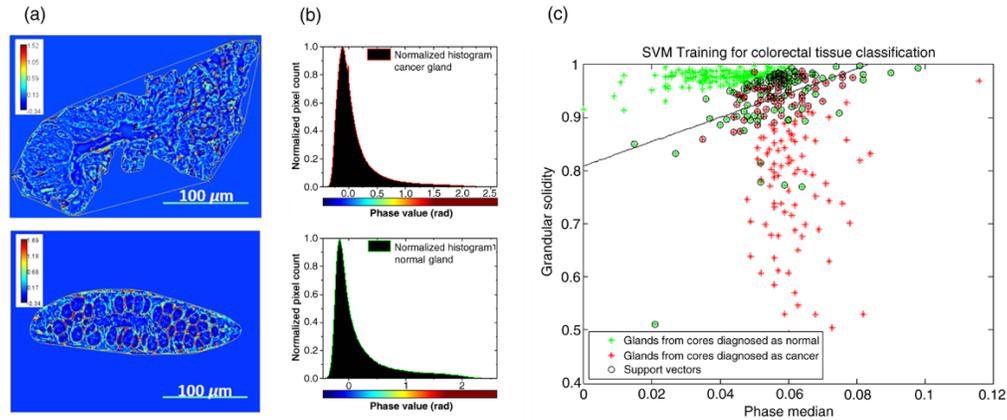

Figure 32 Quantitative parameters for classification. Our classification method augments phase information with the geometric structure of the gland. Scanned images are manually segmented into glands, whose solidity ("glandular solidity") and median phase value are used to train the SVM-based classifier used in this work. (a) Gland identification, (b) feature extraction, and (c) classification. From Kandel, Mikhail E., et al. "Label-free tissue scanner for colorectal cancer screening." Journal of biomedical optics 22.6 (2017): 066016.

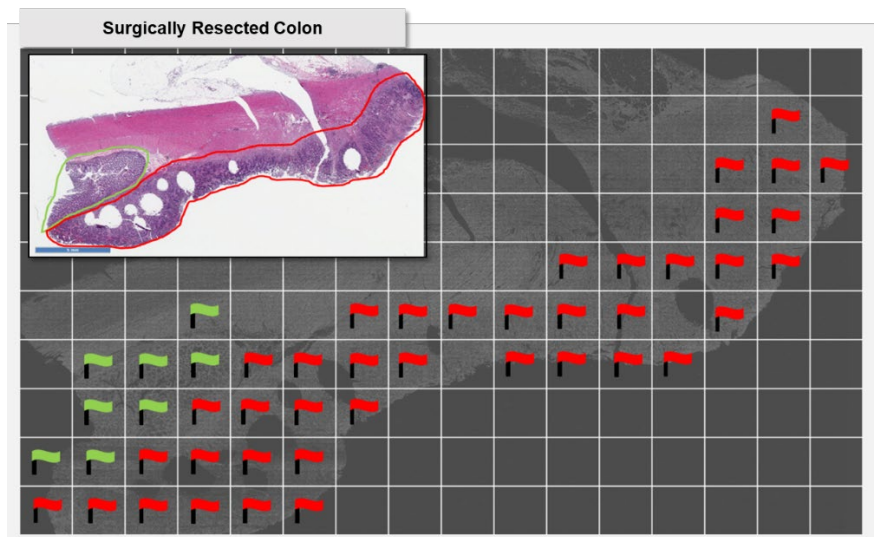

Figure 33 Biopsy flagged with regions of "high" (red) and "low"' suspicion (green). After assembly, the image was analyzed in chunks of 10;000 × 10;000 pixel regions with glands in each region evaluated according to the SVM model. The consensus of constituent glands is represented with a "green" or "red" flag indicating low or high index of suspicion, respectively. Inset: H&E stained parallel section, showing the red-bordered tumor and green-border benign region, as indicated by the pathologist. From Kandel, Mikhail E., et al. "Label-free tissue scanner for colorectal cancer screening." Journal of biomedical optics 22.6 (2017): 066016.

### 4.2.2. Cancer diagnosis

Normal and diseased tissues are characterized by different scattering parameters. Thus, SLIM can serve as a scattering-based diagnosis tool by measuring the scattering mean free path $l_s$ and anisotropy factor $g$, defined as [185]

$$l_s = \frac{L}{\langle \Delta\phi^2(\mathbf{r})\rangle_{\mathbf{r}}}, \qquad (4.2.1)$$

$$g = 1 - \left(\frac{l_s}{L}\right)^2 \frac{\langle |\nabla[\phi(\mathbf{r})]^2|\rangle_{\mathbf{r}}}{2k_0^2}. \qquad (4.2.2)$$

Where $L \ll l_s$ is the tissue slice thickness, $\langle \Delta\phi^2(\mathbf{r})\rangle_{\mathbf{r}} = \langle [\phi(\mathbf{r}) - \langle\phi(\mathbf{r})\rangle_{\mathbf{r}}]^2 \rangle_{\mathbf{r}}$ is the phase variance, $k_0$ is the wavenumber of the illumination. The definition of $g$ (anisotropy) is the average cosine of the scattering angle associated with a slice of thickness $l_s$. This way, the assumption that the tissue is made of discrete particles is removed. Figure 34 (a) shows the SLIM image of a tissue slice cut from a three-month-old rat liver. The scattering mean-free-path map and the anisotropy factor map are shown in Fig. 34 (b) and (c).

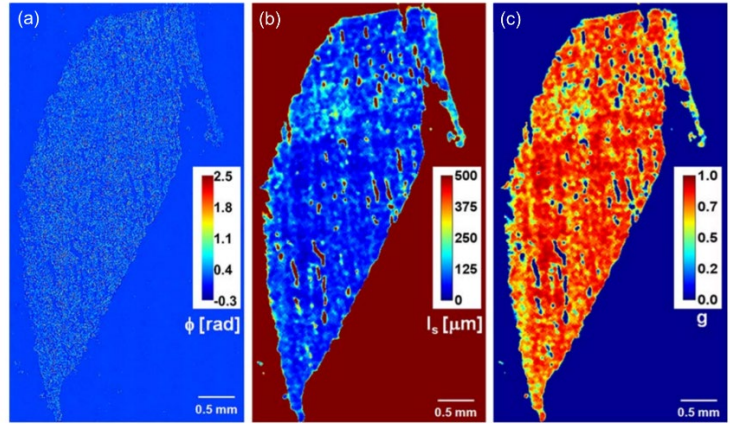

Figure 34 Maps of (a) ϕ, (b) ls, and (c) g for a tissue slice across an entire rat liver; the g map is thresholded to show g = 0 for background. Color bars show ϕ, ls and g, as indicated. From Ding, Huafeng, et al. "Measuring the scattering parameters of tissues from quantitative phase imaging of thin slices." Optics letters 36.12 (2011): 2281-2283.

Another metric that can potentially be an intrinsic cancer marker is the spatial autocorrelation length, defined as the variance of the autocorrelation function of the phase from SLIM images [187]. Figure 35 presents the SLIM images, the spatial autocorrelation maps, and the spatial autocorrelation maps after filtering.

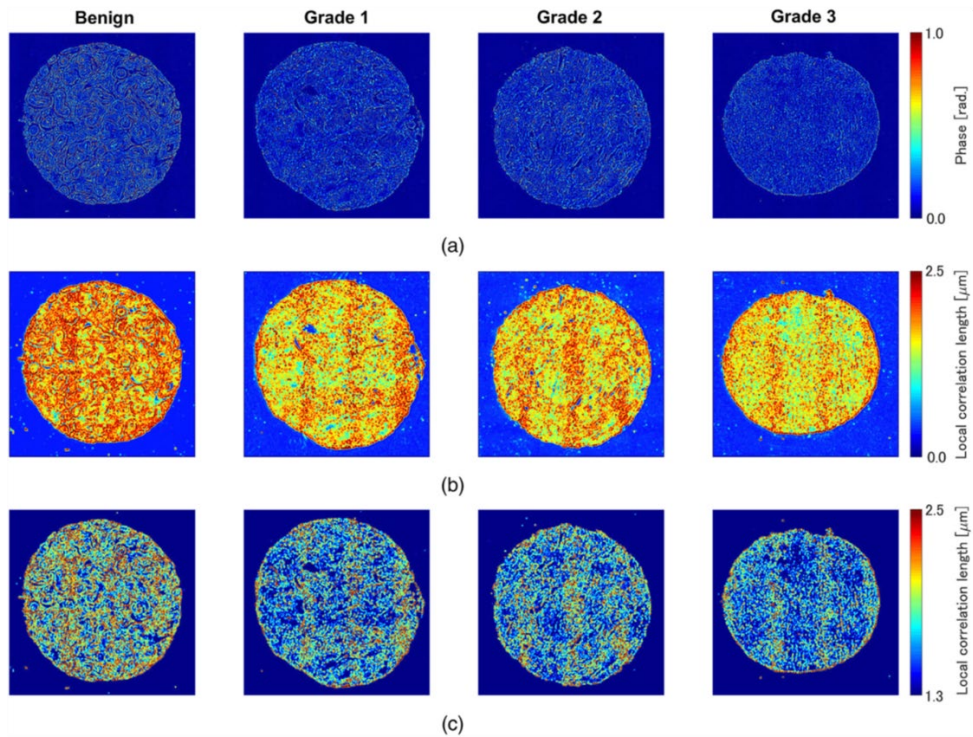

Figure 35 Example of local spatial autocorrelation length maps. (a) Quantitative phase images, (b) the local correlation length maps, and (c) the local correlation length maps. From Takabayashi, Masanori, et al. "Tissue spatial correlation as cancer marker." Journal of biomedical optics 24.1 (2019): 016502.

Figure 36 shows the comparison between the H&E stained bright-field microscopy and SLIM images [188]. Pathologists can be trained with SLIM images to diagnose cancer with about 88% agreement with the results diagnosed with H&E images [188]. With machine learning techniques, the accuracy of diagnosis can be improved, which we will discuss in detail in Chapter 5.

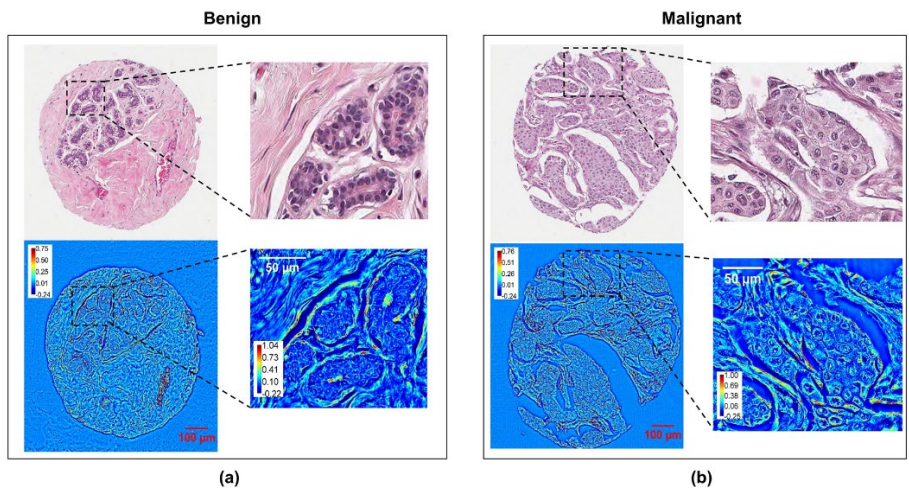

Figure 36 Comparison between H&E stained bright-field microscopy (top row) and SLIM (bottom row) images in their respective abilities to resolve tissue morphology for (a) benign and (b) malignant cases. The H&E images were obtained from stained sections that were adjacent to the unstained sections used for SLIM imaging. Color bars are in radians. From Majeed, Hassaan, et al. "Breast cancer diagnosis using spatial light interference microscopy." Journal of biomedical optics 20.11 (2015): 111210.

### 4.2.3. Cancer prognosis

Cancer prognosis is an estimate of how the disease will proceed. It includes the recovery rate, the recurrence rate, and the clinicians' predictions for the course of the disease. Cancer prognosis can be affected by many factors such as the type of cancer, the stage of cancer, cancer's grade, certain traits of the cancer cells, etc. SLIM can help with cancer prognosis in the aspects of cancer recurrence rate and tumor progression [189, 190].

Prediction of recurrence risk of prostate cancer is after prostatectomy critical for determining whether the patient would benefit from adjuvant treatments. The method based on the optical anisotropy $g$ (see Eq. (4.2.2)) of SLIM images can identify recurrent cases with 73% sensitivity and 72% specificity, which is superior for the same sample set to that of CAPRA-S, a current state of the art method. Optical anisotropy $g$ was calculated in the single stromal layer adjoining 6–18 glands from each of the 33 patients with post-prostatectomy biochemical recurrence of prostate cancer and 159 patients who did not have a recurrence. The calibrated anisotropy value in the recurrent cases (0.913±0.028; median=0.92) was lower than that in the non-recurrent cases (0.932±0.023; median=0.938). The difference in anisotropy values in the cancer-adjacent stroma from the recurrent and non-recurrent groups was statistically significant (one-way ANOVA, $p=7.05\times 10^{-5}$). Figure 37(a) summarized these results. Kaplan-Meier survival analysis was performed to test the utility of anisotropy for predicting biochemical recurrence as the end-point. The anisotropy ranges tested were 0.68–0.93 (67 patients) and 0.93–0.97 (125 patients) and the results (Fig. 37(b)). It shows that patients with low anisotropy values had a higher likelihood of disease progression. The 3-year and 5-year recurrence-free probability dropped from 95% and 90% respectively for patients with high anisotropy values to 70% and 65%, for patients with low anisotropy values. The comparison of the ability of anisotropy to other methods is shown in Fig. 37(c) using the receiver-operating curve analysis.

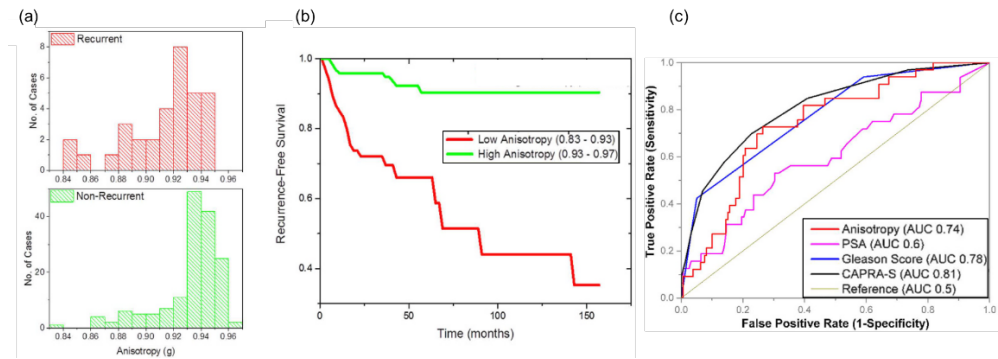

Figure 37 (a) Histograms of the distribution of anisotropy in the single layer of stroma surrounding 6–18 glands from 33 patients with post-prostatectomy biochemical recurrence of prostate cancer and 159 non-recurrent patients. The bin-size on the histogram was set at 0.02. The anisotropy value is lower in the recurrent patients, compared to the non-recurrent patients (One-way ANOVA, $p = 7.05 \times 10^{-5}$) (b) Kaplan-Meier survival curve with end-point as disease recurrence for 67 patients with low anisotropy values (0.83–0.93) and 125 patients with high anisotropy values (0.93–0.97). (c) Comparison of recurrence prediction metrics. The performance of anisotropy measured on quantitative phase images, pre-surgical prostate-specific antigen (PSA) levels, Gleason score and CAPRA-S as post-prostatectomy biochemical recurrence predictors was studied in 192 prostatectomy cases (33 recurrent, 159 non-recurrent). The best performance was observed with CAPRA-S (AUC 0.81) and Gleason scores (AUC 0.78). The discriminatory ability of anisotropy (AUC 0.74) was lower than that of CAPRA-S and Gleason score. However, at the optimal performance point, anisotropy had a sensitivity of 72.7% and specificity of 73.6% compared to the 69.6% sensitivity and 77.4% specificity of CAPRA-S. Pre-surgical PSA level (AUC 0.6) was a poor predictor of recurrence. From Sridharan, Shamira, et al. "Prediction of prostate cancer recurrence using quantitative phase imaging: Validation on a general population." Scientific reports 6.1 (2016): 1-9.

SLIM can also assist with the pancreatic ductal adenocarcinoma (PDAC) prognosis by quantitative analysis of the PDAC fiber structures [169]. Fibrillar collagen in PDAC exhibits an inverse relationship between survival data and fiber width and length ($p < 0.05$). PDAC patients with high alignment per length of segmented fibers show significantly reduced survival rates (Fig. 38(a)). In low survival cases, fiber width and length were greater (Fig. 38(b) and (c)). The straightness has less predictive value in survival rates (Fig. 38 (d)). SLIM was also used for the prognosis of breast cancer [168].

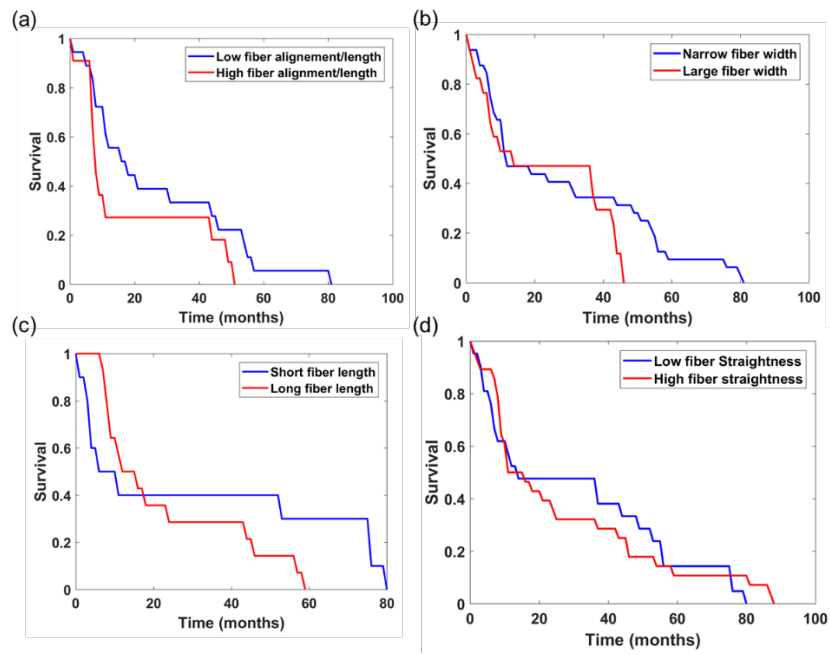

Figure 38 Kaplan-Meier survival curves for pancreatic ductal adenocarcinoma (PDAC), comparing different grades of fiber alignment/length (a), fiber width (b), fiber length (c), and fiber straightness (d). Log-Rank "χ" ^2 of 50.7 (str), 37.43 (width), 25.7 (al), 50.8 (length). From Fanous, Michael, et al. "Quantitative phase imaging of stromal prognostic markers in pancreatic ductal adenocarcinoma." Biomedical Optics Express 11.3 (2020): 1354-1364.

### 4.2.4. SLIM as assisted reproductive technology

The high incidence of male factor infertility affects human and animal reproduction. The ability to evaluate sperm at the microscopic level, at high throughput, is valuable for assisted reproductive technologies (ARTs), as it can allow specific selection of sperm cells for in vitro fertilization (IVF). SLIM, as a non-invasive label-free imaging modality with high sensitivity, can assist with this task [191].

To introduce specificity to SLIM images, we trained a deep-convolutional neural network to perform semantic segmentation, which we will discuss more in detail in Chapter 6. This method can efficiently analyze thousands of sperm cells and identify correlations between dry-mass content and artificial-reproduction outcomes. Specifically, we found that the dry-mass content ratios between the head, midpiece, and tail of the cells can predict the rate of success for zygote cleavage and embryo blastocyst formation. Fig. 39 (a) shows a mosaic made from 27,000 tiles, covering an area of 7.1 cm$^2$. After the halo removal procedure (Fig. 39 (b)), the 3D reconstructed sperm cell is shown in Fig. 39 (c). We can see that deep learning with SLIM imaging tracks subtle but significant differences in sperm morphology (Fig. 39 (d)). The summary of the outcomes with this method will be presented in detail in Chapter 5.

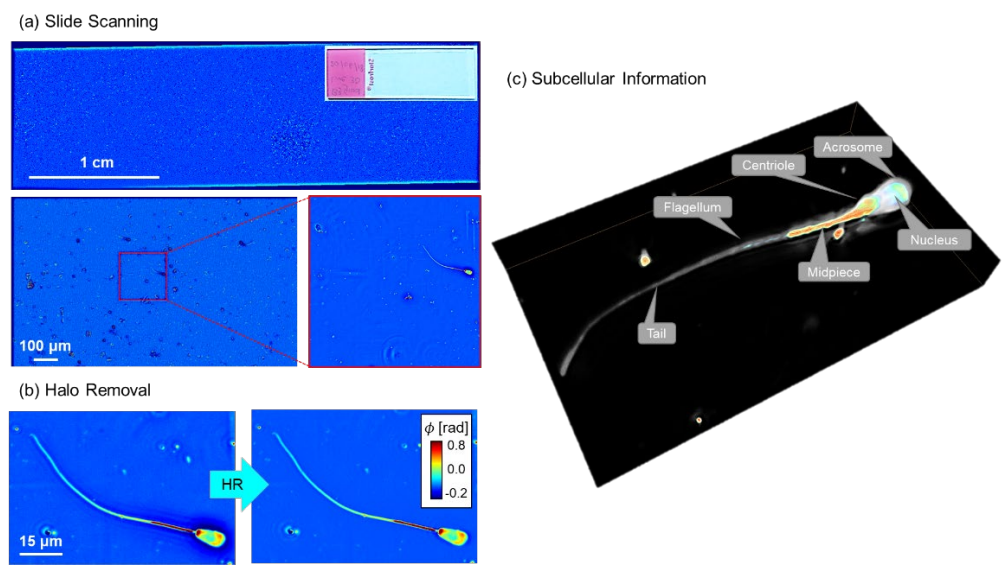

Figure 39 SLIM can image sperm as a fully-automated slide scanner, with thousands of samples on each slide. (a) A large number of samples in each slide motivates the use of automated segmentation techniques. (b) The superior sensitivity of SLIM images is, in part, due to the use of spatially and temporally broadband fields. The partially coherent illumination corrupts the low frequencies, evident as a halo glow surrounding the cell. The halos are corrected by solving a non-linear inverse problem. (c) Tomographic rendering of a spermatozoon using SLIM. The mitochondria-rich midpiece appear as substantially higher in dry mass density. Rendering of the tomogram was performed using AMIRA with the "physics" colormap corresponding to high phase values and a grayscale colormap corresponding to the lower phase values in the nucleus and tail. From Kandel, Mikhail E., et al. "Reproductive outcomes predicted by phase imaging with computational specificity of spermatozoon ultrastructure." Proceedings of the National Academy of Sciences 117.31 (2020): 18302-18309.

### 4.2.5. Blood testing

Impedance counters and flow cytometers are common laboratory methods to detect, identify, and count specific cells from blood samples. However, those methods are often limited to population-level statistics for morphology and bulk measurement in the case of hemoglobin concentration. In some cases, the morphological properties of a single blood cell are necessary for diagnosis such as leukemia [192-194]. The morphological properties are also important to assess the banked blood because stored red blood cells (RBCs) undergo numerous biochemical, structural, and functional changes, commonly referred to as storage lesions. SLIM can measure the thickness, refractive index, the membrane fluctuation of the RBCs, which can report further on the cell stiffness [195]. This property directly affects the cell's ability to transport oxygen in the microvasculature.

Figure 40 (a) presents the temporal standard deviation ($\sigma_T$) map from the 128 SLIM images. The histogram of the $\sigma_T$ map is shown in Fig. 40 (b). The arrow points to the spatial average of the $\sigma_T$ map, which we use as the representative displacement parameter for the particular RBC. The histograms of the $\sigma_T$ for different weeks are summarized in Fig. 40 (c). The arrows show the mean phase fluctuation of N = 110 ± 15 cells. It can be seen that the position of these arrows consistently shifts toward lower values with time, indicating the cell stiffness increases over time.

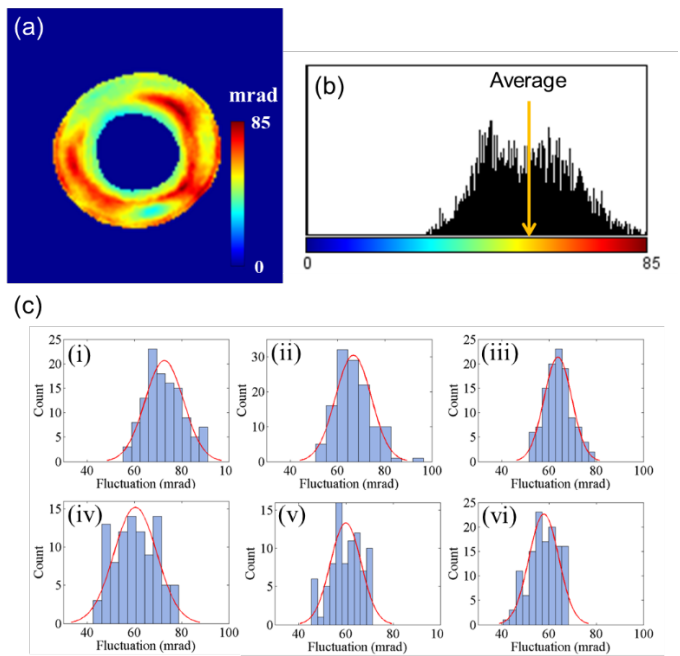

Figure 40 RBC fluctuation: (a) Temporal standard deviation map of a single RBC. (b) Histogram of the STD map in (a); representative average of the STD map is shown by the arrow. (c) Histogram of the average STD values for 110 6 15 RBCs at different weeks. From Bhaduri, Basanta, et al. "Optical assay of erythrocyte function in banked blood." Scientific reports 4 (2014): 6211.

The mean cell hemoglobin (MCH) can be obtained by calculating the dry mass of the RBCs. Figure 41 depicts the MCH as a function of storage time for all the samples. Note that the MCH does not change with storage time. This result indicates that, while cells can undergo volume changes during storage, they do not lose hemoglobin into the storage solution.

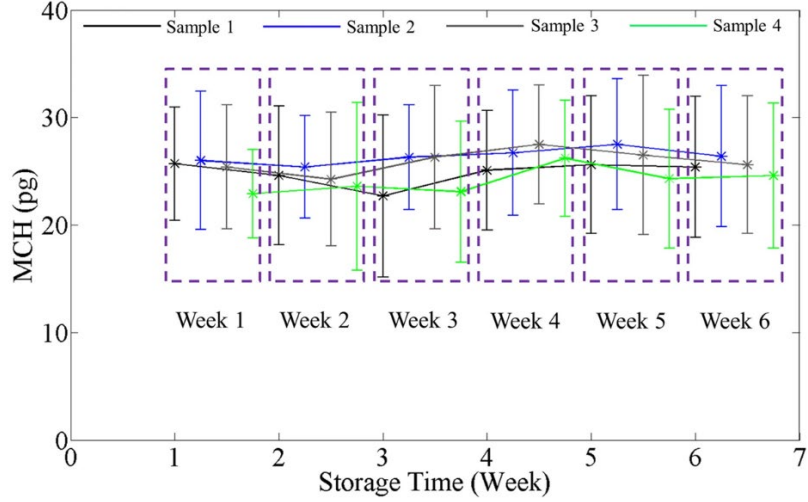

Figure 41 Variation in mean cell hemoglobin with storage time. Data points for different samples are shifted to distinguish them from each other and grouped by the week of the measurement. The error bars in the plots are twice the standard deviation of the MCH calculated over the groups of cells (N 5 110 6 15). Form Bhaduri, Basanta, et al. "Optical assay of erythrocyte function in banked blood." Scientific reports 4 (2014): 6211.

### 4.3. Diffraction tomography using SLIM

### 4.3.1. White-light diffraction tomography (WDT)

White light tomography extends the theory of diffraction tomography to white-light illumination [122]. Using SLIM images, the scattering potential is solved by deconvolving with the impulse response for white light under Born approximation. The axial dimension of the object is reconstructed by scanning the focus through the object. This method is capable of rendering three-dimensional (3D) tomograms for unlabeled live cells with 350 nm transverse and 950 nm axial resolution.

The scattered field for the incident plane wave, $U_i = A(\omega)e^{i\beta(\omega)z}$, under Born approximation, can be calculated as (Fig. 42(a))

$$U_s(\mathbf{k}_\perp, z; \omega) = -\frac{\beta_0^2(\omega)A(\omega)e^{i\gamma z}}{2\gamma}\chi\left[\mathbf{k}_\perp, \gamma - \beta(\omega)\right], \quad (4.3.1)$$

(see definitions of variables in Chapter 2) The scattering potential can be reconstructed with the knowledge of the coherent transfer function and the correlation function between the incident and scattered field $\Gamma_{is}(\mathbf{r}_\perp, z; \tau) = \langle U_i^*(z, t+\tau)U_s(\mathbf{r}_\perp, z; t)\rangle$ as

$$\chi(\mathbf{k}) = \frac{\Gamma_{is}(\mathbf{k}; 0)}{\Sigma(\mathbf{k})}, \quad (4.3.2)$$

where the coherent transfer function (CTF) $\Sigma(\mathbf{k})$ can be calculated as

$$\Sigma(\mathbf{k}) = \frac{1}{8\bar{n}^2} \frac{(Q^2 + k_\perp^2)^2}{Q^3} S\left(-\frac{Q^2 + k_\perp^2}{2Q}\right), \qquad (4.3.3)$$

where $S$ is the spectral density and $Q = \sqrt{\beta^2 - \mathbf{k}_\perp^2} - \beta$. The Fourier transform of the coherent transfer function gives the point spread function (PSF). We can see from this relation that broader spectral density and higher numerical aperture (NA) will give a narrower PSF. Figure. 42 (b,c) illustrates the transfer function $\Sigma(\mathbf{k})$ for the system. It can be seen that the width of $k_z$ coverage increases with larger $k_x$, meaning that the sectioning is stronger for finer structures. The transverse and longitudinal cross-sections of the calculated and measured $\Sigma(x, y, z)$ are shown in Fig. 42 (d,e). The structure of the object is finally recovered via a sparse deconvolution algorithm. Figure 43 presents the WDT of HT 29 cells.

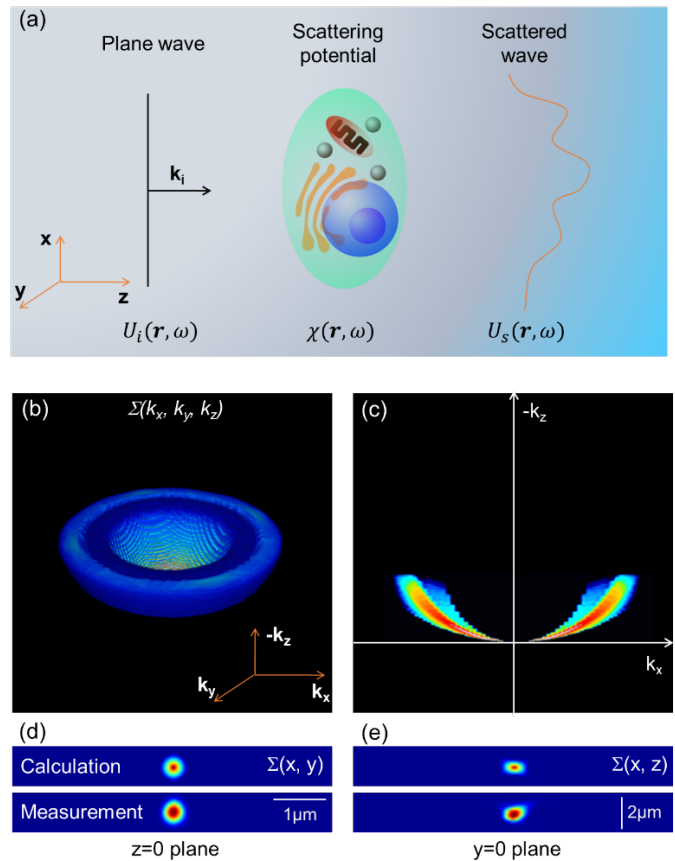

Figure 42 The scattering problem. (a) Illustration of light scattering under the first-order Born approximation where a plane wave's wavefront is perturbed by the object. (b) Three-dimensional rendering of the instrument transfer function, using the proposed WDT calculation. (c) Cross-section of the transfer function at the ky = 0 plane. (d) Calculated and measured PSF at the z = 0 plane. (e) Calculated and measured PSF in the y = 0 plane. From Kim, Taewoo, et al. "White-light diffraction tomography of unlabeled live cells." Nature Photonics 8.3 (2014): 256-263.

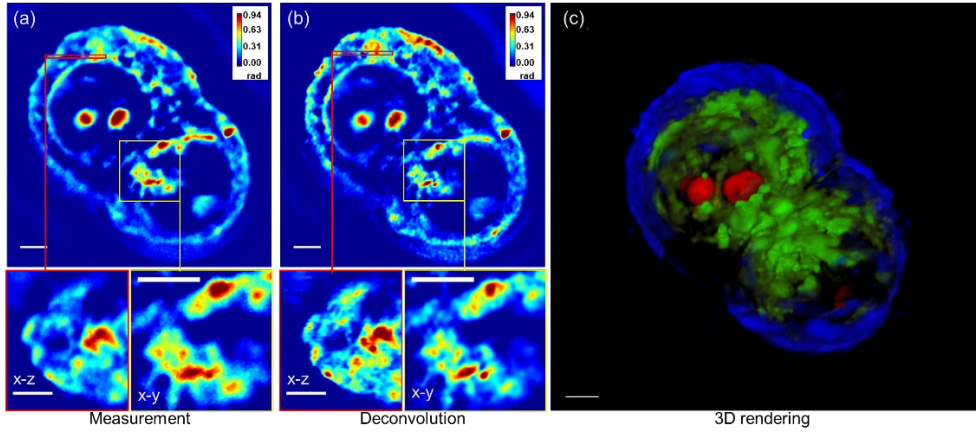

Figure 43 WDT of HT29 cells. (a) A measured z-slice (top), a cross-section at the area indicated by the red box (bottom left) and a zoomed-in image of the area indicated by the yellow box (bottom right), measured using a ×63/1.4 NA oil immersion objective. (b) A deconvolved z-slice corresponding to the measurement shown in a (top), a cross-section at the area indicated by the red box (bottom left) and a zoomed-in image of the area indicated by the yellow box (bottom right). By comparing a and b, the resolution increase can be clearly seen. (c) False-colour three-dimensional rendering of the deconvolution result. We used z-stacks of 140 images, each with a dimension of 640 × 640. Owing to the large image dimension, the image is split into 25 sub-images for faster deconvolution. Overall, the deconvolution process took approximately an hour. Scale bars in all panels, 5 microns. From Kim, Taewoo, et al. "White-light diffraction tomography of unlabelled live cells." Nature Photonics 8.3 (2014): 256-263.

### 4.3.2. Wolf phase tomography (WPT)

Wolf phase tomography (WPT) [79] is a fast 3D refractive index (RI) reconstruction method, based on the Wolf equations for propagating correlations of partially coherent light. This approach involves minimal computational steps, renders high-resolution RI tomograms, without time-consuming deconvolution operations. WPT decouples the refractive index distribution from the thickness of the sample directly in the space-time domain, without the need for Fourier transformation. From three independent intensity measurements corresponding to each phase shift in SLIM, the RI distribution is reconstructed right away from the Laplacian and second time derivative of the complex correlation functions. WPT is capable of extracting intrinsic refractive index changes in live cells with a sensitivity on the order of $10^{-5}$. The 3D RI can be reconstructed by

$$n(\mathbf{r}) = \sqrt{\frac{m(\mathbf{r}) - n_0^2 [1 - g(\mathbf{r})]}{1 + g(\mathbf{r})}}, \qquad (4.3.4)$$

where the functions $m$ and $g$ are defined as

$$m(\mathbf{r}) = \frac{c^2 \left( \nabla^2 \Re[\Gamma_{is}(\mathbf{r},\mathbf{r},\tau)] + \zeta(\mathbf{r}) \right)}{\dfrac{\partial^2 \Re[\Gamma_{is}(\mathbf{r},\mathbf{r},\tau)]}{\partial \tau^2}} \Bigg|_{\tau = -\pi/\langle\omega\rangle}, \qquad (4.3.5)$$

$$\zeta(\mathbf{r}) = -2\Re \int_0^\infty \langle \nabla U_i^*(\mathbf{r},\omega) \cdot \nabla U_s(\mathbf{r},\omega) \rangle e^{i\omega\pi/\langle\omega\rangle} d\omega, \qquad (4.3.6)$$

$$g(\mathbf{r}) = \frac{\left.\frac{\partial^2 \Re[\Gamma_{ii}(\mathbf{r},\mathbf{r},\tau)]}{\partial \tau^2}\right.}{\left.\frac{\partial^2 \Re[\Gamma_{is}(\mathbf{r},\mathbf{r},\tau)]}{\partial \tau^2}\right|_{\tau=-\pi/\langle\omega\rangle}}, \quad (4.3.7)$$

$c$ is the speed of light in vacuum, $n_0$ is the refractive index of the background. The correlation functions are defined as $\Gamma_{pq}(\mathbf{r}_1,\mathbf{r}_2,\tau) = \langle U_p^*(\mathbf{r}_1,t)U_q(\mathbf{r}_2,t+\tau)\rangle_t$, $p,q = \{i,s\}$. $\Re$ denotes the real part. $\Gamma_{is}$ is the correlation function between the incident and scattered fields, $\Gamma_{ii}$ is the autocorrelation function of the incident fields. The term in Eq. (4.3.6) does not substantially contribute to the final RI and can be omitted for faster construction.

The normalized spectrum of the halogen source measured by the spectrometer (Ocean Optics) is shown in Fig. 44 (a). The real part of the normalized autocorrelation $\Gamma_{ii}$ is obtained by taking the Fourier transform of the spectrum. The second-order time derivative of $\Gamma_{ii}$ is depicted in Fig. 44 (a). The real part of $\Gamma_{is}$ can be solved from SLIM images (see Fig. 44 (b) and (c)). The Laplacian in Eq. (4.3.5) is calculated using three images of $\Re[\Gamma_{is}]$ (shown in Fig. 44 (c)) with the first-order finite difference approximation. The z-component of the Laplacian was computed using three axially distributed frames, separated by a distance that matches the x-y pixel sampling and is much smaller than the diffraction spot. The second-order derivatives in Eq. (4.3.5) and (4.3.7) are calculated in MATLAB, using three phase-shifted frames. The WPT algorithm requires 40 ms to reconstruct the refractive index map at one z-position, with a 3-megapixels field of view.

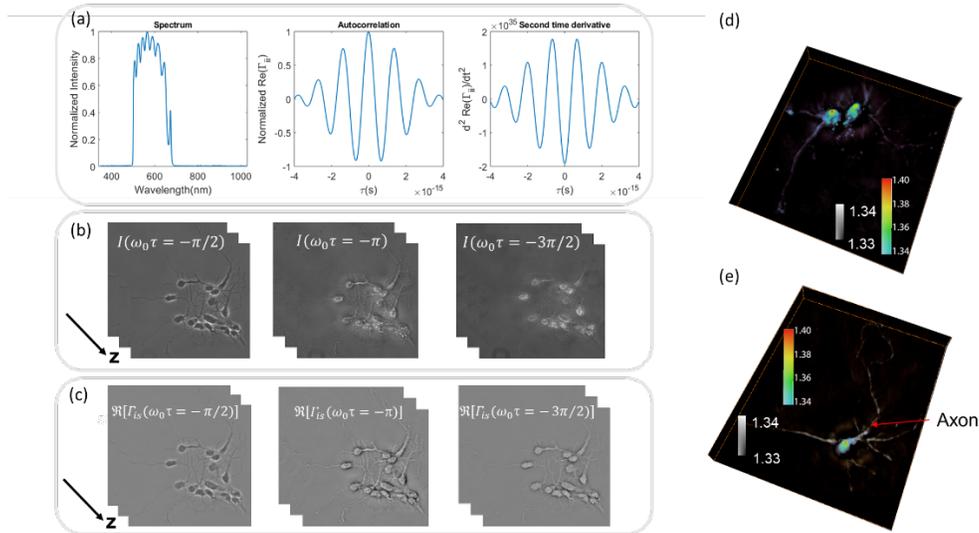

Figure 44 Working principle of WPT. (a) Spectrum, Autocorrelation and Second-order derivative of the autocorrelation of the halogen source measured by the spectrometer. (b) Three phase-shifted frames of hippocampal neurons (40x/0.75NA objective). (c) The real part of the correlation function at three different time-lags. (d, e) 3D rendering of RI tomograms of the hippocampal neurons. Two colormaps as indicated are used to enhance the dendrites and axons. The axon is pointed with a red arrow. From Chen, X., et al., Wolf phase tomography (WPT) of transparent structures using partially coherent illumination. Light: Science & Applications, 2020.

The 3D RI tomography of neurons is presented in Fig. 44 (d) and (e). The 3D rendering of a bovine sperm cell is displayed in Fig. 45 (a). In the sperm head, the acrosome and the nucleus can be identified with RI values between 1.35 and 1.37. The centriole and mitochondria-rich midpiece of the sperm cell yield high refractive index values (Fig. 45 (b)). The tail of the sperm has an RI value of 1.35 and the axial filament inside the tail with a slightly higher RI value of 1.36 can be recognized. The end piece of the sperm has the lowest RI value of approximately 1.34.

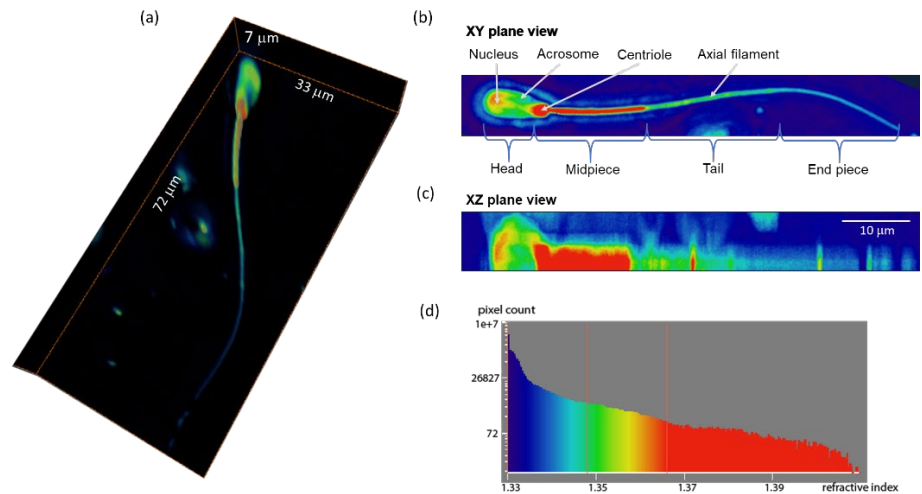

Figure 45 WPT of sperm cells. (a) 3D RI tomogram of a spermatozoon (40x/0.75NA objective). (b) XY plane projection view. The nucleus, acrosome, centriole, and axial filament of the sperm cell are pointed by the white arrows. (c) XZ plane projection view. (d) Histogram of the RI of the sperm cell. From Chen, X., et al., Wolf phase tomography (WPT) of transparent structures using partially coherent illumination. Light: Science & Applications, 2020.

Due to the high throughput, low phototoxicity, absence of photobleaching, and easy sample preparation, WPT is capable of studying real-time volumetric biological events in living cells. We imaged the growth and proliferation of hippocampal neurons over the course of several days, in 6-well plates. The RI distribution of the whole well of neurons is displayed in Fig. 46 (a). One tile zoom-in of the whole well and its distribution of RI are shown in Fig. 46 (b). Figure 46 (c) describes the average of the RI within this tile versus time. Figure 46 (d) shows that the variance of the RI for this tile increases with time as well.

Figure 46 (e) is the zoomed-in image of the red box in Fig. 46 (b) containing two neurons. The neurons spread out into two regions at around t=16 hours, kept growing until around t=53 hours, and then died. We can see that both the average and variance of the RI show three different stages (Figs. 46 (f-g)). One significant change in the average and variance of the RI appears when the two neurons were separated (red arrows). Another change is visible when the two neurons died (green arrows). The death event was accompanied by a decrease in the mean RI, likely due to the membrane permeability, which allowed for water influx. Figure 46 (h) is a zoom-in image of the yellow box in Fig. 46 (b) containing one neuron. The neuron dendrites started to appear at approximately t=13 hours time point, resulting in a jump in the average RI (Fig. 46 (i)). The neuron kept growing until approximately t=62 hours and then died, leading to a decrease in the average RI. Some oscillations in the variance (Fig. 46 (j)) of the RI appear before the neuron died, while exhibiting a clear change after the neuron died.

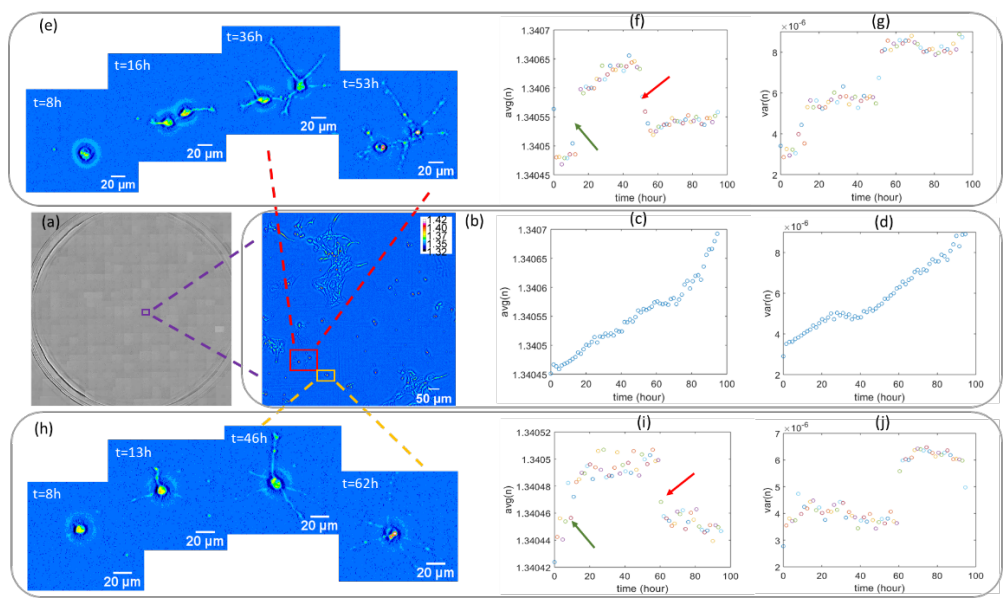

Figure 46 Dynamic WPT of live cells across multi-well plates. (a) RI map across a whole well of living hippocampal neurons (10x/0.3NA objective) is composed of 20 x 21 mosaic tiles, each of 214 x 204 µm2 area. (b) Zoom-in RI map of the purple box in (a) with average (c) and variance (d) of the RI vs. time. (e) Zoom-in RI map of the red box in (b) with average (f) and variance (g) of the RI. The green arrow indicates the increase of the RI when the two neurons separated and their dendrites appeared, the red arrow shows the decrease in RI when the two neurons died. (h) Zoom-in RI map of the yellow box in (b) with average (i) and variance (j) of the RI. The green arrow indicates the jump of the RI when the dendrites appeared, the red arrow shows the decrease in RI when the neuron died. From Chen, X., et al., Wolf phase tomography (WPT) of transparent structures using partially coherent illumination. Light: Science & Applications, 2020.

## 5. Emerging trends in SLIM imaging

### 5.1. PICS: Phase imaging with computational specificity

Phase imaging with computational specificity (PICS) aims to use recent advances in artificial intelligence to introduce specificity for structures or biological processes with the ultimate goal of simplifying the analysis of label-free data [196]. This is typically accomplished by providing a semantic segmentation feature map where a class label is assigned to each pixel in the phase map.

Central to these AI approaches was the development of neural networks capable of efficiently integrating local textural information with contextual shape information such as the U-Net architecture [197].

In general, deep convolutional neural networks consist of a series of pixel-wise nonlinear operations that remap the values in the input image into another form such as a single number, class probability vector, or whole image [198, 199]. The architecture shown in Fig. 47 is divided into a "contracting" and "expanding" path that is linked by a "bottleneck" at the bottom. Importantly, the U-Net architecture includes concatenation layers that copy the output from previous levels to stabilize the training procedure which in turn results in an improvement in resolution.

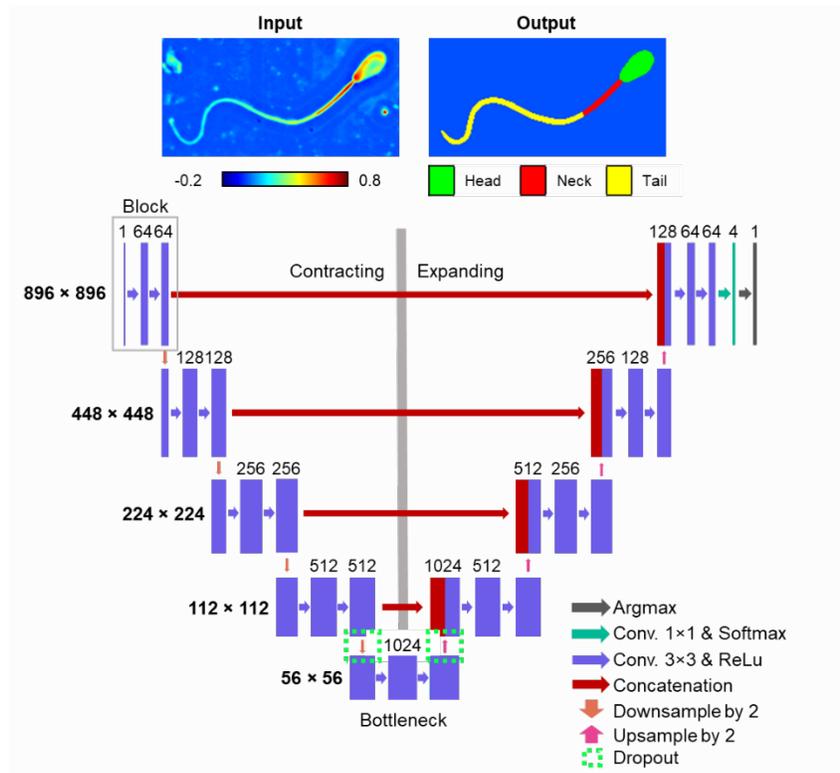

Figure 47 The U-Net architecture consists of several nonlinear operations that map one image into another. In this case a SLIM image is mapped into an annotated image. The series of operations, or architecture, consists of a contracting path that decreases the height and width of the image while increasing the number of channels. After passing the "bottleneck" the image is upsampled and the number of channels is reduced. In U-Net the contracting path is connected to the expanding path at each layer with a concatenation operation that improves accuracy and provides numerical stability during training. During training random convolution at the bottleneck are removed (dropout) to introduce a degree of redundancy and regularization in the network. From Kandel, Mikhail E., et al. "Reproductive outcomes predicted by phase imaging with computational specificity of spermatozoon ultrastructure." Proceedings of the National Academy of Sciences 117.31 (2020): 18302-18309.

Each "block" of the contracting path results in a smaller version of the input image that has been processed in a non-linear way. We note that without nonlinear operations, the entire neural network transformation could reduce to a single convolution. The first block consists of a filter bank of randomly initialized convolution kernels that expand the number of channels in the input image [200]. The values of the convolution kernels are updated at each training step so that the network output will, hopefully, converge to the training data after some number of optimization steps. To introduce nonlinearity, the convolution is paired with a nonlinear "activation" function such as a threshold that removes negative outputs [201] or otherwise has some saturation effect [202]. After application of the convolution and activation, the resulting images resemble a distorted version of the input, and the operation is repeated a second time with different weights. The values are propagated to the next block by downsampling the image such that successive blocks decrease the image size while increasing the number of channels. After passing the "bottleneck" - the point in the network where the data has the lowest resolution but the greatest number of channels - the image is successively upsampled by a series of blocks where the resolution is increased, and the number of channels decreased. In those blocks, an upsampling operation is used to increase the resolution, and convolution is used to decrease the number of channels. These operations are followed by non-linear activation. To improved numerical stability, the input to the filter bank is combined with the output of layers that have matching dimensionality. The result of this procedure is a many channel image equal to the size of the input, and a final convolution and activation result in the target output.

## 5.2. SLIM and AI in cell biology

Fluorescence microscopy addresses the principle deficit of scattered light imaging, by highlighting stains or proteins that are specific to molecular structures or cellular chemistry [203]. This modality is compatible with SLIM imaging as the modulating element in SLIM can be easily configured to behave like a mirror and, therefore, it is straightforward to co-localize fluorescence and phase images on the same camera. Alternatively, in non-commercial designs, which are prone to ambient light leakage, separate light paths have been employed to measure more challenging specimens such as fluorescently conjugated antibodies and proteins [92, 174]. While straightforward, multiplexing fluorescence microscopy and phase imaging negate many of the advantages of SLIM by introducing contrast agents, increasing phototoxic stress, and decreasing acquisition rates.

Fortunately, recent developments in artificial intelligence (AI) offer a way to perform label-free imaging while maintaining the specificity advantage associated with fluorescence microscopy. In phase imaging with computational specificity (PICS), these computationally generated annotations are used to analyze the SLIM data for parameters such as the cellular dry mass [145].

One of the biggest challenges in live-cell imaging is the automated analysis of high-content time-lapse sequences used in drug discovery [204]. In general, these sequences consist of several experimental conditions segregating into wells that are imaged at fixed intervals (Fig. 48). With thousands of individual cells monitored over weeks, data volumes and observational bias have motivated the development of purely computational analysis strategies. For these applications, co-localized fluorescence and phase images are readily available to generate semantic segmentation maps for PICS-style experiments.

This approach is illustrated in (Fig. 48) and is motivated by the ability of deep convolutional neural networks [205] to perform image-to-image translation [206-208]. In short, a timelapse sequence consisting of the biologically relevant portion of the experiment is acquired without labels. Then, the cells are fixed and stained followed by co-localized fluorescence and QPI imaging to produce a training corpus that estimates the fluorescent signal. After training, the resulting neural network is applied to the unstained timelapse sequence. In this way, staining is avoided during the biologically relevant portion of the experiment.

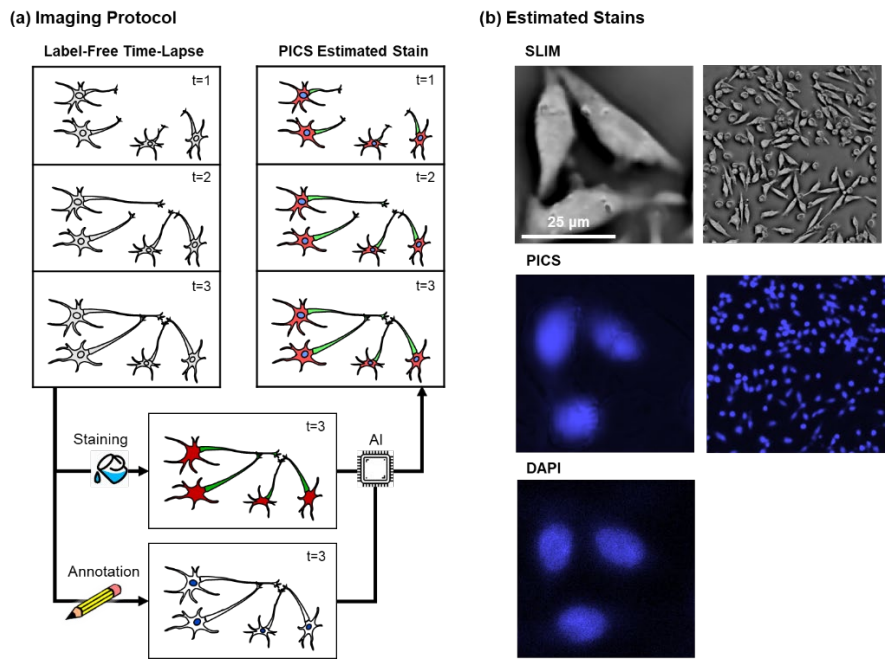

Figure 48 PICS imaging for estimating fluorescent signals. (a) Live cell imaging with computational specificity is performed in two imaging steps. During the first step, unmodified cells are imaged using a label-free technique such as SLIM. After the experiment is over, the cells are stained and imaged with co-localized fluorescence producing a raining corpus for the neural network. After the network is trained, it can be used to digitally stain the experiment, thus avoiding a toxic chemical staining. (b) PICS performance closely matches chemical analogs (20x/0.3). From Kandel, Mikhail E., et al. "Phase Imaging with Computational Specificity." Nature Comm. (2020).

When this procedure is applied to SLIM imaging, the SLIM system can estimate the DAPI stain with only a small loss of resolution that plays little role in semantic segmentation (Fig. 48). The associated semantic maps are then used to perform a per-cellular compartment analysis of the data [145]. While the principal motivation for PICS is to avoid toxicity or destructive fixation due to fluorescent stains, a further advantage comes from the digital nature of the procedure. The transmitted light signal is often more intense than in fluorescence microscopy meaning that the PICS estimated signal is typically an order of magnitude faster to acquire compared with its true chemical counterpart [145, 196]. Additionally, as PICS staining is a digital process, a larger number of fluorescent stains can be estimated from a single label-free image than what is otherwise possible with conventional spectral based fluorescence multiplexing strategies.

Although an emerging technology, PICS has been shown relevant for a variety of cellular systems. The initial PICS publication [145] included SW480 and SW620 cells which are cancer cells derived from the same patient and are frequently used to study disease progression. In that work, it was shown that PICS could replace DiI and DAPI stains. Importantly, PICS provided an automated and label-free approach to estimate the ratio of nuclear and cytoplasmic dry-mass which is a prognostic marker for cancer progression (for example in [209]).

When PICS was first presented, two important observations were made: 1) that training could be performed on fixed (dead) cells and evaluated on living (unmodified) specimens and 2) that the neural network was surprisingly tolerant to variations in cellular morphology. The latter was evidenced by the ability of the network to be trained on more round, stressed, and confluent cells at the end of the experiment with the inference still accurate for intact fibroblast shapes observed at the start of the experiment. This observation further contributes to the trend where computational neural networks are understood to learn more general rules that are applicable beyond the dataset used to train them. Further, the same training corpus was used for both cell lines (SW480 & SW620). Additionally, due to the high sectioning and suppression of multiple scattering in QPI, it was possible to perform PICS style imaging in liver cancer spheroids using a purely 2D training approach. This approach should be relatively straightforward to extend to 2.5D and 3D network architectures [210].

As an emerging technique, the limitations of PICS style imaging are the subject of current investigations and are usually phrased in terms of the ability of the underlying network to learn the image remapping. The preferred quality scores are derived from the Pearson correlation between actual and imputed fluorescence, as well as more wholistic measurements such as cell counts performed on training pairs. When considering a nuclear stain such as DAPI, it was observed the neural network had "learned" (or more formally responded to) both local textural information as well as higher-order features adjacent to the cells. Therefore, the network incorporated morphological information from sounding regions to estimate the expected DAPI signal.

The extent to which morphology contributes to protein concentration estimation was experimentally evaluated in subsequent work [211]. There, PICS was used to impute fluorophores associated with antibody staining – specifically Tau & MAP2 concentration levels; a pair of proteins whose differential expression distinguished between axons and dendrites [212]. The associated semantic maps were then used to measure dry-mass traffic inside the annotated structures using the DPS method mentioned in previous sections [177].

The observation that morphology could serve as a proxy for staining levels was further validated in [213] where PICS was used to overcome the toxicity of a common cellular viability assay. In that work, the commonly used, but toxic, system of Hoechst 33342 and SYTOX™ Green provided a chemically motivated fluorescent marker for cellular viability. To produce a network capable of matching the stain, the U-Net architecture was revised to use MobileNet blocks following Google's EfficientNet proposal [214]. Next, transfer learning from ImageNet was used to boost accuracy. As cell viability is primarily a function of membrane integrity and is only readily identifiable in the most pathological of cases - within a broader context, PICS continues a trend in machine learning where computers can identify features with more accuracy than human annotators.

### 5.3. SLIM and AI in pathology

SLIM imaging in conjunction with artificial intelligence offers a solution to outstanding challenges in digital pathology. As histopathological resected tissue is most often imaged with the aid of exogenous contrast agents such as H&E [92, 215], variability in staining procedures can frustrate comparison across specimens and instruments. In addition to variation introduced during the staining procedure, diagnosis is performed by a pathologist, which leads to observational bias and errors. Lastly, many procedures in pathology are labor-intensive and the non-trivial time to analyze samples often puts a practical limitation on morphological analysis.

In this context, it is not surprising that early efforts were made to automate diagnosis on SLIM images using what are today considered "classical" methods. A step towards removing observational bias was taken in [155], where the phase values inside the glands were used to identify tissue as potentially cancerous. In that work, features derived from tissue glandular solidity (fill factor) and structural anisotropy were combined in a non-linear fashion using a support vector machine [216]. A similar procedure was developed, where a VGG16 convolutional neural network [217] was used to implicitly generate and combine those classical features (Fig. 49, 50) [218].

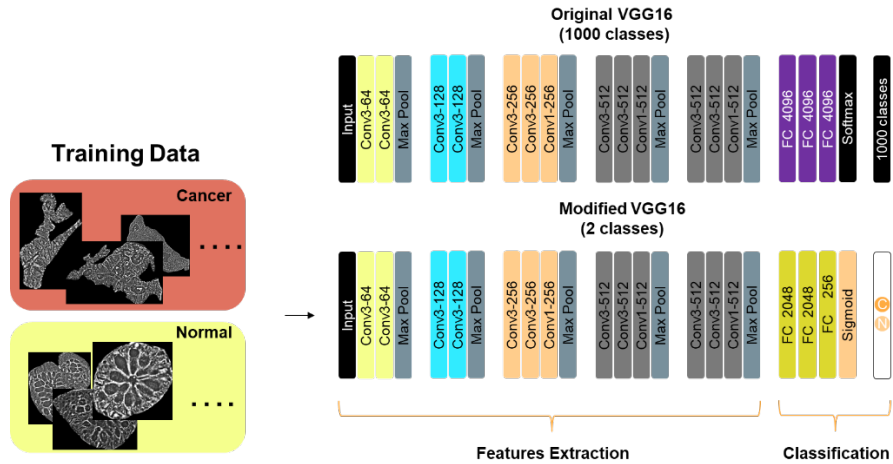

Figure 49 Modified VGG16 network. Input image size is 256 x 256 x3. A pad of length 1 is added before each Max Pool layer. Conv1 : Convolutional layer with 1x1 filter; Conv3 : Convolutional layer with 3x3 filter; Max Pool: Maximum pooling layer over 2x2 pixels (stride=2); All hidden layers are followed by RELU activation. First FC layer is followed by 0.5 dropout. From Zhang, Jingfang K., et al. "Label-free colorectal cancer screening using deep learning and spatial light interference microscopy (SLIM)." APL Photonics 5.4 (2020): 040805.

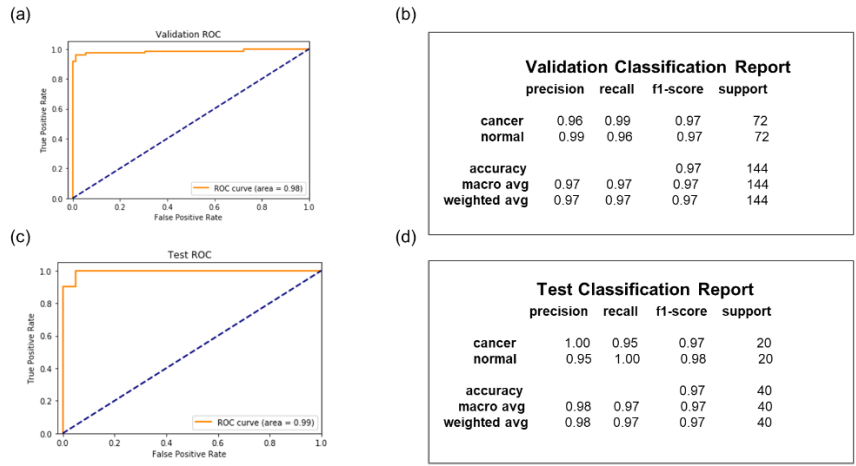

Figure 50. ROC (Receiver Operating Characteristics) curve, with AUC (Area Under the Curve), and classification reports for the validation dataset and the test dataset respectively. The AUC score is 0.98 for the validation dataset and 0.99 for the test dataset, as indicated. The two classes, cancer and normal, have balanced support for both the validation dataset and the test dataset, with the validation dataset providing 72 actual occurrences for each class and the test dataset providing 20 actual occurrences for each class. The accuracy hits 97% for both the validation and the test. From Zhang, Jingfang K., et al. "Label-free colorectal cancer screening using deep learning and spatial light interference microscopy (SLIM)." APL Photonics 5.4 (2020): 040805.

While these methods helped address observational bias, they nevertheless required a pathologist to manually circumscribe the gland. A step towards fully automated analysis was taken in [219] where "texons" were used to produce a semantic segmentation map. In that approach, much like in convolutional neural networks, a filter bank is used to expand the SLIM image into a series of variations, which are non-linearly composited to form a semantic segmentation map. This procedure was, ultimately used to grade prostate cancer biopsies [219].

The tradeoff between intrinsic imaging and external contrast agents is particularly acute in reproductive pathology. When inspecting gametes such as sperms cells, the use of fluorescence labels has enabled new cell-sorting strategies and given insights into developmental biology. Nevertheless, these methods are considered too invasive for use in a clinical setting. Thus, artificial reproduction procedures such as intracytoplasmic sperm injection rely on transmitted light imaging [220]. These microscopes are most often boosted with label-free contrast enhancement techniques such as DIC, phase-contrast, or Hoffman modulation contrast. Invariably, similar difficulties arise as those encountered when analyzing cancer biopsies, namely that the data is qualitative and subject to acquisition specific variation, and the analysis is subject to the judgment of a pathologist. In addition to observational bias, relying on human observers makes it time-consuming to ascertain morphological features such as the dimensions of the organelles within a population of cells [221]. While a tedious morphological annotation can be performed by hand for fixed specimens [211], it is difficult to imagine manual annotation for time-critical decisions such as when selecting live, moving, sperm cells.

In these cases, SLIM imaging can be used to perform interferometrically normalized data acquisition, while artificial intelligence can automatically annotate the samples. This approach is exemplified by the efforts in [191] to relate morphological differences in sperm cells with reproductive outcomes (Fig. 51). In that work, the authors digitized a large number of slides generated from animals with known fertility rates. A large number of samples motivated an end-to-end analysis strategy based on machine learning. A U-Net architecture was trained to segment the spermatozoan into compartments [197]. The training was performed unconventionally, with a two-step procedure resembling a generative adversarial network [222]. This procedure aimed to produce a larger training corpus in a time-efficient manner and consisted of a fine annotation performed on a limited dataset. Then, a coarse correction was performed, followed by a final training round. It was found that the neural network training procedure was able to merge differences between annotators which in turn, helped to mitigate intraobserver variation. The semantic segmentation was used to address a biologically relevant question, specifically, the effect of morphology on cleavage and blastocyst rates. The results (Fig. 52) confirmed the well-known theory that fertilization success is connected to hydrodynamic properties while providing further, although somewhat indirect, evidence that blastocyst formation rates are connected to male-factor cytoskeletal structures such as the centriole.

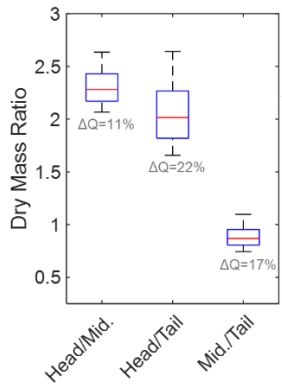 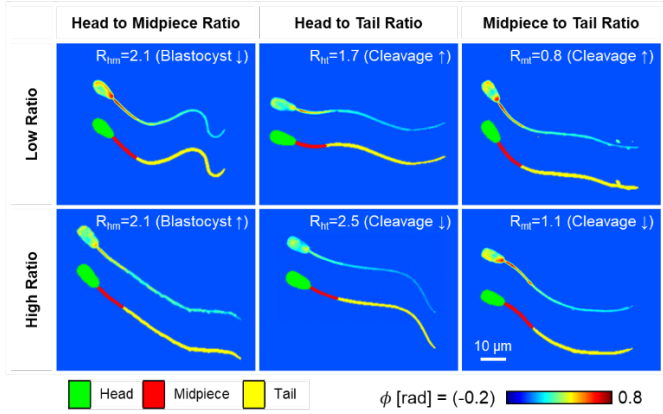

Figure 51 Deep learning tracks subtle but significant differences in sperm morphology. (A) Histogram of the distribution of sperm dry mass ratios shows that structural differences between sperm cells are relatively narrow as evidenced by the percentage difference between the first and third quartile. For the head to midpiece, head to tail, midpiece to tail are (only) 11%, 24%, and 17% respectively. (B) Dry mass maps of representative sperm cells along with semantic segmentation are labeled with their dry mass ratios (Rhm, Rht, Rmt). Additionally, ↑ denotes an increase or ↓ a decrease in ART outcome as determined in Fig. 5 and SI Appendix. These differences are especially difficult to visualize with conventional techniques as typical microscope images are not proportional to dry mass and the naked eye is unable to segment, integrate, and divide portions of an image (40x/0.75, SLIM). From Kandel, Mikhail E., et al. "Reproductive outcomes predicted by phase imaging with computational specificity of spermatozoon ultrastructure." Proceedings of the National Academy of Sciences 117.31 (2020): 18302-18309.

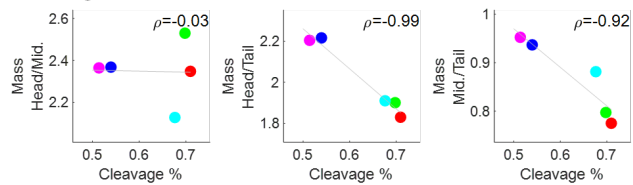

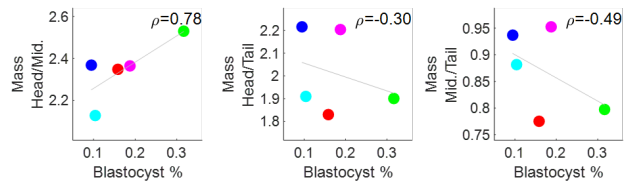

(c) Summary of ART Outcomes

| Embryo Cleavage | Pearson's R | P-Value |
|---|---|---|
| Head / Midpiece | -0.03 | >0.36 |
| **Head / Tail** | **-0.99** | **<0.01** |
| **Midpiece / Tail** | **-0.98** | **<0.01** |
| Blastocyst Formation | | |
| **Head / Midpiece** | **+0.78** | **0.05** |
| Head / Tail | -0.30 | 0.31 |
| Midpiece / Tail | -0.49 | 0.22 |

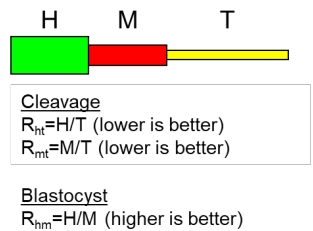

Cleavage
$R_{ht}$=H/T (lower is better)
$R_{mt}$=M/T (lower is better)

Blastocyst
$R_{hm}$=H/M (higher is better)

Figure 52 Summary of outcomes. (a) Cleavage is strongly favored by a more massive tail, while (b) blastocyst development is favored by a heavier head. (c) Summary across the five bulls for cleavage and blastocyst rates. From Kandel, Mikhail E., et al. "Reproductive outcomes predicted by phase imaging with computational specificity of spermatozoon ultrastructure." Proceedings of the National Academy of Sciences 117.31 (2020): 18302-18309.

In the broader context of AI algorithms, it was observed that semantic segmentation-based scoring was able to inform better on the underlying biological process when compared to end-to-end metrics. Specifically, by having an annotated image rather than a single number for each cell's "fertility" potential, the authors were able to interpret their results in terms of hydrodynamic properties as well as identify important cellular ultrastructures within the semantic maps (such as the centriole). A further motivation for using semantic segmentation was the lack of a direct correspondence between individual sperm cells and reproductive fate. Unsurprisingly, constructing a training corpus, for example, one where individual sperm cells are matched to fertility outcomes, remains the greatest challenge for using AI in reproductive medicine. However, the ability of SLIM to generate pairs of phase and fluorescence images from the same field of view greatly simplifies the problem of generating ground truth data. For several applications, it has been shown that these data can be generated automatically, with no manual intervention.

## 6. Summary and Outlook

In sum, we described SLIM as a label-free, common-path, white-light, phase-shifting interferometer, which upgrades existing phase-contrast microscopes. The common-path interferometry and white-light illumination allow for speckle-free and sub-nanometer path-length stability (Chapters 1, 2). The phase in the SLIM image is reconstructed with four intensity frames corresponding to each phase shift. The halo effect from the phase-contrast microscope is removed with our halo removal algorithm. The software developed in-house enables high-throughput acquisition, whole slide scanning, mosaic tile registration, and imaging with a color camera (Chapter 3).

Particularly because of its inherent stability, SLIM enables a large number of basic science and clinical applications (Chapter 4). SLIM can study cell dynamics, cell growth and proliferation, cell migration, and mass transport, etc. In clinical settings, SLIM can assist with cancer studies, reproductive technology, and blood testing, etc. WDT and WPT are two 3D tomographic methods developed with SLIM.

Recently, the development of deep learning brings new exciting opportunities for SLIM imaging (Chapter 5). AI adds computational specificity to SLIM data and allows more applications in cell biology and pathology. However, SLIM images degrade for highly scattering samples, such as embryos (Fig. 53). The recently developed gradient light interference microscopy (GLIM) is capable of suppressing multiple scattering significantly, and, thus, extend QPI to thick specimens.

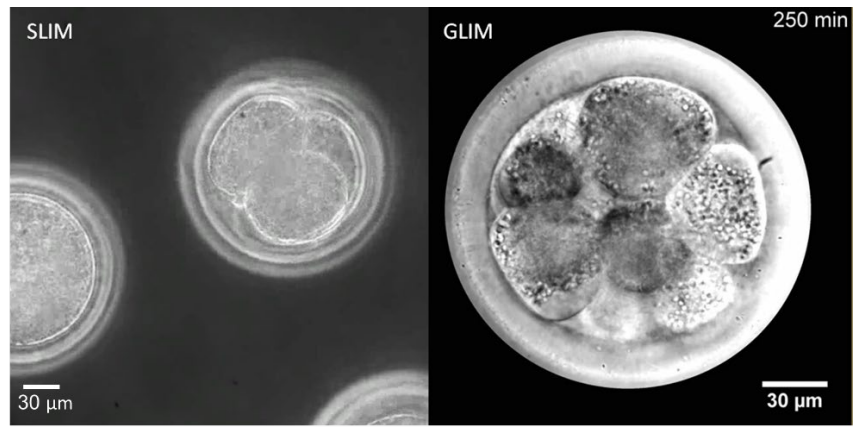

Figure 53 Imaging embryos with SLIM (left) versus GLIM (right). Here, bovine embryos are imaged with both techniques. While SLIM brings information at various depths into focus, GLIM is more depth selective. The middle (thickest) part of the embryo imaging using SLIM lost contrast due to multiple scattering. This is not the with the GLIM image. In summary, GLIM has better depth sectioning compared to SLIM. From "Nguyen, Tan H., et al. "Gradient light interference microscopy for 3D imaging of unlabeled specimens." Nature communications 8.1 (2017)".

As an upgrading module to existing microscopes and seamless integration with the fluorescence channels, SLIM has a low threshold for adoption in the field. We anticipate that this adoption process will be boosted by the recent availability of the commercial product (CellVista SLIM and SLIM Pro, Phi Optics., Inc.). It is certain that computational tools will continue to play an increasingly important role in microscopy in general, and SLIM in particular. It is the software tools that will exploit SLIM data to solve specific problems in biomedicine. In combination with AI, SLIM holds exciting opportunities for removing the drawbacks associated with fluorescence, while maintaining the chemical specificity. This capability will likely enable important nondestructive, longitudinal studies on live cells and cellular systems, including cell growth, cell viability, various organelle identification and cell differentiation. In the clinical space of applications, SLIM and AI will be able to provide screening, diagnosis and prognosis in unlabeled biopsies, as well as sperm selection.

*Funding*

This work is supported by National Science Foundation (CBET0939511 STC, NRT-UtB 173525), National Institute of General Medical Sciences (GM129709); National Cancer Institute (CA238191).

*Disclosures*

The authors declare the following competing interests: G.P. has a financial interest in Phi Optics, Inc., a company developing quantitative phase imaging technology for materials and life science applications.

# References

1. Combs, C.A. and H. Shroff, *Fluorescence Microscopy: A Concise Guide to Current Imaging Methods.* Curr Protoc Neurosci, 2017. **79**(1): p. 2 1 1-2 1 25.
2. Hoebe, R.A., et al., *Controlled light-exposure microscopy reduces photobleaching and phototoxicity in fluorescence live-cell imaging.* Nat Biotechnol, 2007. **25**(2): p. 249-53.
3. Laissue, P.P., et al., *Assessing phototoxicity in live fluorescence imaging.* Nat Methods, 2017. **14**(7): p. 657-661.
4. Jensen, E.C., *Use of fluorescent probes: their effect on cell biology and limitations.* Anat Rec (Hoboken), 2012. **295**(12): p. 2031-6.
5. Piston, D.W. and G.J. Kremers, *Fluorescent protein FRET: the good, the bad and the ugly.* Trends Biochem Sci, 2007. **32**(9): p. 407-14.
6. Waters, J.C., *Accuracy and precision in quantitative fluorescence microscopy*. 2009, The Rockefeller University Press.
7. Zacharias, D.A., et al., *Partitioning of lipid-modified monomeric GFPs into membrane microdomains of live cells.* Science, 2002. **296**(5569): p. 913-6.
8. Seniya, C., C. Towers, and D. Towers. *Improvements in low-cost label-free QPI microscope for live cell imaging*. in *European Conference on Biomedical Optics*. 2017. Optical Society of America.
9. Joshi, B., et al., *Label-free route to rapid, nanoscale characterization of cellular structure and dynamics through opaque media.* Sci Rep, 2013. **3**(1): p. 2822.
10. Cherkezyan, L., et al., *Interferometric spectroscopy of scattered light can quantify the statistics of subdiffractional refractive-index fluctuations.* Phys Rev Lett, 2013. **111**(3): p. 033903.
11. Girshovitz, P. and N.T. Shaked, *Generalized cell morphological parameters based on interferometric phase microscopy and their application to cell life cycle characterization.* Biomed Opt Express, 2012. **3**(8): p. 1757-73.
12. Zangle, T.A. and M.A. Teitell, *Live-cell mass profiling: an emerging approach in quantitative biophysics.* Nat Methods, 2014. **11**(12): p. 1221-8.
13. Carney, P.S., et al., *Phase in nanooptics.* ACS Nano, 2012. **6**(1): p. 8-12.
14. Park, Y., C. Depeursinge, and G. Popescu, *Quantitative phase imaging in biomedicine.* Nature Photonics, 2018. **12**(10): p. 578-589.
15. Gabriel, P., *Quantitative Phase Imaging of Cells and Tissues*. 2011, New York: McGraw-Hill Education.
16. Kim, T., et al., *Breakthroughs in Photonics 2013: Quantitative Phase Imaging: Metrology Meets Biology.* IEEE Photonics Journal, 2014. **6**(2): p. 1-9.
17. Mertz, J., *Introduction to Optical Microscopy*. 2 ed. 2019, Cambridge: Cambridge University Press.
18. Zernike, F., *Phase contrast, a new method for the microscopic observation of transparent objects.* Physica, 1942. **9**(7): p. 686-698.
19. Nomarski, G., *Differential microinterferometer with polarized waves.* J. Phys. Radium Paris, 1955. **16**: p. 9S.
20. Kim, M.K., *Principles and techniques of digital holographic microscopy.* SPIE reviews, 2010. **1**(1): p. 018005.
21. Tychinsky, V., et al., *Quantitative phase imaging of living cells: application of the phase volume and area functions to the analysis of "nucleolar stress".* J Biomed Opt, 2013. **18**(11): p. 111413.
22. Bon, P., et al., *Fast label-free cytoskeletal network imaging in living mammalian cells.* Biophys J, 2014. **106**(8): p. 1588-95.
23. Gannavarpu, R., et al., *Spatiotemporal characterization of a fibrin clot using quantitative phase imaging.* PLoS One, 2014. **9**(11): p. e111381.
24. Li, Z., et al., *Optical properties of tissues quantified using morphological granulometry from phase-contrast images of thin tissue samples.* J Xray Sci Technol, 2015. **23**(1): p. 111-8.
25. Descloux, A., et al., *Combined multi-plane phase retrieval and super-resolution optical fluctuation imaging for 4D cell microscopy.* Nature Photonics, 2018. **12**(3): p. 165-172.
26. Nguyen, T.H., et al., *Gradient light interference microscopy for 3D imaging of unlabeled specimens.* Nat Commun, 2017. **8**(1): p. 210.
27. Wolf, E., *What kind of phases does one measure in usual interference experiments?* Optics Communications, 2011. **284**(19): p. 4235-4236.
28. Kashani, A.H., et al., *Optical coherence tomography angiography: A comprehensive review of current methods and clinical applications.* Prog Retin Eye Res, 2017. **60**: p. 66-100.
29. Drexler, W. and J.G. Fujimoto, *Optical coherence tomography: technology and applications*. 2008: Springer Science & Business Media.
30. Edwards, C., et al., *Epi-illumination diffraction phase microscopy with white light.* Opt Lett, 2014. **39**(21): p. 6162-5.
31. Chowdhury, S. and J. Izatt, *Structured illumination quantitative phase microscopy for enhanced resolution amplitude and phase imaging.* Biomed Opt Express, 2013. **4**(10): p. 1795-805.
32. Choi, Y., et al., *Full-field and single-shot quantitative phase microscopy using dynamic speckle illumination.* Opt Lett, 2011. **36**(13): p. 2465-7.

33. Platt, B.C. and R. Shack, *History and principles of Shack-Hartmann wavefront sensing.* J Refract Surg, 2001. **17**(5): p. S573-7.
34. Soldevila, F., et al., *Phase imaging by spatial wavefront sampling.* Optica, 2018. **5**(2): p. 164-174.
35. Li, J., et al., *Efficient quantitative phase microscopy using programmable annular LED illumination.* Biomed Opt Express, 2017. **8**(10): p. 4687-4705.
36. Kellman, M.R., et al., *Physics-Based Learned Design: Optimized Coded-Illumination for Quantitative Phase Imaging.* IEEE Transactions on Computational Imaging, 2019. **5**(3): p. 344-353.
37. Isikman, S.O., et al., *Modern Trends in Imaging VIII: Lensfree Computational Microscopy Tools for Cell and Tissue Imaging at the Point-of-Care and in Low-Resource Settings.* Analytical Cellular Pathology, 2012. **35**(4): p. 229-247.
38. Gerchberg, R.W., *A practical algorithm for the determination of phase from image and diffraction plane pictures.* Optik, 1972. **35**: p. 237-246.
39. Zheng, G., R. Horstmeyer, and C. Yang, *Wide-field, high-resolution Fourier ptychographic microscopy.* Nat Photonics, 2013. **7**(9): p. 739-745.
40. Zuo, C., et al., *Transport of intensity equation: a tutorial.* Optics and Lasers in Engineering, 2020. **135**: p. 106187.
41. Hoover, E.E. and J.A. Squier, *Advances in multiphoton microscopy technology.* Nat Photonics, 2013. **7**(2): p. 93-101.
42. Campagnola, P.J. and L.M. Loew, *Second-harmonic imaging microscopy for visualizing biomolecular arrays in cells, tissues and organisms.* Nat Biotechnol, 2003. **21**(11): p. 1356-60.
43. Witte, S., et al., *Label-free live brain imaging and targeted patching with third-harmonic generation microscopy.* Proc Natl Acad Sci U S A, 2011. **108**(15): p. 5970-5.
44. Zhang, C., D. Zhang, and J.X. Cheng, *Coherent Raman Scattering Microscopy in Biology and Medicine.* Annu Rev Biomed Eng, 2015. **17**: p. 415-45.
45. Cheng, J.-X. and X.S. Xie, *Coherent Raman scattering microscopy.* 2016: CRC press.
46. Wu, X., et al., *Label-free detection of breast masses using multiphoton microscopy.* PLoS One, 2013. **8**(6): p. e65933.
47. Galli, R., et al., *Intrinsic indicator of photodamage during label-free multiphoton microscopy of cells and tissues.* PLoS One, 2014. **9**(10): p. e110295.
48. Quinn, K.P., et al., *Diabetic Wounds Exhibit Distinct Microstructural and Metabolic Heterogeneity through Label-Free Multiphoton Microscopy.* J Invest Dermatol, 2016. **136**(1): p. 342-344.
49. Zhuo, S., et al., *Label-free monitoring of colonic cancer progression using multiphoton microscopy.* Biomed Opt Express, 2011. **2**(3): p. 615-9.
50. van Munster, E.B. and T.W. Gadella, *Fluorescence lifetime imaging microscopy (FLIM)*, in *Microscopy techniques*. 2005, Springer. p. 143-175.
51. Li, L.L., et al., *BODIPY-Based Two-Photon Fluorescent Probe for Real-Time Monitoring of Lysosomal Viscosity with Fluorescence Lifetime Imaging Microscopy.* Anal Chem, 2018. **90**(9): p. 5873-5878.
52. Jenkins, J., et al., *Sulforhodamine Nanothermometer for Multiparametric Fluorescence Lifetime Imaging Microscopy.* Anal Chem, 2016. **88**(21): p. 10566-10572.
53. Margineanu, A., et al., *Screening for protein-protein interactions using Förster resonance energy transfer (FRET) and fluorescence lifetime imaging microscopy (FLIM).* Scientific reports, 2016. **6**: p. 28186.
54. Zheng, W., et al., *Autofluorescence of epithelial tissue: single-photon versus two-photon excitation.* J Biomed Opt, 2008. **13**(5): p. 054010.
55. Doherty, J., G. Cinque, and P. Gardner, *Single-cell analysis using Fourier transform infrared microspectroscopy.* Applied Spectroscopy Reviews, 2016. **52**(6): p. 560-587.
56. Shakya, B.R., et al., *The use of Fourier Transform Infrared (FTIR) spectroscopy in skin cancer research: a systematic review.* Applied Spectroscopy Reviews, 2020: p. 1-33.
57. Yu, M.C., et al., *Label Free Detection of Sensitive Mid-Infrared Biomarkers of Glomerulonephritis in Urine Using Fourier Transform Infrared Spectroscopy.* Sci Rep, 2017. **7**(1): p. 4601.
58. Xia, F., et al., *In vivo label-free confocal imaging of the deep mouse brain with long-wavelength illumination.* Biomed Opt Express, 2018. **9**(12): p. 6545-6555.
59. Wu, T.F., et al., *A light-sheet microscope compatible with mobile devices for label-free intracellular imaging and biosensing.* Lab Chip, 2014. **14**(17): p. 3341-8.
60. Gibson, A. and H. Dehghani, *Diffuse optical imaging.* Philos Trans A Math Phys Eng Sci, 2009. **367**(1900): p. 3055-72.
61. Wang, L.V. and J. Yao, *A practical guide to photoacoustic tomography in the life sciences.* Nat Methods, 2016. **13**(8): p. 627-38.
62. Wang, L.V. and S. Hu, *Photoacoustic tomography: in vivo imaging from organelles to organs.* Science, 2012. **335**(6075): p. 1458-62.
63. Chen, X. and O. Korotkova, *Scattering of light from hollow and semi-hollow 3D scatterers with ellipsoidal, cylindrical and cartesian symmetries.* Computer Optics, 2016. **40**(5): p. 635-641.
64. Chen, X. and O. Korotkova, *Probability density functions of instantaneous Stokes parameters on weak scattering.* Optics Communications, 2017. **400**: p. 1-8.
65. Born, M. and E. Wolf, *Principles of optics: electromagnetic theory of propagation, interference and diffraction of light.* 2013: Elsevier.

66. Wolf, E., *Principles and development of diffraction tomography*, in *Trends in Optics*. 1996, Elsevier. p. 83-110.
67. Hu, C. and G. Popescu, *Physical significance of backscattering phase measurements.* Opt Lett, 2017. **42**(22): p. 4643-4646.
68. Ledwig, P. and F.E. Robles, *Epi-mode tomographic quantitative phase imaging in thick scattering samples.* Biomed Opt Express, 2019. **10**(7): p. 3605-3621.
69. Matlock, A., et al., *Inverse scattering for reflection intensity phase microscopy.* Biomed Opt Express, 2020. **11**(2): p. 911-926.
70. Mandel, L. and E. Wolf, *Optical coherence and quantum optics*. 1995: Cambridge university press.
71. Kandel, M.E., et al., *Epi-illumination gradient light interference microscopy for imaging opaque structures.* Nat Commun, 2019. **10**(1): p. 4691.
72. Chen, X. and O. Korotkova, *Phase structuring of 2D complex coherence states.* Opt Lett, 2019. **44**(10): p. 2470-2473.
73. Chen, X. and O. Korotkova, *Complex degree of coherence modeling with famous planar curves.* Opt Lett, 2018. **43**(24): p. 6049-6052.
74. Korotkova, O. and X. Chen, *Phase structuring of the complex degree of coherence.* Opt Lett, 2018. **43**(19): p. 4727-4730.
75. Goodman, J.W., *Statistical optics*. 2015: John Wiley & Sons.
76. Li, J., et al., *Mitigation of atmospheric turbulence with random light carrying OAM.* Optics Communications, 2019. **446**: p. 178-185.
77. Chen, X., et al., *Synthesis of Im-Bessel correlated beams via coherent modes.* Opt Lett, 2018. **43**(15): p. 3590-3593.
78. Korotkova, O., X. Chen, and T. Setälä, *Electromagnetic Schell-model beams with arbitrary complex correlation states.* Optics letters, 2019. **44**(20): p. 4945-4948.
79. Chen, X., et al., *Wolf phase tomography (WPT) of transparent structures using partially coherent illumination.* Light Sci Appl, 2020. **9**(1): p. 142.
80. Chen, X. and O. Korotkova, *Optical beam propagation in soft anisotropic biological tissues.* Osa Continuum, 2018. **1**(3): p. 1055-1067.
81. Chen, X., J. Li, and O. Korotkova, *Light scintillation in soft biological tissues.* Waves in Random and Complex Media, 2020. **30**(3): p. 481-489.
82. Gabor, D., *Holography, 1948-1971.* Science, 1972. **177**(4046): p. 299-313.
83. Lu Rong, 戎., et al., *Speckle noise reduction in digital holography by use of multiple polarization holograms.* Chinese Optics Letters, 2010. **8**(7): p. 653-655.
84. Hennelly, B.M., et al., *Review of twin reduction and twin removal techniques in holography.* 2009.
85. Matoba, O., et al., *Review of three-dimensional imaging of dynamic objects by parallel phase-shifting digital holography.* Optical Engineering, 2018. **57**(06): p. 061613.
86. Cuche, E., F. Bevilacqua, and C. Depeursinge, *Digital holography for quantitative phase-contrast imaging.* Opt Lett, 1999. **24**(5): p. 291-3.
87. Han, J., et al., *Slightly off-axis interferometry for microscopy with second wavelength assistance.* Appl Opt, 2011. **50**(17): p. 2793-8.
88. Hussain, A., et al., *Super resolution imaging achieved by using on-axis interferometry based on a Spatial Light Modulator.* Opt Express, 2013. **21**(8): p. 9615-23.
89. Singh Mehta, D. and V. Srivastava, *Quantitative phase imaging of human red blood cells using phase-shifting white light interference microscopy with colour fringe analysis.* Applied Physics Letters, 2012. **101**(20): p. 203701.
90. Zhang, Q., et al., *Quantitative refractive index distribution of single cell by combining phase-shifting interferometry and AFM imaging.* Sci Rep, 2017. **7**(1): p. 2532.
91. Bhaduri, B., K. Tangella, and G. Popescu, *Fourier phase microscopy with white light.* Biomed Opt Express, 2013. **4**(8): p. 1434-41.
92. Wang, Z., et al., *Spatial light interference microscopy (SLIM).* Opt Express, 2011. **19**(2): p. 1016-26.
93. Rockward, W.S., et al., *Quantitative phase measurements using optical quadrature microscopy.* Appl Opt, 2008. **47**(10): p. 1684-96.
94. Majeed, H., et al., *Magnified Image Spatial Spectrum (MISS) microscopy for nanometer and millisecond scale label-free imaging.* Opt Express, 2018. **26**(5): p. 5423-5440.
95. Pfau, P.R., et al., *Criteria for the diagnosis of dysplasia by endoscopic optical coherence tomography.* Gastrointest Endosc, 2003. **58**(2): p. 196-202.
96. Remmersmann, C., et al., *Phase noise optimization in temporal phase-shifting digital holography with partial coherence light sources and its application in quantitative cell imaging.* Appl Opt, 2009. **48**(8): p. 1463-72.
97. Phillips, Z.F., M. Chen, and L. Waller, *Single-shot quantitative phase microscopy with color-multiplexed differential phase contrast (cDPC).* PLoS One, 2017. **12**(2): p. e0171228.
98. Shan, M.G., et al., *White-light diffraction phase microscopy at doubled space-bandwidth product.* Optics Express, 2016. **24**(25): p. 29034-29040.
99. Choi, W., et al., *Tomographic phase microscopy.* Nat Methods, 2007. **4**(9): p. 717-9.
100. Cotte, Y., et al., *Marker-free phase nanoscopy.* Nature Photonics, 2013. **7**: p. 113.

101. Kim, T., et al., *Solving inverse scattering problems in biological samples by quantitative phase imaging.* Laser & Photonics Reviews, 2016. **10**(1): p. 13-39.
102. Robles, F.E., et al., *Molecular imaging true-colour spectroscopic optical coherence tomography.* Nat Photonics, 2011. **5**(12): p. 744-747.
103. Charriere, F., et al., *Cell refractive index tomography by digital holographic microscopy.* Opt Lett, 2006. **31**(2): p. 178-80.
104. Wang, X., et al., *Image reconstruction of effective Mie scattering parameters of breast tissue in vivo with near-infrared tomography.* J Biomed Opt, 2006. **11**(4): p. 041106.
105. Dierolf, M., et al., *Ptychographic X-ray computed tomography at the nanoscale.* Nature, 2010. **467**(7314): p. 436-9.
106. Wolf, E., *Three-dimensional structure determination of semi-transparent objects from holographic data.* Optics Communications, 1969. **1**(4): p. 153-156.
107. Lee, K., et al., *Low-coherent optical diffraction tomography by angle-scanning illumination.* J Biophotonics, 2019. **12**(5): p. e201800289.
108. Habaza, M., et al., *Tomographic phase microscopy with 180 degrees rotation of live cells in suspension by holographic optical tweezers.* Opt Lett, 2015. **40**(8): p. 1881-4.
109. Jung, J., et al., *Label-free non-invasive quantitative measurement of lipid contents in individual microalgal cells using refractive index tomography.* Sci Rep, 2018. **8**(1): p. 6524.
110. Simon, B., et al., *Tomographic diffractive microscopy with isotropic resolution.* Optica, 2017. **4**(4): p. 460-463.
111. Macias-Garza, F., K. Diller, and A. Bovik, *Missing Cone Of Frequencies And Low-Pass Distortion In Three-Dimensional Microscopic Images.* Optical Engineering, 1988. **27**(6): p. 276461.
112. Lim, J., et al., *Comparative study of iterative reconstruction algorithms for missing cone problems in optical diffraction tomography.* Opt Express, 2015. **23**(13): p. 16933-48.
113. Habaza, M., et al., *Rapid 3D refractive-index imaging of live cells in suspension without labeling using dielectrophoretic cell rotation.* Advanced Science, 2017. **4**(2): p. 1600205.
114. Jenkins, M.H. and T.K. Gaylord, *Three-dimensional quantitative phase imaging via tomographic deconvolution phase microscopy.* Appl Opt, 2015. **54**(31): p. 9213-27.
115. Maschek, D., et al., *A new approach for the study of the chemical composition of bordered pit membranes: 4Pi and confocal laser scanning microscopy.* Am J Bot, 2013. **100**(9): p. 1751-6.
116. Bassi, A., B. Schmid, and J. Huisken, *Optical tomography complements light sheet microscopy for in toto imaging of zebrafish development.* Development, 2015. **142**(5): p. 1016-20.
117. Araya-Polo, M., et al., *Deep-learning tomography.* The Leading Edge, 2018. **37**(1): p. 58-66.
118. Yoon, J., et al., *Identification of non-activated lymphocytes using three-dimensional refractive index tomography and machine learning.* Sci Rep, 2017. **7**(1): p. 6654.
119. Li, J., et al., *Three-dimensional tomographic microscopy technique with multi-frequency combination with partially coherent illuminations.* Biomed Opt Express, 2018. **9**(6): p. 2526-2542.
120. Nguyen, T.H., et al., *Quantitative phase imaging with partially coherent illumination.* Opt Lett, 2014. **39**(19): p. 5511-4.
121. Soto, J.M., J.A. Rodrigo, and T. Alieva, *Label-free quantitative 3D tomographic imaging for partially coherent light microscopy.* Opt Express, 2017. **25**(14): p. 15699-15712.
122. Kim, T., et al., *White-light diffraction tomography of unlabelled live cells.* Nature Photonics, 2014. **8**(3): p. 256-263.
123. Israelsen, N.M., et al., *Real-time high-resolution mid-infrared optical coherence tomography.* Light Sci Appl, 2019. **8**(1): p. 11.
124. Chen, Z., et al., *Optical doppler tomography.* Ieee Journal of Selected Topics in Quantum Electronics, 1999. **5**(4): p. 1134-1142.
125. Li, J., et al., *Optical diffraction tomography microscopy with transport of intensity equation using a light-emitting diode array.* Optics and Lasers in Engineering, 2017. **95**: p. 26-34.
126. Nguyen, T. and G. Nehmetallah, *Non-Interferometric Tomography of Phase Objects Using Spatial Light Modulators.* Journal of Imaging, 2016. **2**(4): p. 30.
127. Zuo, C., et al., *Lensless phase microscopy and diffraction tomography with multi-angle and multi-wavelength illuminations using a LED matrix.* Opt Express, 2015. **23**(11): p. 14314-28.
128. Horstmeyer, R., et al., *Diffraction tomography with Fourier ptychography.* Optica, 2016. **3**(8): p. 827-835.
129. Hillman, T.R., et al., *High-resolution, wide-field object reconstruction with synthetic aperture Fourier holographic optical microscopy.* Opt Express, 2009. **17**(10): p. 7873-92.
130. Chowdhury, S., et al., *Structured illumination microscopy for dual-modality 3D sub-diffraction resolution fluorescence and refractive-index reconstruction.* Biomed Opt Express, 2017. **8**(12): p. 5776-5793.
131. Kingston, A.M., et al., *Ghost tomography.* Optica, 2018. **5**(12): p. 1516-1520.
132. Amiot, C.G., et al., *Ghost optical coherence tomography.* Opt Express, 2019. **27**(17): p. 24114-24122.
133. Wang, L.V., *Multiscale photoacoustic microscopy and computed tomography.* Nat Photonics, 2009. **3**(9): p. 503-509.
134. Hoshi, Y. and Y. Yamada, *Overview of diffuse optical tomography and its clinical applications.* J Biomed Opt, 2016. **21**(9): p. 091312.

135. Belanger, S., et al., *Real-time diffuse optical tomography based on structured illumination.* J Biomed Opt, 2010. **15**(1): p. 016006.
136. Wang, Z., et al., *Topography and refractometry of nanostructures using spatial light interference microscopy.* Opt Lett, 2010. **35**(2): p. 208-10.
137. Edwards, C., et al., *Effects of spatial coherence in diffraction phase microscopy.* Opt Express, 2014. **22**(5): p. 5133-46.
138. Kim, T., et al., *Deterministic signal associated with a random field.* Opt Express, 2013. **21**(18): p. 20806-20.
139. Nguyen, T.H., et al., *Halo-free Phase Contrast Microscopy.* Sci Rep, 2017. **7**(1): p. 44034.
140. Mehta, S.B. and C.J. Sheppard, *Using the phase-space imager to analyze partially coherent imaging systems: bright-field, phase contrast, differential interference contrast, differential phase contrast, and spiral phase contrast.* Journal of Modern Optics, 2010. **57**(9): p. 718-739.
141. Hopkins, H.H., *On the diffraction theory of optical images.* Proceedings of the Royal Society of London. Series A. Mathematical and Physical Sciences, 1953. **217**(1130): p. 408-432.
142. Kandel, M.E., et al., *Real-time halo correction in phase contrast imaging.* Biomed Opt Express, 2018. **9**(2): p. 623-635.
143. Nguyen, T.H., et al., *Quantitative phase imaging of weakly scattering objects using partially coherent illumination.* Opt Express, 2016. **24**(11): p. 11683-93.
144. Bhaduri, B., et al., *Diffraction phase microscopy: principles and applications in materials and life sciences.* Advances in Optics and Photonics, 2014. **6**(1): p. 57-119.
145. Fanous, M., et al., *Quantifying myelin in brain tissue using color spatial light interference microscopy (cSLIM).* arXiv preprint arXiv:2003.01053, 2020.
146. Rinehart, M., Y. Zhu, and A. Wax, *Quantitative phase spectroscopy.* Biomed Opt Express, 2012. **3**(5): p. 958-65.
147. Pan, F., L. Yang, and W. Xiao, *Coherent noise reduction in digital holographic microscopy by averaging multiple holograms recorded with a multimode laser.* Opt Express, 2017. **25**(18): p. 21815-21825.
148. Savage, N., *Digital spatial light modulators.* Nature Photonics, 2009. **3**(3): p. 170-172.
149. Nguyen, T.H. and G. Popescu, *Spatial Light Interference Microscopy (SLIM) using twisted-nematic liquid-crystal modulation.* Biomedical optics express, 2013. **4**(9): p. 1571-1583.
150. Sarshar, M., T. Lu, and B. Anvari, *Combined optical micromanipulation and interferometric topography (COMMIT).* Biomed Opt Express, 2016. **7**(4): p. 1365-74.
151. Geerts, N., *The Hilbert transform in complex Envelope Displacement Analysis (CEDA).* DCT rapporten, 1996. **1996**.
152. Coquoz, S., et al., *High-speed phase-shifting common-path quantitative phase imaging with a piezoelectric actuator.* J Biomed Opt, 2016. **21**(12): p. 126019.
153. Seniya, C., C.E. Towers, and D.P. Towers, *A flexible quantitative phase imaging microscope for label-free imaging of thick biological specimens using aperture masks.* bioRxiv, 2019: p. 709121.
154. Thalhammer, G., et al., *Speeding up liquid crystal SLMs using overdrive with phase change reduction.* Opt Express, 2013. **21**(2): p. 1779-97.
155. Kandel, M.E., et al., *Label-free tissue scanner for colorectal cancer screening.* J Biomed Opt, 2017. **22**(6): p. 66016.
156. Bhaduri, B., et al., *Cardiomyocyte imaging using real-time spatial light interference microscopy (SLIM).* PLoS One, 2013. **8**(2): p. e56930.
157. Yan, Z., et al., *Three-dimensional mesostructures as high-temperature growth templates, electronic cellular scaffolds, and self-propelled microrobots.* Proc Natl Acad Sci U S A, 2017. **114**(45): p. E9455-E9464.
158. Reynaud, E.G., et al., *Guide to light-sheet microscopy for adventurous biologists.* Nat Methods, 2015. **12**(1): p. 30-4.
159. Stuurman, N. and R.D. Vale, *Impact of New Camera Technologies on Discoveries in Cell Biology.* Biol Bull, 2016. **231**(1): p. 5-13.
160. Kandel, M.E., *High throuput platform for multiscale quantitative phase imaging.* 2016.
161. Fujitsu. *Fujitsu Server PRIMERGYWindows Server 2012 R2 Storage Spaces Performance.* 2014; Available from: https://sp.ts.fujitsu.com/dmsp/Publications/public/wp-windows-storage-spaces-r2-performance-ww-en.pdf.
162. Kandel, M.E., et al., *Three-dimensional intracellular transport in neuron bodies and neurites investigated by label-free dispersion-relation phase spectroscopy.* Cytometry A, 2017. **91**(5): p. 519-526.
163. Sinclair, D.M.M., *Measuring Solid-State Drive Behavior.*
164. Shearer, D., *Samba Architecture: Part II. Samba Basics*, in *SAMBA Developers Guide*, J.R.T.S.T. Vernooij, Editor. 1997.
165. Ou, X., et al., *Quantitative phase imaging via Fourier ptychographic microscopy.* Opt Lett, 2013. **38**(22): p. 4845-8.
166. De Castro, E. and C. Morandi, *Registration of translated and rotated images using finite Fourier transforms.* IEEE Transactions on pattern analysis and machine intelligence, 1987(5): p. 700-703.
167. Preibisch, S., S. Saalfeld, and P. Tomancak, *Globally optimal stitching of tiled 3D microscopic image acquisitions.* Bioinformatics, 2009. **25**(11): p. 1463-5.

168. Majeed, H., et al., *Quantitative Histopathology of Stained Tissues using Color Spatial Light Interference Microscopy (cSLIM).* Sci Rep, 2019. **9**(1): p. 14679.
169. Fanous, M., et al., *Quantitative phase imaging of stromal prognostic markers in pancreatic ductal adenocarcinoma.* Biomed Opt Express, 2020. **11**(3): p. 1354-1364.
170. Li, X., B. Gunturk, and L. Zhang. *Image demosaicing: A systematic survey.* in *Visual Communications and Image Processing 2008.* 2008. International Society for Optics and Photonics.
171. Jiao, Y., et al., *Real-time Jones phase microscopy for studying transparent and birefringent specimens.* Opt Express, 2020. **28**(23): p. 34190-34200.
172. Malvar, H.S., L.-w. He, and R. Cutler. *High-quality linear interpolation for demosaicing of bayer-patterned color images.* in *2004 IEEE International Conference on Acoustics, Speech, and Signal Processing.* 2004. IEEE.
173. Fan, A., et al., *Coupled circumferential and axial tension driven by actin and myosin influences in vivo axon diameter.* Sci Rep, 2017. **7**(1): p. 14188.
174. Mir, M., et al., *Optical measurement of cycle-dependent cell growth.* Proc Natl Acad Sci U S A, 2011. **108**(32): p. 13124-9.
175. Kandel, M.E., et al., *Cell-to-cell influence on growth in large populations.* Biomed Opt Express, 2019. **10**(9): p. 4664-4675.
176. Sridharan, S., M. Mir, and G. Popescu, *Simultaneous optical measurements of cell motility and growth.* Biomed Opt Express, 2011. **2**(10): p. 2815-20.
177. Wang, R., et al., *Dispersion-relation phase spectroscopy of intracellular transport.* Opt Express, 2011. **19**(21): p. 20571-9.
178. Kandel, M.E., et al., *Label-Free Imaging of Single Microtubule Dynamics Using Spatial Light Interference Microscopy.* ACS Nano, 2017. **11**(1): p. 647-655.
179. Mir, M., et al., *Label-free characterization of emerging human neuronal networks.* Sci Rep, 2014. **4**(1): p. 4434.
180. Lee, Y.J., et al., *Quantitative assessment of neural outgrowth using spatial light interference microscopy.* J Biomed Opt, 2017. **22**(6): p. 66015.
181. Cintora, P., et al., *Cell density modulates intracellular mass transport in neural networks.* Cytometry A, 2017. **91**(5): p. 503-509.
182. Yin, C., et al., *Network science characteristics of brain-derived neuronal cultures deciphered from quantitative phase imaging data.* Sci Rep, 2020. **10**(1): p. 15078.
183. Robbin, S., et al., *Pathological basis of disease.* 1999, Har Cocert Asia: Saunders Company.
184. Wang, Z., et al., *Tissue refractive index as marker of disease.* J Biomed Opt, 2011. **16**(11): p. 116017.
185. Ding, H., et al., *Measuring the scattering parameters of tissues from quantitative phase imaging of thin slices.* Opt Lett, 2011. **36**(12): p. 2281-3.
186. Ding, H., et al., *Optical properties of tissues quantified by Fourier-transform light scattering.* Opt Lett, 2009. **34**(9): p. 1372-4.
187. Takabayashi, M., et al., *Tissue spatial correlation as cancer marker.* J Biomed Opt, 2019. **24**(1): p. 1-6.
188. Majeed, H., et al., *Breast cancer diagnosis using spatial light interference microscopy.* J Biomed Opt, 2015. **20**(11): p. 111210.
189. Sridharan, S., et al., *Prediction of prostate cancer recurrence using quantitative phase imaging: Validation on a general population.* Sci Rep, 2016. **6**(1): p. 33818.
190. Sridharan, S., et al., *Prediction of prostate cancer recurrence using quantitative phase imaging.* Sci Rep, 2015. **5**(1): p. 9976.
191. Kandel, M.E., et al., *Reproductive outcomes predicted by phase imaging with computational specificity of spermatozoon ultrastructure.* Proc Natl Acad Sci U S A, 2020. **117**(31): p. 18302-18309.
192. Evans, A.A., et al., *Geometric localization of thermal fluctuations in red blood cells.* Proc Natl Acad Sci U S A, 2017. **114**(11): p. 2865-2870.
193. Park, Y., et al., *Measurement of red blood cell mechanics during morphological changes.* Proc Natl Acad Sci U S A, 2010. **107**(15): p. 6731-6.
194. Park, Y., et al., *Refractive index maps and membrane dynamics of human red blood cells parasitized by Plasmodium falciparum.* Proc Natl Acad Sci U S A, 2008. **105**(37): p. 13730-5.
195. Bhaduri, B., et al., *Optical assay of erythrocyte function in banked blood.* Sci Rep, 2014. **4**(1): p. 6211.
196. Kandel, M.E., et al., *Phase imaging with computational specificity (PICS) for measuring dry mass changes in sub-cellular compartments.* Nature Communications, 2020. **11**(1): p. 6256-6256.
197. Ronneberger, O., P. Fischer, and T. Brox. *U-net: Convolutional networks for biomedical image segmentation.* in *International Conference on Medical image computing and computer-assisted intervention.* 2015. Springer.
198. Krizhevsky, A., I. Sutskever, and G.E. Hinton. *Imagenet classification with deep convolutional neural networks.* in *Advances in neural information processing systems.* 2012.
199. Isola, P., et al. *Image-to-Image Translation with Conditional Adversarial Networks.* in *2017 IEEE Conference on Computer Vision and Pattern Recognition (CVPR).* 2017.
200. Kumar, S.K., *On weight initialization in deep neural networks.* arXiv preprint arXiv:1704.08863, 2017.
201. Nair, V. and G.E. Hinton. *Rectified linear units improve restricted boltzmann machines.* in *Proceedings of the 27th international conference on machine learning (ICML-10).* 2010.

202. Ramachandran, P., B. Zoph, and Q.V. Le, *Swish: a self-gated activation function.* arXiv preprint arXiv:1710.05941, 2017. **7**.
203. Lichtman, J.W. and J.A. Conchello, *Fluorescence microscopy.* Nat Methods, 2005. **2**(12): p. 910-9.
204. Moffat, J.G., J. Rudolph, and D. Bailey, *Phenotypic screening in cancer drug discovery—past, present and future.* Nature reviews Drug discovery, 2014. **13**(8): p. 588-602.
205. Rawat, W. and Z. Wang, *Deep Convolutional Neural Networks for Image Classification: A Comprehensive Review.* Neural Comput, 2017. **29**(9): p. 2352-2449.
206. Isola, P., et al. *Image-to-image translation with conditional adversarial networks.* in *Proceedings of the IEEE conference on computer vision and pattern recognition*. 2017.
207. Christiansen, E.M., et al., *In Silico Labeling: Predicting Fluorescent Labels in Unlabeled Images.* Cell, 2018. **173**(3): p. 792-803 e19.
208. Rivenson, Y., et al., *Virtual histological staining of unlabelled tissue-autofluorescence images via deep learning.* Nat Biomed Eng, 2019. **3**(6): p. 466-477.
209. Sung, W.W., et al., *High nuclear/cytoplasmic ratio of Cdk1 expression predicts poor prognosis in colorectal cancer patients.* BMC Cancer, 2014. **14**: p. 951.
210. Guo, S.M., et al., *Revealing architectural order with quantitative label-free imaging and deep learning.* Elife, 2020. **9**: p. e55502.
211. Rubessa, M., et al., *Morphometric analysis of sperm used for IVP by three different separation methods with spatial light interference microscopy.* Syst Biol Reprod Med, 2020. **66**(1): p. 26-36.
212. Dehmelt, L. and S. Halpain, *The MAP2/Tau family of microtubule-associated proteins.* Genome Biol, 2005. **6**(1): p. 204.
213. Hu, C., et al., *Label-free cell viability assay using phase imaging with computational specificity.* bioRxiv, 2020: p. 2020.10.28.359554.
214. Tan, M. and Q.V. Le, *Efficientnet: Rethinking model scaling for convolutional neural networks.* arXiv preprint arXiv:1905.11946, 2019.
215. Paxton, S., M. Peckham, and A. Knibbs, *The Leeds Histology Guide.* 2003.
216. Berwick, R., *An idiot's guide to support vector machines.* Advanced Computer Vision. University of Central Florida, 2003.
217. Simonyan, K. and A. Zisserman, *Very deep convolutional networks for large-scale image recognition.* arXiv preprint arXiv:1409.1556, 2014.
218. Zhang, J.K., et al., *Label-free colorectal cancer screening using deep learning and spatial light interference microscopy (SLIM).* APL Photonics, 2020. **5**(4): p. 040805.
219. Nguyen, T.H., et al., *Automatic Gleason grading of prostate cancer using quantitative phase imaging and machine learning.* J Biomed Opt, 2017. **22**(3): p. 36015.
220. Van Steirteghem, A.C., et al., *High fertilization and implantation rates after intracytoplasmic sperm injection.* Hum Reprod, 1993. **8**(7): p. 1061-6.
221. Kovac, J.R. and L.I. Lipshultz, *Sperm morphology and reproductive success.* Asian J Androl, 2016. **18**(3): p. 402.
222. Yi, X., E. Walia, and P. Babyn, *Generative adversarial network in medical imaging: A review.* Med Image Anal, 2019. **58**: p. 101552.